\documentclass[12pt]{article}

\ifx\pdfoutput\undefined
\usepackage[dvips,bookmarks]{hyperref}
\else
\usepackage{hyperref}
\fi
\hypersetup{colorlinks=false,bookmarksopen,bookmarksnumbered,citecolor=blue,
   pdfstartview=FitH}

\usepackage[dvips]{graphicx}
\usepackage{latexsym}
\usepackage{amssymb,amsfonts,amsmath}
\usepackage{graphicx}
\usepackage{indentfirst}
 \usepackage{bbm}
\usepackage{mathrsfs}

\usepackage{amsmath, amsthm,amssymb}
\usepackage{mathrsfs}
\usepackage{hyperref}
\usepackage{amsfonts}
\usepackage{dsfont}
\usepackage{slashed, tensor}
\usepackage{booktabs}
\usepackage{graphicx}
\usepackage{mathrsfs}

\parskip=\medskipamount

\newcommand {\cA}{{\cal A}}
\newcommand {\cB}{{\cal B}}

\newcommand {\cI}{{\cal I}}
\newcommand {\cJ}{{\cal J}}

\newcommand {\cL}{{\cal L}}

\newcommand {\cN}{{\cal N}}
\newcommand {\cO}{{\cal O}}

\def\a{\alpha}
\def \bi{\bibitem}

\def\b{\beta}

\def\d{\delta}

\def\f{\phi}
\def\g{\gamma}

\def\l{\lambda}

\def\p{\pi}
\def\q{\theta}
\def\r{\rho}
\def\s{\sigma}
\def\t{\tau}

\def\x{\xi}
\def\z{\zeta}

\def\F{\Phi}
\def\J{\Psi}
\def\L{\Lambda}

\def\P{\Pi}

\def\S{\Sigma}
\def\U{\Upsilon}

\def\tr{{\rm tr}}
\def\rd{{\rm d}}
\def\ri{{\rm i}}
\def\re{{\rm e}}

\newcommand{\ve}{\varepsilon}                            

\newcommand{\pa}{\partial}                           
\newcommand{\hf}{\frac12}

\newcommand{\vf}{\varphi}

\newcommand{\be}{\begin{equation}}
\newcommand{\ee}{\end{equation}}
\newcommand{\bea}{\begin{eqnarray}}
\newcommand{\eea}{\end{eqnarray}}
\newcommand{\non}{\nonumber}
\newcommand{\ba}{\begin{array}}
\newcommand{\ea}{\end{array}}

\newcommand{\bm}[1]{\mbox{\boldmath$#1$}}

\def\double #1{#1{\hbox{\kern-2pt $#1$}}}

\newcommand{\sSU}{\mathsf{SU}}

\newcommand{\sSO}{\mathsf{SO}}
\newcommand{\sO}{\mathsf{O}}
\newcommand{\sU}{\mathsf{U}}

\newcommand{\sOSp}{\mathsf{OSp}}

\newcommand{\bsubeq}{\begin{subequations}}
\newcommand{\esubeq}{\end{subequations}}

\numberwithin{equation}{section}

\begin{document}

\begin{titlepage}
\begin{flushright}
July, 2015
\end{flushright}
\vspace{2mm}

\begin{center}
{\Large \bf Implications of $\bm{\cN=4}$ superconformal symmetry\\
in three spacetime dimensions
}\\
\end{center}

\begin{center}

{\bf
Evgeny I. Buchbinder, Sergei M. Kuzenko and Igor B.
Samsonov\footnote{On leave from Tomsk Polytechnic University, 634050
Tomsk, Russia.}
}

{\footnotesize{
{\it School of Physics M013, The University of Western Australia\\
35 Stirling Highway, Crawley W.A. 6009, Australia}} ~\\
}

\end{center}

\begin{abstract}
\baselineskip=14pt

We study implications of $\cN=4$ superconformal symmetry 
in three dimensions, thus 
extending our earlier results in arXiv: 1503.04961 devoted to the $\cN \leq 3$
cases. We show that the three-point function of the supercurrent in ${\cal N}=4$ superconformal field theories contains two linearly independent forms. 
However, only one of these structures contributes to the 
three-point function of the energy-momentum tensor and 
the other one is present in those $\cN=4$ superconformal theories 
which are not invariant under the mirror map. We point out that 
general  ${\cal N} = 4$ superconformal field theories 
 admit two inequivalent flavour current multiplets
and show that the three-point function of each of them
is determined by
one tensor structure. As an example, we compute the two- and three-point functions of the conserved currents 
in ${\cal N}=4$ superconformal models of free hypermultiplets. 
We also derive the {\it universal} relations between 
the coefficients appearing in  the two- and three-point correlators
of the supercurrent and flavour current multiplets in all superconformal theories 
with ${\cal N} \leq 4$ supersymmetry. 
Our derivation is based on the use of Ward identities in conjunction 
with superspace reduction techniques. 
\end{abstract}

\vspace{1cm}
\begin{flushright}
{\it Dedicated to the memory of Professor Boris M. Zupnik}\\[5mm]
\end{flushright}

\end{titlepage}

\newpage
\renewcommand{\thefootnote}{\arabic{footnote}}
\setcounter{footnote}{0}

\tableofcontents
\vspace{1cm}
\bigskip\hrule

\section{Introduction}

In our recent work \cite{BKS},
the two- and three-point correlation functions of
the supercurrent and flavour current multiplets have been computed
for three-dimensional (3D) $\cN$-extended superconformal field theories
with $1\leq \cN \leq 3$.
Here we extend the analysis of \cite{BKS} to the $\cN=4$ case.
We also study the reduction of correlation functions in $\cN$-extended
superconformal field theory to $(\cN-1)$-extended superspace.

In two dimensions,
$\cN=3$ supersymmetry automatically implies $\cN=4$
\cite{A-GF,Bagger}
for nonlinear $\s$-models.\footnote{The
proof given in  \cite{A-GF,Bagger} is as follows. 
It is known that  $\cN$-extended supersymmetry requires
the existence of $\cN-1$ anti-commuting complex structures for the
$\s$-model target space.
In the case $\cN=3$, the target space has two such structures, $I$ and $J$.
Their product, $K:=I \,J$,
is a third complex structure which anti-commutes with $I$ and $J$, and therefore
the $\sigma$-model is  $\cN=4$ supersymmetric.}
What about three dimensions?
As far as the supersymmetric nonlinear $\s$-models are concerned,
3D $\cN=3$ supersymmetry again implies $\cN=4$.
Indeed, the proof given in \cite{A-GF,Bagger}
remains valid in three dimensions.
Moreover,  off-shell $\cN=3$ supersymmetric
$\s$-models can be shown to possess off-shell $\cN=4$ supersymmetry \cite{KPT-MvU}.
Analogous results hold for $\cN=3$ super Yang-Mills theories with matter \cite{Zupnik2010}.
However,  a rather counter-intuitive situation
occurs with parity odd Chern-Simons terms.
The $\cN=3$ Chern-Simons action\footnote{The 
$\cN=3$ Chern-Simons action was constructed for the first time
by Zupnik and Hetselius  \cite{ZH} in 3D $\cN=3$ harmonic superspace,
and several years later
it was re-discovered \cite{KaoLee,Kao} at the component level.}
exists for any  gauge
group \cite{ZH,KaoLee,Kao}.
On the other hand, it is well known that no  $\cN=4$
supersymmetric Chern-Simons action can be constructed (for a
recent proof, see \cite{KN14}), although abelian $\cN=4$ BF
couplings are abundant \cite{BrooksG}.

The $R$-symmetry group is (locally isomorphic to) $\sSU(2) $
in $\cN=3$ supersymmetry and $\sSU(2)_{\rm L} \times \sSU(2)_{\rm R} $ for $\cN=4$.
This difference implies that there are two inequivalent $\cN=4$ gauge multiplets
\cite{BrooksG,Zupnik98,Zupnik99}
and two inequivalent $\cN=4$ hypermultiplets \cite{Zupnik98,Zupnik99},
as compared with a single vector multiplet and a single hypermultiplet in
$\cN=3$ supersymmetry. The inequivalent $\cN=4$ vector multiplets
obey different off-shell constraints and transform in different representations of
the $R$-symmetry group,
and similarly for the inequivalent hypermultiplets.\footnote{The inequivalent
vector multiplets and hypermultiplets can be described in terms of superfields
that are defined on two different supersymmetric subspaces of the $\cN=4$ harmonic superspace
 \cite{Zupnik98,Zupnik99}.}

The doubling of gauge and matter multiplets in $\cN=4$ supersymmetry
has important implications on the structure of $\cN=4$  superconformal field theory.
First of all, there are two inequivalent $\cN=4$ flavour current
multiplets whereas there is only one in superconformal models with $\cN=1,2,3$
supersymmetry which were considered in \cite{BKS}.
Secondly, we will demonstrate that the three-point function of the $\cN=4$
supercurrent has two independent structures, as compared with
a single structure in the  $\cN=1,2,3$ cases studied in \cite{BKS}.

The zoo of $\cN=4$ superconformal field theories in three dimensions is pretty large.
The trivial examples of such theories are provided by models of free  hypermultiplets.
More interesting are interacting models of hypermultiplets coupled to 
vector multiplets, with BF couplings for the vector multiplets.
The non-abelian $\cN=4$ superconformal field theories include the
Gaiotto-Witten models \cite{GW} and their generalisations \cite{HLLLP}.
For all abelian  $\cN=4$ superconformal field theories, there  exist off-shell realisations.
As concerns the non-abelian $\cN=4$ superconformal theories proposed in
\cite{GW,HLLLP}, it is not yet known how to formulate them in
$\cN=4$ superspace, which is an interesting open problem.
There also exist Chern-Simons-matter theories with  $\cN=6$ 
\cite{ABJM,HLLLP2,ABJ} and $\cN=8$  \cite{BLG1,BLG2,Gustavsson} 
superconformal symmetry. 
The correlation functions of certain conserved currents in these theories
can be studied using  the $\cN=4$ superfield methods developed in the
present paper.

This paper is organised as follows.
In section 2 we give a brief review of the superconformal
building blocks  for the two- and three-point correlation functions 
in 3D $\cN$-extended superspace following the
conventions and notation used in \cite{BKS}. We also elaborate on those
properties of the building blocks which are specific to the 
$\cN=3$ and $\cN=4$ cases. 
In section 3 we develop a new representation for  the correlation functions of
the $\cN=3$ flavour current multiplets 
originally computed in \cite{BKS}.
This representation allows us to easily upgrade the $\cN=3$ flavour current
correlators to the $\cN=4$ ones which are derived in section 4. 
Here we also construct two- and three-point functions of the $\cN=4$
supercurrent and demonstrate that the latter involves two independent
tensor structures which distinguish the $\cN=4$ supercurrent correlators
from those in the  $\cN=1,2,3$ cases. In section 5
we consider a particular example of $\cN=4$ superconformal field theories 
given by the model of free $\cN=4$
hypermultiplets for which we  explicitly compute the correlation
functions of the supercurrent and the flavour current multiplets. For this
model we find important relations between the coefficients in the two- and
three-point functions which are interpreted as the manifestations of Ward
identities for these correlators. We argue that though these relations
between the coefficients are found for the particular model of free
hypermultiplets, they hold for generic $\cN=4$ superconformal field theories as
well. Section 6 is devoted to the derivation of 
the Ward identities for the
$1\leq\cN\leq 4$ flavour current multiplets. In section 7 we uncover
various relations between the coefficients in the  two-point and
three-point functions both for the supercurrents and flavour current
multiplets for all $\cN\leq4$. 
Finally, in section 8 we discuss the results and some open problems.

The main body of the paper is accompanied by three technical appendices. 
Appendix A is devoted  to a  brief review of 3D off-shell $\cN=4$ multiplets.
In appendix B we use the 3D  $\cN=4$ harmonic superspace approach
to derive a new representation for the $q^+$ hypermultiplet propagator which is
important in studying the implications of the Ward identity for the
correlation functions of the $\cN=4$ flavour current multiplets. 
In appendix C we collect some details of the reduction
of the $\cN=4$ correlation functions computed in this paper to 
the $\cN=3$ and $\cN=2$ superspaces. 


\section{Superconformal building blocks} \label{section2}

This section contains a brief summary of those results in \cite{BKS}
which are necessary for our subsequent analysis. In addition, we
elaborate on specific technical features of the $\cN=3$ and $\cN=4$ cases.

\subsection{Superconformal transformations and primary superfields}

Consider $\cN$-extended Minkowski superspace
${\mathbb M}{}^{3|2\cN}$
parametrised by real bosonic $(x^a)$ and fermionic $(\q^\a_I)$
coordinates
$$
z^A=(x^a,\theta^{\alpha}_I), \qquad a =0,1,2~, \qquad
\a = 1,2~, \qquad I = 1, \dots, \cN~.
$$
Here the indices `$a$' and `$\a$' are Lorentz and spinor ones, respectively,
while  `$I$' is the $R$-symmetry index.
The 3D $\cN$-extended superconformal group $\sOSp(\cN|4;{\mathbb R})$
cannot be realised to act by smooth transformations on ${\mathbb M}{}^{3|2\cN}$.
However, a transitive  action of
$\sOSp(\cN|4;{\mathbb R})$
is naturally defined on the so-called compactified Minkowski superspace
$\overline{\mathbb M}{}^{3|2\cN}$ in which
${\mathbb M}{}^{3|2\cN}$ is embedded as a dense open domain \cite{KPT-MvU}.
In general, only infinitesimal superconformal transformations are  well defined on
${\mathbb M}{}^{3|2\cN}$.
Such a transformation
\bea
\d z^A = \x z^A \quad \Longleftrightarrow \quad
\d x^a = \x^a (z) +{\rm i}\, \x^\a_I (z) \q^\b_I  (\g^a)_{\a\b} ~, \qquad
\d \q^\a_I = \x^\a_I (z)
\eea
is associated with
an even real supervector field on ${\mathbb M}{}^{3|2\cN}$~,
\bea
\x  = \x^A  D_A = \x^{a}   \pa_{ a}
+  \x^{\a}_I  D_{\a}^I = -\hf \x^{\a\b} \pa_{\a\b} +  \x^{\a}_I  D_{\a}^I ~,
\qquad \overline{\x^A} =\x^A~,
\label{xi}
\eea
which obeys the equation $[\x, D_\a^I ] \propto D_\b^J$. All solutions of this
equation are called the conformal Killing supervector fields of Minkowski superspace.
They span a Lie superalgebra (with respect to the standard Lie bracket
$[\x_1, \x_2]$) that is isomorphic to
the superconformal algebra $\frak{osp}(\cN|4;{\mathbb R})$.

Explicit expressions for the components $\x^A = (\x^a , \x^\a_I)$
of the most general conformal Killing supervector field
are given by eq.\ (4.4) in
\cite{BKS}. Equivalent results were derived earlier by Park \cite{Park3}
and later in \cite{KPT-MvU}. In the present paper, we will not need these
explicit expressions.
For our analysis, it suffices to use the relation
\be
[\x, D_\a^I ] = -(D^I_\a \x^\b_J) D^J_\b = \l_\a{}^\b (z)D_\b^I
+ \L^{IJ}(z) D_\a^J -\hf \s (z) D^I_\a~.
\label{master1}
\ee
Here the superfield parameters on the right
are expressed in terms of $\x^A$ as follows:
\bea
\l_{\alpha\beta} (z) =-\frac1{\cN}D^I_{(\alpha}\xi^I_{\beta)}~,\quad
\Lambda^{IJ}(z) =-2D_\alpha^{[I}\xi^{J]\alpha}~,\quad
\sigma (z)
=\frac1{\cN}D^I_\alpha\xi_I^\alpha=\frac13\partial_a\xi^a~.
\label{2.4par}
\eea
One may think of $\l_{\alpha\beta} (z) $, $\Lambda^{IJ}(z)$
and $\s(z)$
as the parameters of special local Lorentz, $R$-symmetry
and scale transformations, respectively,
due to their  action on the covariant derivative given by \eqref{master1}.
The same interpretation is supported by
the explicit expressions for $\l_{\alpha\beta} (z) $, $\Lambda^{IJ}(z)$
and $\s(z)$ as polynomials in $z^A$:
\begin{subequations}\label{Local-param}
\bea
\l^{\alpha\beta} (z)&=&\lambda^{\alpha\beta}-x^{\gamma(\alpha}b_\gamma^{\beta)}
-\frac{\ri}2 b^{\alpha\beta}\theta^\gamma_I \theta_{I\gamma}
+2\ri\eta_I^{(\alpha}\theta_I^{\beta)}~,\\
\Lambda_{IJ} (z) &=&\L_{IJ}
+4 \ri\eta_{[I}^\alpha \theta_{J]\alpha}
+2\ri b_{\alpha\beta}\theta_I^\alpha \theta_J^\beta~, \\
\sigma (z)&=&\s+b_{\alpha\beta}x^{\beta\alpha}+2\ri\theta^\alpha_I
\eta_{I\alpha}~.
\eea
\end{subequations}
Here the constant parameters $\l_{\alpha\beta}  $, $\Lambda^{IJ}$ and $\s$
correspond to the Lorentz, $R$-symmetry and scale transformations
from $\sOSp(\cN|4;{\mathbb R})$, while  $b^{\a \b} $ and $\eta_{I\a}$
generate the special conformal and $S$-supersymmetry transformations.

It is the $z$-dependent parameters \eqref{2.4par} which
appear, along with $\x$ itself, in the superconformal transformation law\footnote{The
transformation law \eqref{primaryTL} is a 3D super-extension of the
Mack-Salam construction \cite{MackSalam}.}
of
a primary tensor superfield of dimension $q$
\bea
\delta\Phi_{\cal A}^{\cal I} =
-\xi\Phi_{\cal A}^{\cal I}-q\,\sigma(z)\Phi_{\cal A}^{\cal I}
+\lambda^{\alpha\beta}(z) (M_{\alpha\beta})_{\cal A}{}^{\cal B}
\Phi_{\cal B}^{\cal I}
+\Lambda_{IJ}(z)(R^{IJ})^{\cal I}{}_{\cal J}\Phi_{\cal A}^{\cal
J}~.
\label{primaryTL}
\eea
Here
$\Phi_{\cal A}^{\cal I}$
is assumed to transform in some representations of the Lorentz and $R$-symmetry
groups with respect to its indices `$\cA$' and `$\cI$', respectively.
The matrices $M_{\alpha\beta}$ and $R^{IJ}$ in \eqref{primaryTL}
are the Lorentz and $\sSO(\cN)$ generators, respectively.
It should be mentioned that the $R$-symmetry subgroup of
 $\sOSp(\cN|4;{\mathbb R})$ is $\sO(\cN)$. In what follows,
 its connected component of the identity, $\sSO(\cN)$, will be referred to as
 the $R$-symmetry group.

Consider a correlation function
$ \langle \F_1 (z_1) \dots  \F_n(z_n)\rangle$ of  several primary superfields
 $\F_1$, $\dots $, $\F_n$   (with their indices suppressed)
that originate in some superconformal field theory.  In terms of this
correlation function,
the statement of superconformal invariance is
\bea
\sum_{k=1}^n \langle \F_1 (z_1) \dots \d \F_k (z_k) \dots \F_n(z_n)\rangle = 0~.
\label{2.77}
\eea

\subsection{Two-point functions}

In ordinary conformal field theory in $d$ dimensions, a comprehensive discussion of
the building blocks for the two- and three-point correlation functions of primary fields
was given by Osborn and Petkou \cite{OP} who built on the earlier works by Mack \cite{Mack}
and others \cite{FGG,Koller,TMP,FP,Stanev}.
Their analysis was extended
to superconformal field theories formulated in superspace by Osborn and Park
\cite{Osborn,Park6,Park4,Park3}.

In the case of 3D superconformal field theories,
the building blocks for the two- and three-point correlation functions
were derived first in  \cite{Park3} using the coset construction for
$\sOSp(\cN|4;{\mathbb R})$ and
more recently in \cite{BKS} using the supertwistor approach.
All building blocks are composed of the following
two-point structures:
\begin{subequations}
\bea
{\bm x}_{12}^{\alpha\beta} &=& (x_{1}-x_{2})^{\alpha\beta}
+2\ri \theta_{1I}^{(\alpha}\theta_{2I}^{\beta)}
-\ri\theta_{12I}^\alpha \theta_{12I}^\beta~,\label{super-interv-X}\\
\theta_{12I}^\alpha &=& (\theta_{1}-\theta_2)^\alpha_{I}~.
\label{super-interv-Theta}
\eea
\label{super-interv}
\end{subequations}
The former transforms homogeneously at $z_1$ and $z_2$,
\begin{subequations}\label{2pt-transf}
\bea
\widetilde{\delta} {\bm x}_{12}^{\alpha\beta} &=& \left(\frac12\delta^\alpha{}_\gamma
 \sigma(z_1) -\l^\alpha{}_\gamma(z_1) \right){\bm x}_{12}^{\gamma\beta}
+{\bm x}_{12}^{\alpha\gamma}\left(
\frac12\delta_\gamma{}^\beta \sigma(z_2) -\l_\gamma{}^\beta(z_2)
\right)~,
\label{2.9aa}
\eea
while the latter involves an inhomogeneous piece in its transformation law,
\bea
\widetilde{\delta}\theta^\alpha_{12I}&=&
\left(\frac12\delta^\alpha{}_\beta\sigma(z_1)
 -\l^\alpha{}_\beta(z_1)\right)\theta^\beta_{12I}
+ \Lambda_{IJ}(z_2)\theta^\alpha_{12J}
 -{\bm x}_{12}^{\alpha\beta}\eta_{I\beta}(z_2)
 ~,
 \label{2.7b}
\eea
\end{subequations}
with $\eta_{I\alpha}(z):=-\frac\ri2 D^I_\alpha \sigma(z)
=\eta_{I\alpha} - b_{\alpha\beta}\theta^\beta_I$.
Here the variation $\widetilde{\delta}$  is defined to act on
an arbitrary $n$-point function  $\cO(z_1, \dots , z_n) $
by the rule
\bea
\widetilde{\delta} \cO (z_1, \dots, z_n) = \sum_{i=1}^n \x_{z_i} \cO (z_1, \dots, z_n)
~.
\eea

As follows from \eqref{primaryTL}, each primary superfield $\F$ is determined
by the following data: (i) its dimension $q$;
(ii) the representation $T$ of the Lorentz group to which $\F$ belongs;
and (iii) the representation $D$ of $\sSO(\cN)$ in which $\F$ transforms.
There are three building blocks, which are descendants of \eqref{super-interv}
and which take care of the above data in the correlation functions
of primary superfields.

Firstly,  using (\ref{super-interv-X}) we define the scalar two-point function
\be
{\bm x}_{12}{}^2 := -\frac12{\bm x}_{12}^{\alpha\beta}{\bm
x}_{12\alpha\beta}
\label{4.133}
\ee
with the transformation law
\be
\widetilde{\delta} {\bm x}_{12}{}^2 = \Big(\sigma(z_1)+ \sigma(z_2) \Big) {\bm
x}_{12}{}^2~.
\label{delta-x2}
\ee
In general, the correlation functions contain multiplicative factors
proportional to  powers of ${\bm x}_{12}{}^2 $
in such a way as to guarantee the right scaling properties.

Secondly, the Lorentz structure of the primary fields
in correlation functions is taken care of by the $2\times 2$ matrix
\bea
\underline{\hat{\bm x}}_{12}&:=&\frac{\hat{\bm x}_{12}}{
\sqrt{-{\bm x}_{12}{}^2}}~, \qquad
(\ve \underline{\hat{\bm x}}_{12})^2 = {\mathbbm 1}_2~,
\eea
where we have used the matrix notation
$\hat{\bm x}_{12} = ({\bm x}_{12}^{\a\b})$ and
 $\ve =(\ve_{\a \b})$. Its transformation law is
 \bea
 \widetilde\delta \underline{\bm x}_{12}^{\alpha\beta}
&=& -\lambda^\alpha{}_\gamma(z_1)\, \underline{\bm
x}_{12}^{\gamma\beta}-\underline{\bm x}_{12}^{\alpha\gamma}\,
\lambda_\gamma{}^\beta(z_2)~.
\eea

Thirdly, the $\sSO(\cN)$ structure of the primary fields
in correlation functions is taken care of by
the $\cN\times \cN$ matrix
\be
u_{12} =(u_{12}^{IJ})~, \qquad
u_{12}^{IJ} = \delta^{IJ} + 2\ri
\theta^{ \a I}_{12}({\bm x}_{12}^{-1})_{\alpha\beta}
\theta^{\b J}_{12}~,
\label{two-point-u}
\ee
where
\be
({\bm x}_{12}^{-1})_{\alpha\beta} = -\frac{
{\bm x}_{12\beta\alpha}}{{\bm x}_{12}{}^2}
\ee
is the inverse for $({\bm x}_{12})^{\alpha\beta}$,
that is $({\bm x}_{12}^{-1})_{\alpha\beta}({\bm x}_{12})^{\beta\gamma}
=\delta_\alpha^\gamma$.
One may check that the matrix
$u_{12}$ is orthogonal and unimodular,
\be
u_{12}^{\rm T}u_{12} = {\mathbbm 1}_\cN~,\qquad
\det u_{12} = 1~.
\label{2.17}
\ee
It follows from (\ref{2pt-transf}) that
\be
\widetilde{\delta} u_{12}^{IJ} = \Lambda^{IK}(z_1)u_{12}^{KJ} - u_{12}^{IK}\Lambda^{KJ}(z_2)~.
\label{deltaU}
\ee
The above properties provide the rationale why
$ u_{12}^{IJ} $ naturally arises in correlation functions
of primary superfields with $\sSO(\cN)$ indices.

The two-point correlation function of the primary superfield
$\Phi_{\cal A}^{\cal I}$ and its conjugate $\bar \F^\cA_\cI $
is fixed by the superconformal symmetry up to a single coefficient $c$ and has the form
\bea
\langle
\Phi^{\cal I}_{\cal A}(z_1)
\bar{\Phi}_{\cal J}^{\cal B}(z_2)
\rangle=c\frac{
T_{\cal A}{}^\cB(\ve \underline{\hat{\bm x}}_{12})
{D}^{\cal I}{}_\cJ(u_{12})
}{({\bm x}_{12}{}^2)^q}~
\label{OO}
\eea
provided the representations $T$ and $D$ are irreducible.
The denominator in (\ref{OO}) is
fixed by the dimension of $\F$.

Before turning to three-point building blocks, it should be pointed out
that the two-point structure ${\bm x}_{12}^{\alpha\beta}$
defined by \eqref{super-interv-X} has the following
symmetry property
\be
{\bm x}_{21}^{\alpha\beta} = -{\bm x}_{12}^{\beta\alpha}~.
\label{antisymmetry}
\ee
It can be decomposed into its symmetric and antisymmetric parts,
\be
{\bm x}_{12}^{\alpha\beta}=
x_{12}^{\alpha\beta}+ \frac{\ri}2\varepsilon^{\alpha\beta} \theta_{12}{}^2~,
\label{x12-sym}
\ee
where
\be
\theta_{12}{}^2 := \theta_{12I}^\alpha \theta_{12I\alpha}~, \qquad
x_{12}^{\alpha\beta} = x_{12}^{\beta \a}
:=(x_{1}-x_{2})^{\alpha\beta}+2\ri\theta_{1I}^{(\alpha}\theta_{2I}^{\beta)}
~.
\label{x12-sym1}
\ee
As is seen from \eqref{2.9aa}, the two-point structure $x_{12}^{\alpha\beta}$
does not transform homogeneously, unlike ${\bm x}_{12}^{\alpha\beta}$.
However,  in practice it is often useful to deal with
${x}_{12}^{\alpha\beta}$ for concrete calculations.

\subsection{Three-point functions}

Associated with three superspace points $z_1$, $z_2$ and $z_3$
are the following three-point structures:
\begin{subequations}
\label{three-points}
\bea
{\bm X}_{1\alpha\beta}&=&-({\bm x}^{-1}_{21})_{\alpha\gamma}
{\bm x}_{23}^{\gamma\delta}
({\bm x}^{-1}_{13})_{\delta\beta}~,\label{3ptX}\\
\Theta^I_{1\alpha}&=& ({\bm x}^{-1}_{21})_{\alpha\beta}\theta^{I\beta}_{12}
-({\bm x}^{-1}_{31})_{\alpha\beta}\theta_{13}^{I\beta}~,\label{3ptTheta}\\
U_{1}^{IJ}&=&u_{12}^{IK}u_{23}^{KL}u_{31}^{LJ}~.
\label{U}
\eea
\end{subequations}
They transform as tensors
at the  point $z_1$
\begin{subequations}
\bea
\widetilde{\delta}
{\bm X}_{1\alpha\beta}&=&\lambda_\alpha{}^\gamma(z_1) {\bm X}_{1\gamma\beta}
+{\bm X}_{1\alpha\gamma}\lambda^\gamma{}_\beta(z_1)-
\sigma(z_1) {\bm X}_{1\alpha\beta} ~,\\
\widetilde{\delta} \Theta_{1\alpha}^I
&=&\left(\lambda_\alpha{}^\beta(z_1)-\frac12\delta_\alpha{}^\beta\sigma(z_1)
\right)\Theta_{1\beta}^I
+\Lambda^{IJ}(z_1)\Theta^J_{1\alpha}~,\\
\widetilde{\delta} U^{IJ}_{1} &=& \Lambda^{IK}(z_1)U_{1}^{KJ} - U_{1}^{IK}\Lambda^{KJ}(z_1)~.
\eea
\end{subequations}
These objects have many properties resembling those
of the two-point functions. In particular, the tensor
(\ref{3ptX}) can be decomposed into symmetric and antisymmetric
parts similar to (\ref{x12-sym}),
\be
{\bm X}_{1\alpha\beta} = X_{1\alpha\beta}
-\frac{\ri}2\varepsilon_{\alpha\beta}\Theta_{1}{}^2~,
\label{XX}
\ee
where  the symmetric spinor
$X_{1\alpha\beta}=X_{1\beta \a}$ is equivalently represented as
a three-vector $X_{1m}=-\frac12 \gamma_m^{\alpha\beta}X_{1\alpha\beta}$.

Next, the matrix (\ref{U}) can be expressed in terms of
(\ref{3ptX}) and (\ref{3ptTheta}) similarly to
(\ref{two-point-u}):
\be
U_{1}^{IJ}=\delta^{IJ}
+2\ri \Theta^I_{1\alpha}({\bm X}_{1}^{-1})^{\alpha\beta}\Theta^J_{1\beta}
=\delta^{IJ}-2\ri\frac{\Theta^I_{1\alpha}{\bm X}_{1}^{\beta\alpha}\Theta^J_{1\beta}}{{\bm X}_{1}{}^2}
~.
\label{U-explicit}
\ee
The matrix $U_1 = (U_{1}^{IJ})$ is  orthogonal and unimodular.

We point out that in (\ref{three-points}) we have defined the
three-point structures which transform as tensors at the
 point $z_1$. Performing cyclic permutations of the
superspace points $z_1$, $z_2$ and $z_3$ in (\ref{three-points})
 one obtains similar
objects which transform as tensors at the superspace points $z_2$
and $z_3$. The three-point structures at different superspace
points are related to each other as follows
\begin{subequations}
\bea
{\bm x}_{13}^{\alpha\alpha'}{\bm X}_{3\alpha'\beta'}{\bm x}_{31}^{\beta'\beta}
&=&-({\bm X}_{1}^{-1})^{\beta\alpha}=\frac{{\bm X}_{1}^{\alpha\beta}}{{\bm
X}_{1}{}^2}~,\label{222a}\\
\Theta^I_{1\gamma} {\bm x}_{13}^{\gamma\delta}{\bm X}_{3\delta\beta}
&=&u^{IJ}_{13} \Theta^J_{3\beta}~,\label{222b}\\
U_{3}^{IJ} &=& u_{31}^{IK}U_{1}^{KL}u_{13}^{LJ}~.
\label{222c}
\eea
\end{subequations}

Various primary superfields, including the supercurrent, obey certain differential
constraints.
In order to take into account these constraints
in correlation functions, we need rules to evaluate covariant derivatives
of the variables (\ref{3ptX}) and (\ref{3ptTheta}) and also those obtained from
them by cyclic permutations of the superspace points $z_1$, $z_2$ and $z_3$.
Given a function $f({\bm X}_{3},\Theta_{3})$,
one can prove the following differential identities:
\begin{subequations}
\label{useful-prop}
\bea
D_{(1)\gamma}^I f({\bm X}_{3},\Theta_{3}) &=&
({\bm x}^{-1}_{13})_{\alpha\gamma}u^{IJ}_{13}
 {\cal D}^{J\alpha}_{(3)} f({\bm X}_{3},\Theta_{3})~,
 \label{useful-prop-a}\\
D_{(2)\gamma}^I f({\bm X}_{3},\Theta_{3})
&=& \ri ({\bm x}^{-1}_{23})_{\alpha\gamma}u^{IJ}_{23}
 {\cal Q}^{J\alpha}_{(3)} f({\bm X}_{3},\Theta_{3})~,
\label{useful-prop-b}
\eea
\end{subequations}
where
we have introduced the operators
\be
{\cal D}^I_{(3)\alpha} = \frac\partial{\partial \Theta_{3I}^\alpha }
 +\ri\gamma^m_{\alpha\beta}\Theta_{3}^{I\beta} \frac\partial{\partial X^m_{3}}~,\qquad
{\cal Q}^I_{(3)\alpha} =\ri\frac\partial{\partial \Theta_{3I}^\alpha }
 +\gamma^m_{\alpha\beta} \Theta_{3}^{I\beta} \frac\partial{\partial X^m_{3}}
 ~.
\label{generalized-DQ}
\ee

Let
$\Phi$, $\J$ and $\P$
be primary superfields (with indices suppressed)
of dimensions $q_1$, $q_2$
and $q_3$, respectively.
The three-point correlation function for
these superfields can be found with the use of the ansatz
\bea
\langle
\Phi^{{\cal I}_1}_{{\cal A}_1}(z_1)
\J_{{\cal A}_2}^{{\cal I}_2}(z_2)
\P_{{\cal A}_3}^{{\cal I}_3}(z_3) \rangle
&=&\frac{T^{(1)}{}_{{\cal A}_1}{}^{{\cal B}_1}( \ve \underline{ \hat{\bm x}}_{13})
 T^{(2)}{}_{{\cal A}_2}{}^{{\cal B}_2}( \ve\underline{\hat{\bm x}}_{23})
D^{(1)}{}^{{\cal I}_1}{}_{{\cal J}_1}(u_{13})
D^{(2)}{}^{{\cal I}_2}{}_{{\cal J}_2}(u_{23})}{({\bm x}_{13}{}^2)^{q_1} ({\bm x}_{23}{}^2)^{q_2}}
~~~~ \non\\&&
\times H^{{\cal J}_1 {\cal J}_2 {\cal I}_3}_{{\cal B}_1 {\cal B}_2 {\cal A}_3}
 ({\bm X}_{3},\Theta_{3},U_{3})~,
\label{OOO}
\eea
where $H^{{\cal J}_1 {\cal J}_2 {\cal I}_3}_{{\cal B}_1 {\cal B}_2 {\cal
A}_3}$ is a tensor constructed in terms of the three-point
functions (\ref{three-points}). The functional form of this
tensor is highly constrained by the following conditions:
\begin{itemize}

\item[(i)] It should obey the  scaling property
\bea
H^{{\cal J}_1 {\cal J}_2 {\cal I}_3}_{{\cal B}_1 {\cal B}_2 {\cal A}_3}
 (\lambda^2{\bm X},\lambda\Theta,U)
 =(\lambda^2)^{q_3-q_2-q_1}
 H^{{\cal J}_1 {\cal J}_2 {\cal I}_3}_{{\cal B}_1 {\cal B}_2 {\cal A}_3}
 ({\bm X},\Theta,U) 
\eea
in order for the correlation function to have the correct transformation law
under the superconformal group.

\item[(ii)] When some of
the superfields $\F$, $\J$ and $\P$
obey differential equations such as the conservation
conditions of conserved current multiplets, the tensor
$H^{{\cal J}_1 {\cal J}_2 {\cal I}_3}_{{\cal B}_1 {\cal B}_2 {\cal A}_3}$
is constrained by certain differential equations as well.
In deriving such equations the identities
\eqref{useful-prop}
may be useful.

\item[(iii)] When two  of
the superfields $\F$, $\J$ and $\P$  (or all of them) coincide,
the tensor $H$
should obey certain constraints originating from the symmetry
under permutations of the superfields, e.g.
\be
\langle \Phi_{{\cal I}}^{{\cal A}}(z_1) \Phi_{{\cal J}}^{{\cal B}}(z_2)
\P_{{\cal K}}^{{\cal C}}(z_3) \rangle =
(-1)^{\epsilon(\Phi)}
\langle \Phi_{{\cal J}}^{{\cal B}}(z_2) \Phi_{{\cal I}}^{{\cal A}}(z_1)
\P_{{\cal K}}^{{\cal C}}(z_3) \rangle~,
\ee
where $\epsilon(\Phi)$ is the Grassmann parity of $\Phi_{{\cal I}}^{{\cal A}}$.
\end{itemize}
These constraints fix the functional form of the tensor $H$
(and, hence, the three-point correlation function) up to
a few arbitrary constants.

\subsection{Specific features of the $\cN=3$ case}
\label{sect2.2}

An important feature of the $\cN=3$ case is that
the $R$-symmetry group $\sSO(3)$ is related to $\sSU(2)$ by the isomorphism
$\sSO(3) \cong \sSU(2)/ {\mathbb Z}_2$.
This isomorphism
makes it possible
to convert
the $\sSO(3)$  index of every isovector $Z_I$ into a pair  of isospinor ones,
\bea
Z_I ~\to ~ Z_i{}^j :=\frac{\rm i}{\sqrt 2} (\vec{Z} \cdot \vec{\s})_i{}^j
\equiv Z_I (\t_I)_i{}^j~, \qquad Z_i{}^i=0~,
\label{N=3conversion}
\eea
with $\vec \s$ the Pauli matrices.\footnote{Our definition of the $\t$-matrices agrees with the one adopted in \cite{KPT-MvU}.}

The isospinor indices will be raised and lowered using the
$\sSU(2) $ invariant antisymmetric tensors $\ve_{ij}$ and $\ve^{ij}$ (normalised as $\ve^{12}=\ve_{21}=1$). The rules for raising and lowering
the isospinor indices are
\bea
\psi^i=\ve^{ij}\psi_j~, \qquad \psi_i=\ve_{ij}\psi^j~.
\eea
In particular, associated with the matrices $(\t)_i{}^j$, eq. (\ref{N=3conversion}),
are the symmetric matrices $(\t_I)_{ij} =(\t_I)_{ji} $ and $(\t_I)^{ij} =(\t_I)^{ji} $
which are related to each other by complex conjugation:
\bea
\overline{ (\t_I)_{ij} } =(\t_I)^{ij}  ~.
\eea
If $A_I$ and $B_I$ are $\sSO(3)$ vectors and $A_{ij}$ and $B_{ij}$ are the associated symmetric isotensors, then
\bea
 A_I  =A_{ij} (\t_I)^{ij}~, \qquad
A_I B_I = A_{ij} B^{ij}~,
\eea
in accordance with the
identities
\be
(\tau_I)_i{}^k (\tau_J)_k{}^j = -\frac1{\sqrt2}
\varepsilon_{IJK}(\tau_K)_i{}^j -\frac12 \delta_{IJ}\delta_i{}^j
~,\qquad
(\tau_I)_{ij}(\tau_I)_{kl} = \frac12(\varepsilon_{ik}\varepsilon_{jl}
+\varepsilon_{il}\varepsilon_{jk})~.
\ee
Given an antisymmetric second-rank $\sSO(3)$ tensor,  $\L^{IJ} =- \L^{JI}$,
for its  counterpart with isospinor indices
$\L^{ijkl}=-\L^{klij}= \L^{IJ}(\t_I)^{ij}(\t_J)^{kl}$
we have
\bea
\L^{IJ} =- \L^{JI} \quad \Longrightarrow \quad
\L^{IJ}(\t_I)^{ij}(\t_J)^{kl}
= -\ve^{jl}\L^{ik}- \ve^{ik}\L^{jl}~,\qquad
\L^{ij}=\L^{ji}~.
\label{2.388}
\eea

Applying the above conversion to
the Grassmann coordinates $\q_I^\a$ and the spinor covariant derivatives
$D_\a^I$
gives
\be
\theta_{ij}^\alpha =  \theta_{I}^\alpha(\tau_I)_{ij}
~,\qquad
D^{ij}_\alpha = (\tau_I)^{ij} D^I_\alpha~,
\label{convert}
\ee
and similarly for the two- and tree-point functions
(\ref{super-interv-Theta}) and (\ref{3ptTheta})
\be
\theta_{12\alpha}^{ij} = (\tau_I)^{ij} \theta_{12\alpha}^I
~,\qquad
\Theta_{1\alpha}^{ij} = (\tau_I)^{ij} \Theta_{1\alpha}^I~.
\ee
The covariant derivatives $D_\a^{ij}$ obey the anti-commutation relation
\bea
\{D_\a^{ij},D_\b^{kl}\}=-2\ri\ve^{i(k}\ve^{l)j}\pa_{\a\b}~.
\eea
In terms of the superspace coordinates $z^A=(x^a,\q^\a_{ij})$,
the explicit realisation of the covariant derivatives is
\bea
D_\a^{ij}=\frac{\pa}{\pa\q^\a_{ij}}+\ri\,\q^{\b \,ij}\pa_{\a\b}~.
\eea

The isomorphism $\sSO(3)\cong \sSU(2)/{\mathbb Z}_2$
implies that associated with the orthogonal unimodular matrix
$u^{IJ}_{12}$ given by (\ref{two-point-u}) is a
unique, up to sign, unitary and unimodular matrix ${\bf u}_{12}^{ij}$ such that
\be
(\tau_I)^{ii'} (\tau_J)^{jj'} u^{IJ}_{12}=\frac12({\bf u}_{12}^{i j}{\bf u}_{12}^{i'j'}
+{\bf u}_{12}^{i' j}{\bf u}_{12}^{i j'})~.
\label{bf u}
\ee
The matrix ${\bf u}_{12}^{ij}$ can be chosen as
\be
{\bf u}_{12}^{ij}
 =-\varepsilon^{ij}- \ri  \frac{\theta^{ik}_{12\alpha}{\bm x}_{12}^{\beta\alpha}
 \theta^j_{12k\beta}}{{\bm x}_{12}{}^2}
 +\frac18 \varepsilon^{ij}\frac{\theta_{12}{}^4}{{\bm x}_{12}{}^2} ~, \qquad
\q_{12}{}^4 := (\q_{12}^{\a ij} \q_{12 \a ij})^2~.
 \label{rule}
\ee
It is easy to check that \eqref{rule} is indeed unitary and unimodular,
\be
{\bf u}_{12}^\dag{\bf u}_{12} = {\mathbbm 1}_2 ~,\qquad
\det{\bf u}_{12} =1~,
\ee
and obeys the equation (\ref{bf u}) with $u^{IJ}_{12}$ given by
(\ref{two-point-u}).

The transformation law of the orthogonal matrix $u_{12}$, eq.
(\ref{deltaU}),
has the following counterpart in terms of
the unitary matrix ${\bf u}_{12}$:
\be
\widetilde{\delta}{\bf u}_{12}^{ij} = \Lambda^i_k(z_1){\bf u}^{kj}_{12} +{\bf u}^{ik}_{12}
\Lambda^j_k(z_2)~.
\label{delta bf u}
\ee
Here the symmetric matrix $\Lambda^{ij}(z)$ with isospinor indices
is related to the antisymmetric matrix $\L^{IJ}(z)$ with isovector indices, eq. \eqref{2.4par},  according to the general rule \eqref{2.388}.

Let us introduce
one more $2\times 2$ matrix by the rule
\be
n^{ij}_{12} = \frac{{\bf u}_{12}^{ij}}{{\bm x}_{12}}
=-\frac{\varepsilon^{ij}}{x_{12}}
-\ri\frac{\theta^{ik}_{12\alpha} x_{12}^{\alpha\beta}
\theta^j_{12k\beta}}{x_{12}{}^3}~.
\label{n}
\ee
The second expression for
$n^{ij}_{12}$ is given  in terms
of the symmetric part of ${\bm x}^{\alpha\beta}_{12}$ given by (\ref{x12-sym1}).
It may be shown that the two-point function (\ref{n}) obeys
the analyticity condition
\be
D^{(ij}_{(1)\alpha} n_{12}^{k)l}=0~.
\label{n-analit}
\ee
This is why $n^{ij}_{12}$
appears in the correlation functions of $\cN=3$
flavour current multiplets.

Similarly to (\ref{bf u}), we can represent
(\ref{U-explicit}) as
\be
(\tau)_I^{ii'} (\tau)_J^{jj'} U^{IJ}_{1}=\frac12({\bf U}_{1}^{i j}{\bf U}_{1}^{i'j'}
+{\bf U}_{1}^{i' j}{\bf U}_{1}^{i j'})~,
\label{bf U}
\ee
where we have introduced the matrix
\be
{\bf U}_1^{ij}
=-\varepsilon^{ij}+\ri  \Theta^{ik}_{1\alpha}({\bm X}_1^{-1})^{\alpha\beta}\Theta^j_{1k\beta}
 +\frac18 \varepsilon^{ij}\frac{\Theta_1{}^4}{{\bm X}_1{}^2}~,
 \label{U-N3}
\ee
which can be expressed
as a product of three two-point functions  (\ref{rule})
\be
{\bf U}_1^{ij}= - {\bf u}_{12}^{ik}{\bf u}_{23\,kl}{\bf
u}_{31}^{lj}~.
\label{Uu}
\ee
As a consequence, the transformation law
(\ref{delta bf u}) implies
\be
\widetilde{\delta}{\bf U}_1^{ij} = \Lambda^i_{k}(z_1){\bf U}_1^{kj}
+{\bf U}_1^{ik}\Lambda^j_{k}(z_1)~.
\ee

By analogy with
(\ref{n}) we introduce the matrix
\be
N^{ij}_{1} = \frac{{\bf U}_{1}^{ij}}{{\bm X}_{1}}
=-\frac{\varepsilon^{ij}}{X_{1}}
-\ri\frac{\Theta^{ik}_{1\alpha} X_{1}^{\alpha\beta}
\Theta^j_{1k\beta}}{X_{1}{}^3}~,
\label{N}
\ee
which obeys the analyticity condition
\be
{\cal D}^{(ij}_{(1)\alpha} N_{1}^{k)l}=0~,
\label{N-analyt}
\ee
where the derivative ${\cal D}^{ij}_\alpha$ is related to
(\ref{generalized-DQ}) by the rule (\ref{convert}).

Here we have only considered  the thee-point functions
\eqref{U-N3} and \eqref{N} which transform as
tensors at $z_1$. Performing cyclic
permutations of the superspace points $z_1$, $z_2$ and $z_3$
leads to similar
objects which transform as tensors at $z_2$ and
$z_3$.

\subsection{Specific features of the $\cN=4$ case}
\label{N4-sect}

In the case of $\cN=4$ supersymmetry, the $R$-symmetry group $\sSO(4)$ possesses
the isomorphism
$\sSO(4) \cong  \big( {\sSU}(2)_{\rm L}\times {\sSU}(2)_{\rm R} \big)/{\mathbb Z}_2$
which can be used  to convert
each $\sSO(4)$ vector index into a pair of $\sSU(2)$ ones,\footnote{Our definition of the
$\t$-matrices agrees with the one used in  \cite{KPT-MvU}
and differs from that adopted in \cite{KLT-M11}.}
\bea
Z_I =(\vec Z, Z_4) ~\to ~ Z_{ i}{}^{\tilde k} :=\frac{\rm i}{\sqrt 2} (\vec{Z} \cdot \vec{\s})_{ i}{}^{\tilde k}
+ \frac{1}{\sqrt 2}Z_4 \d_{ i}{}^{\tilde k} \equiv Z_I (\t_I)_i{}^{\tilde k}~.  
\label{N=4conversion}
\eea
The index `$I$' is an $\sSO(4)$ vector one,
while the indices `$i$' and  `$\tilde{i}$'  are, respectively,
$\sSU(2)_{\rm L}$ and $\sSU(2)_{\rm R}$ spinor indices.
Given $\sSU(2)_{\rm L}$ and $\sSU(2)_{\rm R}$ spinors $\psi_i$ and $\chi_{\tilde{i}}$, respectively, we will raise and lower their indices by using the antisymmetric
tensors $\ve^{ij},\ve_{ij}$ and $\ve^{{\tilde i}{\tilde j}},\ve_{{\tilde i}{\tilde j}}$
(normalised by $\ve^{12}=\ve_{21}=\ve^{{\tilde 1}{\tilde 2}}=\ve_{{\tilde 2}{\tilde 1}}=1$)
according to the rules:
\bea
\psi^{i}=\ve^{ij}\psi_j~,~~~
\psi_{i}=\ve_{ij}\psi^j
~,~~~~~~
\chi^{{\tilde i}}=\ve^{\tilde i\tilde j}\chi_{\tilde j}~,~~~
\chi_{\tilde i}=\ve_{\tilde i\tilde j}\chi^{\tilde j}
~.
\eea
The complex conjugation acts on the $\t$-matrices as
\bea
\overline{\, (\t_I)_{i\tilde{i}} \,}=(\t_I)^{i\tilde{i}} =\ve^{ij} \ve^{\tilde i \tilde j}(\t_I)_{j\tilde{j}}~.
\eea
The $\t$-matrices have the following properties:
\bea
(\t_{(I})_{i\tilde{j}}(\t_{J)})^{j\tilde{j}}=\hf\d_{IJ}\d_i^j
~, \qquad
(\t_{(I})_{j\tilde{i}}(\t_{J)})^{j\tilde{j}}=\hf\d_{IJ}\d_{\tilde{i}}^{\tilde{j}}
~, \qquad (\t_I)_{i\tilde{i}}(\t^I)_{j\tilde{j}}=\ve_{ij}\ve_{\tilde{i}\tilde{j}}
~.~~~
\eea

The conversion from  $\sSO(4)$ to  $\sSU(2)$ indices works as follows.
Associated with  an  $\sSO(4)$ vector $A_{I}$
is the second-rank isospinor $A_{i\tilde{i}}$
defined by
\bea
A_{i\tilde{i}}:=(\t_I)_{i\tilde{i}}A^{I} \quad \longleftrightarrow \quad
A_{I}=(\t_I)^{i\tilde{i}}A_{i\tilde{i}}
~,
\eea
such that
\bea
A_{I}B^{I}= A_{i\tilde{i}}B^{i\tilde{i}}
~.
\eea
Given an antisymmetric second-rank $\sSO(4)$ tensor,
$A_{IJ}=-A_{JI}$, its counterpart with isospinor
indices, $A_{IJ}(\t^I)_{i\tilde i}(\t^J)_{j\tilde j}$, can be decomposed as
\begin{subequations} \label{2.611}
\bea
A_{IJ}=-A_{JI}
~ \longrightarrow ~
A_{IJ}(\t^I)_{i\tilde i}(\t^J)_{j\tilde j}
=\ve_{ij}A_{\tilde i\tilde j}+\ve_{\tilde i\tilde j}A_{ij}
~,\quad
A_{ij}=A_{ji}~,~~A_{\tilde i\tilde j}=A_{\tilde j\tilde i}
~.~~~~
\eea
We also have
\bea
A_{IJ}(\t^I)^{i\tilde i}(\t^J)^{j\tilde j}
=-\ve^{ij}A^{\tilde i\tilde j}-\ve^{\tilde i\tilde j}A^{ij}~.
\eea
\end{subequations}

Applying the conversion rule to
the Grassmann variables $\theta^I_\alpha$ and
covariant derivatives $D^I_\alpha$ gives
$ \theta_\alpha^{i\tilde i } = (\tau_I)^{i\tilde i }\theta^I_\alpha$
and
$D^{i\tilde i }_\alpha = (\tau_I)^{i\tilde i }D^I_\alpha$, respectively.
For the two- and tree-point functions
(\ref{super-interv-Theta}) and (\ref{3ptTheta}), the same rule gives
$
\theta_{12\alpha}^{i\tilde i } = (\tau_I)^{i\tilde i} \theta_{12\alpha}^I$
and
$\Theta_{1\alpha}^{i\tilde i} = (\tau_I)^{i\tilde i} \Theta_{1\alpha}^I$.

The covariant derivatives
$D_\a^{i{\tilde k}}$
satisfy the anti-commutation relations
\bea
\{D_\a^{i{\tilde k}},D_\b^{j{\tilde l}}\}=2\ri\ve^{ij}\ve^{{\tilde k}{\tilde l}}\pa_{\a\b}~.
\label{2.622}
\eea
In terms of the superspace coordinates $z^A=(x^a,\q^\a_{k{\tilde l}})$, 
the explicit realisation
of the covariant derivatives is
\bea
D_\a^{k{\tilde l}}=\frac{\pa}{\pa\q^\a_{k{\tilde l}}}+\ri\,\q^{\b k{\tilde l}}\pa_{\a\b}~.
\eea

Due to the isomorphism
$\sSO(4) \cong  \big( {\sSU}(2)_{\rm L}\times {\sSU}(2)_{\rm R} \big)/{\mathbb Z}_2$,
the orthogonal matrix $u_{12}^{IJ}$ given by (\ref{two-point-u}) is
equivalent
to a pair of $\sSU(2)$ matrices ${\bf u}_{12}^{ij}$ and
${\bf u}_{12}^{\tilde i\tilde j}$ constrained by
\be
u^{IJ}_{12} (\tau_I)^{i\tilde i } (\tau_J)^{j\tilde j } = {\bf
u}_{12}^{ij}{\bf u}_{12}^{\tilde i\tilde j}~.
\label{u-uu}
\ee
The solution to this equation is
\begin{subequations}
\label{uu}
\bea
{\bf u}^{ij}_{12}&=&-\varepsilon^{ij}
-\ri \frac{\theta^{\tilde k i}_{12\alpha} x_{12}^{\alpha\beta}\theta^j_{12\beta\tilde k}}{x_{12}{}^2}
-\frac18\frac{\varepsilon^{ij}\theta_{12}{}^4}{x_{12}{}^2}
+\frac18\frac{\varepsilon^{ij}\tilde\theta_{12}^4}{x_{12}{}^4}
\non\\&&
-\frac\ri8 \frac{\theta^{\tilde k i}_\alpha x_{12}^{\alpha\beta}\theta^j_{12\beta\tilde k}\theta_{12}{}^4}{x_{12}{}^4}
+\frac1{128} \frac{\varepsilon^{ij}\theta_{12}{}^8}{x_{12}{}^4}~,
\\
{\bf u}^{\tilde i\tilde j}_{12}&=&-\varepsilon^{\tilde i\tilde j}
-\ri \frac{\theta^{\tilde i k}_{12\alpha} x_{12}^{\alpha\beta}\theta^{\tilde j}_{12\beta k}}{x_{12}{}^2}
-\frac18\frac{\varepsilon^{\tilde i\tilde j}\theta_{12}{}^4}{x_{12}{}^2}
-\frac18\frac{\varepsilon^{\tilde i\tilde j}\tilde\theta_{12}^4}{x_{12}{}^4}
\non\\&&
-\frac\ri8 \frac{\theta^{\tilde i k}_{12\alpha} x_{12}^{\alpha\beta}\theta^{\tilde j}_{12\beta k}\theta_{12}{}^4}{x_{12}{}^4}
+\frac1{128}\frac{\varepsilon^{\tilde i\tilde j}\theta_{12}{}^8}{x_{12}{}^4}~,
\eea
\end{subequations}
where we have used the notation
\begin{subequations}
\label{theta4}
\bea
\theta^4&=&(\theta^{I\alpha}\theta^I_\alpha)^2=(\theta^{\tilde i i \alpha}\theta_{\tilde i i
\alpha})^2~,\\
\tilde\theta^4&=&
\theta^I_\alpha x^{\alpha\beta}\theta^J_\beta
 \theta^K_\mu x^{\mu\nu}\theta^L_\nu \varepsilon_{IJKL}
=\theta^{\tilde k j}_\alpha x^{\alpha\beta} \theta_{i\tilde k\beta}
  \theta^{\tilde l i}_\mu x^{\mu\nu} \theta_{j\tilde l \nu}-
  \theta^{\tilde j k}_\alpha x^{\alpha\beta} \theta_{k \tilde i\beta}
  \theta^{\tilde i l}_\mu x^{\mu\nu} \theta_{l \tilde j \nu}~.~~~~~
\label{tilde-theta}
\eea
\end{subequations}
Both matrices (\ref{uu}) are unitary and unimodular, in particular it holds that
\be
{\bf u}_{12}^{ij} {\bf u}_{12kj} = \delta^i_k~,\qquad
{\bf u}_{12}^{\tilde i\tilde j} {\bf u}_{12\tilde k\tilde j} = \delta^{\tilde i}_{\tilde k}~.
\ee
Note also that the expressions (\ref{uu}) are defined by the
equation (\ref{u-uu}) uniquely, up to an overall sign which we
fix as in (\ref{uu}) for further convenience.

The transformation law (\ref{deltaU}) implies that the matrices ${\bf u}_{12}^{ij}$
and ${\bf u}_{12}^{\tilde i\tilde j}$ defined by (\ref{u-uu}) vary
under the infinitesimal superconformal transformations as
isospinors at $z_1$ and $z_2$,
\be
\widetilde{\delta}{\bf u}_{12}^{ij} = \Lambda^i_k(z_1){\bf u}^{kj}_{12} +{\bf u}^{ik}_{12}
\Lambda^j_k(z_2)~,\qquad
\widetilde{\delta}{\bf u}_{12}^{\tilde i\tilde j}
= \Lambda^{\tilde i}_{\tilde k}(z_1){\bf u}^{\tilde k\tilde j}_{12} +{\bf u}^{\tilde i\tilde k}_{12}
\Lambda^{\tilde j}_{\tilde k}(z_2)~,
\ee
where $\Lambda^{ij}(z)$ and $\Lambda^{\tilde i\tilde j}(z) $ are constructed from
$\Lambda^{IJ}(z)$ by the rule \eqref{2.611}.

Let us define the following matrices:
\begin{subequations}
\label{380}
\bea
\label{380a}
n_{12}^{ij} &=& \frac{{\bf u}_{12}^{ij}}{{\bm x}_{12}}
=-\frac{\varepsilon^{ij}}{x_{12}}
-\ri \frac{\theta^{\tilde k i}_\alpha x_{12}^{\alpha\beta}\theta^j_{12\beta\tilde k}}{x_{12}{}^3}
+\frac18\frac{\varepsilon^{ij}\tilde\theta_{12}^4}{x_{12}{}^5}~,\\
n_{12}^{\tilde i\tilde j} &=& \frac{{\bf u}_{12}^{\tilde i\tilde j}}{{\bm x}_{12}}
=-\frac{\varepsilon^{\tilde i\tilde j}}{x_{12}}
-\ri \frac{\theta^{\tilde i k}_{12\alpha} x_{12}^{\alpha\beta}\theta^{\tilde j}_{12\beta k}}{x_{12}{}^3}
-\frac18\frac{\varepsilon^{\tilde i\tilde j}\tilde\theta_{12}^4}{x_{12}{}^5}~.
\label{380b}
\eea
\end{subequations}
Similarly to (\ref{n-analit}), these matrices obey the
analyticity conditions
\be
D^{\tilde i (i}_{(1)\alpha} n_{12}^{k)l} =0~,\qquad
D^{i(\tilde i }_{(1)\alpha} n_{12}^{\tilde k) \tilde l}=0~.
\label{2.54}
\ee

By analogy with the two-point functions (\ref{u-uu}), we introduce
three-point matrices with $\sSU(2)$ indices
\be
U^{IJ}_{1} (\tau_I)^{i\tilde i } (\tau_J)^{j\tilde j }
= {\bf U}_{1}^{ij}{\bf U}_{1}^{\tilde i\tilde j}~,
\ee
which have the following explicit form
\begin{subequations}\label{U-N4}
\bea
{\bf U}_1^{ij}&=&-\varepsilon^{ij}
-\ri \frac{\Theta^{\tilde k i}_{1\alpha} X_1^{\alpha\beta}\Theta^j_{1\beta\tilde k}}{X_1{}^2}
-\frac18\frac{\varepsilon^{ij}\Theta_1{}^4}{X_1{}^2}
+\frac18\frac{\varepsilon^{ij}\tilde\Theta_1^4}{X_1{}^4}
\non\\&&
-\frac\ri8 \frac{\Theta^{\tilde k i}_{1\alpha} X_1^{\alpha\beta}\Theta^j_{1\beta\tilde k}\Theta_1{}^4}{X_1{}^4}
+\frac1{128} \frac{\varepsilon^{ij}\Theta_1{}^8}{X_1{}^4}~,
\label{U-N4a}\\
{\bf U}_1^{\tilde i\tilde j}&=&-\varepsilon^{\tilde i\tilde j}
-\ri \frac{\Theta^{\tilde i k}_{1\alpha} X_1^{\alpha\beta}\Theta^{\tilde j}_{1\beta k}}{X_1{}^2}
-\frac18\frac{\varepsilon^{\tilde i\tilde j}\Theta_1{}^4}{X_1{}^2}
-\frac18\frac{\varepsilon^{\tilde i\tilde j}\tilde\Theta_1^4}{X_1{}^4}
\non\\&&
-\frac\ri8 \frac{\Theta^{\tilde i k}_{1\alpha} X_1^{\alpha\beta}\Theta^{\tilde j}_{1\beta k}\Theta_1{}^4}{X_1{}^4}
+\frac1{128}\frac{\varepsilon^{\tilde i\tilde j}\Theta_1{}^8}{X_1{}^4}~.~~~~~~
\eea
\end{subequations}
Here the composites $\Theta^4$ and $\tilde\Theta^4$ are defined
by the same rules as in (\ref{theta4}). By construction, the matrices
(\ref{U-N4}) transform as tensors
at the superspace point $z_1$
\be
\widetilde{\delta}{\bf U}_{1}^{ij} = \Lambda^i_k(z_1){\bf U}^{kj}_{1} +{\bf U}^{ik}_{1}
\Lambda^j_k(z_1)~,\qquad
\widetilde{\delta}{\bf U}_{1}^{\tilde i\tilde j}
= \Lambda^{\tilde i}_{\tilde k}(z_1){\bf U}^{\tilde k\tilde j}_{1}
+{\bf U}^{\tilde i\tilde k}_{1}
\Lambda^{\tilde j}_{\tilde k}(z_1)~.
\ee
It is possible to check that the matrices (\ref{U-N4}) can be
expressed as  products of three matrices of the type (\ref{uu})
\be
{\bf U}_1^{ij}= - {\bf u}_{12}^{ik}{\bf u}_{23\,kl}{\bf
u}_{31}^{lj}~,\qquad
{\bf U}_1^{\tilde i\tilde j}= - {\bf u}_{12}^{\tilde i\tilde k}{\bf u}_{23\,\tilde k\tilde l}{\bf
u}_{31}^{\tilde l\tilde j}~.
\label{UUuu}
\ee

The three-point analogs of (\ref{380}) are
\begin{subequations}
\label{U-N-N4}
\bea
N^{ij}_1 &=& \frac{{\bf U}_1^{ij}}{{\bm X}_1}
=-\frac{\varepsilon^{ij}}{X_1}
-\ri \frac{\Theta^{\tilde k i}_{1\alpha} X_1^{\alpha\beta}\Theta^j_{1\beta\tilde k}}{X_1{}^3}
+\frac18\frac{\varepsilon^{ij}\tilde\Theta_1^4}{X_1{}^5}~,\label{2.58a}\\
N^{\tilde i\tilde j}_1 &=& \frac{{\bf U}_1^{\tilde i\tilde j}}{{\bm X}_1}
=-\frac{\varepsilon^{\tilde i\tilde j}}{X_1}
-\ri \frac{\Theta^{\tilde i k}_{1\alpha} X_1^{\alpha\beta}\Theta^{\tilde j}_{1\beta k}}{X_1{}^3}
-\frac18\frac{\varepsilon^{\tilde i\tilde j}\tilde\Theta_1^4}{X_1{}^5}~.
\eea
\end{subequations}
These matrices are analytic with respect to the spinor derivatives (\ref{generalized-DQ})
\be
{\cal D}^{\tilde i( i}_{(1)\alpha} N_1^{k)l}
 =0~,\qquad
{\cal D}^{i (\tilde i}_{(1)\alpha} N_1^{\tilde k) \tilde l}
=0~.
\label{2.59}
\ee

In this section we considered only the three-point functions which
transform as tensors at the superspace point $z_1$.
It is straightforward to obtain
the analogs of these objects transforming covariantly at $z_2$ and
$z_3$ by permuting the superspace points.

The two- and three-point superconformal building blocks constructed
above are very similar to the $\cN=3$ ones given in the previous
subsection. This is not accidental.
It turns out that the latter can be found
from the former by applying the $\cN=4\to \cN=3$ superspace reduction.
Indeed, when we switch off one of the four Grassmann variables $\q_I$ at each
superspace point, say $\theta_4=0$, the expressions (\ref{uu})
prove to coincide with (\ref{rule}),
\be
{\bf u}^{ij}_{12(\cN=4)}|_{\theta_4=0}
={\bf u}^{\tilde i\tilde j}_{12(\cN=4)}|_{\theta_4=0}
={\bf u}^{ij}_{12(\cN=3)}~.
\label{2.60}
\ee
Here we have attached extra subscripts, $(\cN=4)$ and $(\cN=3)$, to the
two-point functions to distinguish them. We usually omit these
labels if no confusion occurs.
For the three-point
functions (\ref{U-N3}) and (\ref{U-N4}) we have similar relations
\be
{\bf U}^{ij}_{1(\cN=4)}|_{\theta_4=0}
={\bf U}^{\tilde i\tilde j}_{1(\cN=4)}|_{\theta_4=0}
={\bf U}^{ij}_{1(\cN=3)}~.
\label{2.61}
\ee
The superspace reduction rules (\ref{2.60}) and (\ref{2.61}) will
be important below when we turn to studying the correlators of
the $\cN=3$ and $\cN=4$ flavour
current multiplets.

\section{Correlation functions for the $\cN=3$ flavour current multiplets revisited}
\label{section3}

Here we obtain a new representation for the correlation functions
of the $\cN=3$ flavour current multiplets computed in \cite{BKS}. 
Such a representation will be more convenient for comparison 
of the $\cN=3$ correlators with $\cN=4$ ones. 

As discussed in \cite{BKS}, 
the $\cN=3$ flavour current multiplet is described
by a primary isovector superfield $L^I$ of dimension 1, 
which is subject to the conservation
equation
\be
D^{(I}_\alpha L^{J)}-\frac13 \delta^{IJ}D^K_\alpha L^{K} =0~.
\label{N3-flavour-conserv}
\ee
Its superconformal transformation law is
\be
\delta L^{I} = -\xi L^{I} -\sigma(z) L^{I}
+\Lambda^{IJ}(z)L^{J}~.
\label{N3-flavour-transf}
\ee
The dimension of $L^I$ is uniquely fixed by requiring the constraint 
\eqref{N3-flavour-conserv} to be invariant under the superconformal 
transformations. 

Consider an $\cN=3$  superconformal field theory possessing 
$n$ flavour current
multiplets $L^{I\bar a}$,  $\bar a =1,\ldots,n$. 
Their two- and three-point functions were found in \cite{BKS} to be
\bea
\langle L^{I\bar a}(z_1) L^{J\bar b}(z_2) \rangle
&=& a_{\cN=3} \frac{u^{IJ}_{12}\delta^{\bar a\bar b}}{{\bm x}_{12}{}^2}~,
\label{2ptL}
\\
\langle L^{I\bar a}(z_1) L^{J\bar b}(z_2) L^{K\bar c}(z_3) \rangle
&=&b_{\cN=3}\, f^{\bar a\bar b\bar c}
\frac{u_{13}^{II'}u_{23}^{JJ'}}{{\bm x}_{13}{}^2{\bm x}_{23}{}^2}
H^{I'J'K}({\bm X}_{3},\Theta_{3})~,
\label{3ptL_}
\eea
where
\bea
H^{IJK}({\bm X},\Theta)&=&\frac{1}{\bm X}\Big[
\varepsilon^{IJK}
-U^{LJ}\varepsilon^{LIK}
+U^{IL}\varepsilon^{LJK}
\non\\&&
-\frac1{16}(\delta^{IJ}\varepsilon^{KMN}U^{MN}
+\varepsilon^{IMN}U^{MN}U^{KJ}
+\varepsilon^{JMN}U^{MN}U^{IK})
\non\\&&
+\frac5{16}(
U^{IJ}\varepsilon^{KMN}U^{MN}
+\delta^{IK}\varepsilon^{JMN}U^{MN}
+\delta^{JK}\varepsilon^{IMN}U^{MN})
\Big]~.
\label{H-flavour-N3}
\eea
In (\ref{3ptL_}), $f^{\bar a \bar b \bar c}$ denotes 
the completely antisymmetric structure constants of 
the Lie algebra of the flavour group which is assumed to be simple.
The tensor $H^{IJK}$ in (\ref{H-flavour-N3}) is expressed
in terms of 
the orthogonal matrix (\ref{U-explicit}).
The correlation functions
(\ref{2ptL}) and (\ref{3ptL_}) are fixed by the superconformal
symmetry and the conservation condition up to arbitrary 
coefficients $a_{\cN=3}$ and $b_{\cN=3}$.

We now switch to the $\sSU(2)$ notation, $L^I \to L^{ij} =(\tau_I)^{ij} L^I$, 
in accordance with the rules 
introduced in subsection \ref{sect2.2}. 
Then, the conservation equation
(\ref{N3-flavour-conserv}) turns into the analyticity condition 
\bea
D^{(ij}_\alpha L^{kl)} &=&0~,
\label{L-conserv}
\eea
and the superconformal transformation
(\ref{N3-flavour-transf}) takes  the form
\bea
\delta L^{ij} &=& -\xi L^{ij} -\sigma(z) L^{ij} + 2\Lambda^{(i}_k(z) L^{j)k}~.
\label{L-transform}
\eea
 Here the symmetric matrix $\Lambda^{ij}(z)$ with isospinor indices
is related to the antisymmetric matrix $\L^{IJ}(z)$ with isovector indices, eq. \eqref{2.4par},  according to the general rule \eqref{2.388}.

Using the relation (\ref{bf u}) between the two-point building blocks with
$\sSO(3)$ and $\sSU(2)$ indices, for (\ref{2ptL}) we immediately get
\be
\langle L^{ij\,\bar a}(z_1) L^{kl\,\bar b}(z_2) \rangle
= \frac{a_{\cN=3}}2 \frac{\delta^{\bar a\bar b}({\bf u}_{12}^{ik} {\bf u}_{12}^{jl}
+{\bf u}_{12}^{jk} {\bf u}_{12}^{il})}{{\bm
x}_{12}{}^2}~.
\label{3.8}
\ee
Contracting the three-point function (\ref{3ptL_}) 
with three $\tau$-matrices leads to 
\be
\langle L^{ij\,\bar a}(z_1) L^{kl\,\bar b}(z_2) L^{mn\,\bar c}(z_3) \rangle
=b_{\cN=3}\, f^{\bar a\bar b\bar c}
\frac{{\bf u}_{13}^{ii'}{\bf u}_{13}^{jj'}
 {\bf u}_{23}^{kk'} {\bf u}_{23}^{ll'} }{{\bm x}_{13}{}^2{\bm x}_{23}{}^2}
H_{i'j'\,k'l'}{}^{mn}({\bm X}_{3},\Theta_{3})~,
\ee
where
\be
H^{ij\,kl\,mn}= H^{(ij)(kl)(mn)}
=(\tau_I)^{ij}(\tau_J)^{kl}(\tau_K)^{mn} H^{IJK}~.
\label{3.10}
\ee

In order to  compute the right-hand side of \eqref{3.10},
it is convenient to rewrite the expression (\ref{H-flavour-N3}) in
the form
\be
H^{IJK}(X,\Theta)= \frac{\varepsilon^{IJK}}{X}
+\frac 12 \frac{\delta^{IJ}\varepsilon^{KMN}A^{MN}}{X^3}
-\frac 12 \frac{\delta^{IK}\varepsilon^{JMN}A^{MN}}{X^3}
-\frac 12 \frac{\delta^{JK}\varepsilon^{IMN}A^{MN}}{X^3}~,
\label{H3.11}
\ee
where
\be
A^{IJ}:=\ri\Theta^{I\alpha}X_{\alpha\beta}\Theta^{J\beta} =
-A^{JI}~.
\ee
For the first term in the right-hand side of (\ref{H3.11}) we apply
the identity
\be
\varepsilon^{IJK} = -\sqrt2\tr(\tau^I \tau^K\tau^J)
=-\sqrt2 \tau^I_{ij} \tau^J_{kl}\tau^K_{mn}
\varepsilon^{mi} \varepsilon^{jk} \varepsilon^{ln}~.
\label{3.13}
\ee
The other terms in (\ref{H3.11}) can be rewritten as
\be
\delta^{IJ}\varepsilon^{KMN}A^{MN}
-\delta^{IK}\varepsilon^{JMN}A^{MN}-
\delta^{KJ}\varepsilon^{IMN}A^{MN}
=-2\sqrt2
\tau^I_{ij} \tau^J_{kl} \tau^K_{mn}
\varepsilon^{mi}A^{jl}\varepsilon^{kn}~,
\label{3.14}
\ee
where
\be
A^{mn }= \varepsilon_{kl} \tau_I^{mk} \tau_J^{nl} A^{IJ}=
\ri \Theta^{mk}_\alpha X^{\alpha\beta} \Theta^n_{k\beta}~.
\label{3.15}
\ee
In deriving (\ref{3.14}), the following identity 
\be
\tau_K^{mn} \varepsilon^{KMN}A_{MN} =-\sqrt2 A^{mn}
\ee
may be useful.
Now, substituting (\ref{3.13}) and (\ref{3.14}) into (\ref{3.10}) we
find
\be
H^{ij\,kl\,mn} =-\frac1{\sqrt2}\left(\frac{\varepsilon^{m(i}\varepsilon^{j)(l}\varepsilon^{k)n}}{X}
+\frac{\varepsilon^{m(i}A^{j)(l}\varepsilon^{k)n}}{X^3}\right)
-\frac1{\sqrt2}\left(\frac{\varepsilon^{n(i}\varepsilon^{j)(l}\varepsilon^{k)m}}{X}
+\frac{\varepsilon^{n(i}A^{j)(l}\varepsilon^{k)m}}{X^3}\right)
~.
\label{410}
\ee
Finally, taking into account the explicit expression for $A^{jl}$ given
by (\ref{3.15}), we note that the tensors in the parentheses in
(\ref{410}) can be rewritten in terms of the matrix $N^{jl}$
introduced in (\ref{N}),
\be
H^{ij\,kl\,mn}({\bm X}_3,\Theta_3) =\frac1{\sqrt2}
\left(
\varepsilon^{m(i}N_3^{j)(l}\varepsilon^{k)n}
+
\varepsilon^{n(i}N_3^{j)(l}\varepsilon^{k)m}\right)~.
\label{H-N}
\ee
We arrive at the final expression for 
the three-point function of the $\cN=3$ flavour
current multiplets 
\begin{subequations}
\label{3pt-flavour-final}
\bea
\langle L^{ij\,\bar a}(z_1) L^{kl\,\bar b}(z_2) L^{mn\,\bar c}(z_3) \rangle
&=&b_{\cN=3}\, f^{\bar a\bar b\bar c}
\frac{{\bf u}_{13}^{ii'}{\bf u}_{13}^{jj'}
 {\bf u}_{23}^{kk'} {\bf u}_{23}^{ll'} }{{\bm x}_{13}{}^2{\bm x}_{23}{}^2}
H_{i'j'\,k'l'}{}^{mn}({\bm X}_{3},\Theta_{3})~,\label{3ptL}\\
H^{ij\,kl\,mn}({\bm X}_3,\Theta_3) &=&\frac1{\sqrt2}
\left(
\frac{\varepsilon^{m(i} {\bf U}_3^{j)(l}\varepsilon^{k)n}}{{\bm X}_3}
+
\frac{\varepsilon^{n(i}{\bf U}_3^{j)(l}\varepsilon^{k)m}}{{\bm X}_3}
\right)~.
\label{3ptH}
\eea
\end{subequations}
Obviously, this three-point function possesses 
the correct superconformal properties since it is built out of the covariant
two- and three-point objects introduced in subsection 
\ref{sect2.2}.

After using the identity (\ref{useful-prop-a}), the 
conservation law (\ref{L-conserv}) implies the
following equation on the tensor $H^{ij\,kl\,mn}$:
\be
{\cal D}_\alpha^{(i'j'}H^{ij)kl\,mn} =0 ~.
\ee
It is easy to see that  (\ref{H-N}) obeys this
equation since the matrix $N^{jl}$ is analytic
(\ref{N-analyt}).

The three-point correlation function (\ref{3ptL}) must have  the
symmetry property
\be
\langle
L^{ij\,\bar a}(z_1) L^{kl\,\bar b}(z_2) L^{mn\,\bar c}(z_3) \rangle
=\langle
L^{mn\,\bar c}(z_3) L^{kl\,\bar b}(z_2) L^{ij\,\bar a}(z_1)
\rangle ~,
\ee
which implies the following constraint for the tensor $H^{ij\,kl\,mn}$
\bea
H_{mn\,pq}{}^{ij}(-{\bm X}_1^{\rm T},-\Theta_1)&=&
-{\bm x}_{13}{}^2 {\bm X}_3{}^2 {\bf u}_{13 mm'}{\bf u}_{13 nn'}{\bf u}_{13pr}
\non\\&&\times
{\bf U}_3^{rr'}
{\bf u}_{13qs}{\bf U}_{3}^{ss'}{\bf u}_{13}^{ii'}{\bf u}_{13}^{jj'}
  H_{i'j'\, r's'}{}^{m'n'}({\bm X}_3,\Theta_3)~.
\eea
Using (\ref{Uu}) one can check that (\ref{3ptH}) does satisfy
this equation.

Finally, 
we point out that
 the explicit form of the correlation function
 (\ref{3pt-flavour-final}) is analogous to the
three-point correlator of flavour current multiplets in 4D $\cN=2$
superconformal theories \cite{KT}.

\section{Correlation functions of conserved $\cN=4$ current multiplets}\label{section4}

In this section we compute the two- and three-point functions of the $\cN=4$
supercurrent and flavour current multiplets. 

\subsection{Correlation functions of flavour current multiplets}

As discussed in \cite{BKS}, 
there are two inequivalent flavour current multiplets,
$L^{IJ}_+$ and $L^{IJ}_-$, in $\cN=4$ superconformal field theories. 
They are described by primary $\sSO(4)$ bivectors,
$L_\pm^{I J} = - L_\pm^{JI}$, subject to the same conservation equation
\be
D_{\a}^{I} L_\pm^{ J K}=
D_{\a}^{[I} L_\pm^{ J K]}
- \frac{2}{3}  D_{\a }^L L_\pm^{ L[J} \d^{K] I}~,
\label{L-N4-conserv}
\ee
which implies that $L^{IJ}_+$ and $L^{IJ}_-$ have dimension 1.
These operators possess the same  superconformal transformation law
\be
\delta L_\pm^{IJ} = -\xi L_\pm^{IJ} -\sigma(z) L_\pm^{IJ}
+2\Lambda^{K[I}(z) L_\pm^{J]K}~.
\label{N4-L-transf}
\ee
However, they have  different algebraic properties,  
\be
 \frac12 \ve^{IJKL}L^{KL}_\pm  = \pm L^{IJ}_\pm~, 
 \label{ASD}
\ee
and thus $L^{IJ}_+$ and $L^{IJ}_-$
belong to inequivalent representations of $\sSO(4)$.

Let us convert the $\sSO(4)$ indices of 
 $L^{IJ}_+$ and $L^{IJ}_-$
into  $\sSU(2)$ ones following the rules described in subsection \ref{N4-sect}, 
and specifically eq. \eqref{2.611}.
The (anti) self-duality conditions \eqref{ASD} imply that  
\be
(\tau_I)^{i \tilde i } (\tau_J)^{j \tilde j} L_+^{IJ} =\varepsilon^{\tilde
i\tilde j}L^{ij}~,\qquad
(\tau_I)^{i\tilde i } (\tau_J)^{j \tilde j} L_-^{IJ} = \varepsilon^{ij} L^{\tilde i
\tilde j}~.
\ee
Here $L^{ij}$ and $L^{\tilde i \tilde j}$ are symmetric,
$L^{ij}=L^{ji}$ and $L^{\tilde i \tilde j}=L^{\tilde j \tilde i}$.
Since  $L^{IJ}_+$ and $L^{IJ}_-$ have different algebraic properties, 
the conservation equation (\ref{L-N4-conserv}) leads to
the two different analyticity conditions: 
\begin{subequations}
\label{4.5}
\bea
D^{\tilde i(i}_\alpha L^{kl)} &=&0~,\label{4.5a}\\
D^{i(\tilde i }_\alpha L^{\tilde k \tilde l)}&=&0~,
\eea
\end{subequations}
where $D^{\tilde i i}_\alpha \equiv D^{i \tilde i }_\alpha := (\tau_I)^{i \tilde i } D_\a^I$. 
It follows from (\ref{N4-L-transf}) and \eqref{ASD} that 
the superconformal transformation laws of 
 $L^{ij}$ and $L^{\tilde i\tilde j}$ are  
\begin{subequations}
\bea
\delta L^{ij} &=& -\xi L^{ij} -\sigma(z) L^{ij} + 2\Lambda^{(i}_k(z) L^{j)k}~,\label{4.6a}\\
\delta L^{\tilde i\tilde j} &=& -\xi L^{\tilde i\tilde j} -\sigma(z) L^{\tilde i\tilde j}
+ 2 \Lambda^{(\tilde i}_{\tilde k}(z) L^{\tilde j)\tilde k}~,
\eea
\end{subequations}
where $\Lambda^{ij}(z)$ and $\Lambda^{\tilde i\tilde j}(z) $ are constructed from
$\Lambda^{IJ}(z)$ by the rule \eqref{2.611}.

We emphasise that the flavour current multiplets $L^{ij}$ and $L^{\tilde i
\tilde j}$ are completely independent and can be studied independently of each other. Since their properties are very similar, here we will 
consider in detail only the correlation functions for $L^{ij}$
and comment on the correlators of $L^{\tilde i \tilde j}$
at the end of this section.

The properties of $L^{ij}$ given by
its conservation equation (\ref{4.5a}) and superconformal transformation (\ref{4.6a}) 
are very similar to those of the $\cN=3$ flavour current multiplet, eqs. 
(\ref{L-conserv}) and  (\ref{L-transform}). 
This similarity is not accidental since 
there proves to exist a unique $\cN=3$ flavour current multiplet  $L^{ij}_{(\cN=3)}$
associated with $L^{ij} \equiv L^{ij}_{(\cN=4)}$.\footnote{Here we have attached 
the labels $(\cN=3)$ and $(\cN=4)$ to these superfields
to distinguish them. Below, when no confusion is possible, these labels are omitted.}
The former is obtained from the latter through the procedure of 
$\cN=4 \to \cN=3$ superspace reduction which has been discussed in the literature
in the cases of $\cN=4$ Minkowski \cite{Zupnik2010} and anti-de Sitter \cite{BKT-M}
superspaces (see also \cite{BKS}). As applied to $ L^{ij}_{(\cN=4)}$, it works as follows. 
For the Grassmann coordinates $\q^\a_I$ of the $\cN=4$ superspace, 
we make  $3+1$ splitting $\q_I \to (\q_{\hat I}, \q_4)$ and then 
consider the $\q_4$-independent 
component of $ L^{ij}_{(\cN=4)}$. It proves to be the desired
$\cN=3$ flavour current multiplet, 
\be
L^{ij}_{(\cN=3)} = L^{ij}_{(\cN=4)} |_{\theta_4=0}~.
\label{L3L4}
\ee
In fact, it is possible to define an inverse correspondence, 
$L^{ij}_{(\cN=3)} \to L^{ij}_{(\cN=4)}$. Specifically, 
given an $\cN=3$ superfield $L^{ij}_{(\cN=3)}$ subject to 
the constraint (\ref{L-conserv}),  there exists a unique
$\cN=4$ superfield $L^{ij}_{(\cN=4)}$ obeying the constraint (\ref{4.5a})
and related to $L^{ij}_{(\cN=3)}$ by (\ref{L3L4}). This means that
all components in the $\theta_4$-expansion of $L^{ij}_{(\cN=4)}$
can be restored from the lowest one given by $L^{ij}_{(\cN=3)}$.

The above simple observation appears crucial for finding
the correlation functions of the $\cN=4$ flavour current multiplets.
Indeed, the expressions (\ref{3.8}) and (\ref{3pt-flavour-final})
can be considered as the lowest components in the
$\theta_4$-expansion of the corresponding
correlators of the $\cN=4$ flavour current multiplets. Moreover,
the full information is encoded in these parts of the correlators
since the higher-order corrections in $\theta_4$ can be uniquely
restored from these lowest components.

Based on these observations, we propose the following ansatz for
the correlation functions of several $\cN=4$ flavour current multiplets $L^{ij\,\bar
a}$,  $\bar a=1,\ldots,n$. The two-point function is
\be
\langle L^{ij\,\bar a}(z_1) L^{kl\,\bar b}(z_2) \rangle
= \frac{a_{\cN=4}}2 \frac{\delta^{\bar a\bar b}({\bf u}_{12}^{ik} {\bf u}_{12}^{jl}
+{\bf u}_{12}^{jk} {\bf u}_{12}^{il})}{{\bm
x}_{12}{}^2}~,
\label{L-N4-2pt}
\ee
while the three-point function reads
\begin{subequations}
\label{3pt-flavour-N4}
\bea
\langle L^{ij\,\bar a}(z_1) L^{kl\,\bar b}(z_2) L^{mn\,\bar c}(z_3) \rangle
&=&b_{\cN=4}\, f^{\bar a\bar b\bar c}
\frac{{\bf u}_{13}^{ii'}{\bf u}_{13}^{jj'}
 {\bf u}_{23}^{kk'} {\bf u}_{23}^{ll'} }{{\bm x}_{13}{}^2{\bm x}_{23}{}^2}
H_{i'j'\,k'l'}{}^{mn}({\bm X}_{3},\Theta_{3})~,\label{3ptL-N4}\\
H^{ij\,kl\,mn}({\bm X}_3,\Theta_3) &=&\frac1{\sqrt2}
\left(
\frac{\varepsilon^{m(i} {\bf U}_3^{j)(l}\varepsilon^{k)n}}{{\bm X}_3}
+
\frac{\varepsilon^{n(i}{\bf U}_3^{j)(l}\varepsilon^{k)m}}{{\bm X}_3}
\right)~.
\label{3ptH-N4}
\eea
\end{subequations}
The two- and three-point building blocks  ${\bf u}^{ij}_{12}$ and ${\bf
U}^{ij}_3$ used in these expressions were introduced in subsection
\ref{N4-sect}. Taking into account the relations (\ref{2.60}) and (\ref{2.61}),
it is clear that (\ref{3.8}) and (\ref{3pt-flavour-final}) are
related to (\ref{L-N4-2pt}) and (\ref{3ptL-N4}) via the superspace
reduction described above 
\begin{subequations}
\label{4.10}
\bea
\langle L^{ij\,\bar a}_{(\cN=4)}(z_1) L^{kl\,\bar b}_{(\cN=4)}(z_2)
\rangle|_{\theta_4=0} &=&
\langle L^{ij\,\bar a}_{(\cN=3)}(z_1) L^{kl\,\bar b}_{(\cN=3)}(z_2)
\rangle~,\\
\langle L_{(\cN=4)}^{ij\,\bar a}(z_1) L_{(\cN=4)}^{kl\,\bar b}(z_2)
L_{(\cN=4)}^{mn\,\bar c}(z_3) \rangle|_{\theta_4=0}
&=&\langle L_{(\cN=3)}^{ij\,\bar a}(z_1) L_{(\cN=3)}^{kl\,\bar b}(z_2)
 L_{(\cN=3)}^{mn\,\bar c}(z_3) \rangle~.~~~~
\eea
\end{subequations}

Let us rewrite the correlation function (\ref{L-N4-2pt}) in terms
of the two-point matrix (\ref{380a})
\be
\langle L^{ij\,\bar a}(z_1) L^{kl\,\bar b}(z_2) \rangle
= \frac{a_{\cN=4}}2\delta^{\bar a\bar b}
(n_{12}^{ik} n_{12}^{jl}
+n_{12}^{jk} n_{12}^{il})~.
\label{4.12}
\ee
Then, owing to (\ref{2.54}), it is obvious that (\ref{4.12}) obeys
the conservation condition
\be
D^{\tilde i'(i'}_\alpha\langle L^{ij)\bar a}(z_1) L^{kl\,\bar b}(z_2)
\rangle=0~\qquad (z_1\ne z_2)~.
\label{4.13}
\ee
In the same manner we express (\ref{3ptH-N4}) in terms of
(\ref{2.58a})
\be
H^{ij\,kl\,mn}({\bm X}_3,\Theta_3) =\frac1{\sqrt2}
\left(
\varepsilon^{m(i} N_3^{j)(l}\varepsilon^{k)n}
+
\varepsilon^{n(i} N_3^{j)(l}\varepsilon^{k)m}
\right)
\ee
and observe that
\be
{\cal D}^{\tilde i'(i'}_\alpha H^{ij)kl\,mn}({\bm X},\Theta) =0~,
\label{4.15}
\ee
as a consequence of (\ref{2.59}). The equations (\ref{4.13}) and
(\ref{4.15}) prove that the correlation functions of $\cN=4$ flavour current multiplets 
constructed in (\ref{L-N4-2pt}) and (\ref{3pt-flavour-N4}) 
do obey the necessary conservation laws.

As concerns 
the correlation functions of the flavour current
multiplets $L^{\tilde i \tilde j}$, they  have the same form as eqs. (\ref{L-N4-2pt}) and
(\ref{3pt-flavour-N4}) but with the indices $i,j,\ldots$ replaced with
$\tilde i,\tilde j\,\ldots$ 
Note also that all mixed two- and and three-point correlators 
involving both $L^{ij}$ and $L^{\tilde i \tilde j}$ vanish. 

\subsection{Correlation functions of the supercurrent}
\label{sectJJJ}

In accordance with \cite{BKNT-M,KNT-M} (see also \cite{BKS}),
the $\cN=4$ supercurrent is described by a primary real 
scalar $J$ subject to 
the conservation equation
\be
D^{I\alpha}D^{K}_\alpha J =\frac14\delta^{IK}D^{L\alpha}
D^L_{\alpha}J~.
\label{N4-J-conserv}
\ee
Its superconformal transformation law is
\be
\delta J= -\xi J -\sigma(z) J~.
\ee
The constraint \eqref{N4-J-conserv} uniquely fixes the dimension of $J$ to be 1.

Since the supercurrent $J$ is a scalar superfield, its two-point
correlation function has a simple form
\be
\langle J(z_1) J(z_2) \rangle
=\frac{c_{\cN=4}}{{\bm x}_{12}{}^2}~,
\label{JJ}
\ee
where $c_{\cN=4}$ is a free coefficient. Using (\ref{x12-sym}) it
is easy to check that (\ref{JJ}) obeys the conservation law
(\ref{N4-J-conserv}).

The three-point correlation function of the $\cN=4$ supercurrent can be found
by making the following ansatz
\be
\langle J(z_1)J(z_2) J(z_3)\rangle =
\frac1{{\bm x}_{13}{}^2{\bm x}_{23}{}^2} H({\bm X}_3,\Theta_3)~,
\label{JJJ}
\ee
where the function $H$ has the homogeneity property
\be
H(\lambda^2 {\bm X},\lambda\Theta) =
\lambda^{-2}H({\bm X},\Theta)~,
\label{5.5}
\ee
for a real positive $\lambda$. With the use of
(\ref{useful-prop-a}), one can check that the supercurrent conservation
condition (\ref{N4-J-conserv}) implies a similar equation for $H$,
\be
{\cal D}^{I\alpha}{\cal D}^{K}_\alpha H =\frac14\delta^{IK}{\cal D}^{L\alpha}
{\cal D}^L_{\alpha}H~,
\label{H-conserv}
\ee
where ${\cal D}^I_\alpha$ is the generalized
spinor covariant derivative (\ref{generalized-DQ}).

The general solution of (\ref{5.5}) can be 
represented 
as a
$\Theta$-expansion
\be
H({\bm X},\Theta)=\frac{c_1}{{\bm X}}
+c_2 \frac{\Theta^2}{{\bm X}^2}
+c_3 \frac{\Theta^4}{{\bm X}^3}
+c_4\frac{\Theta^6}{{\bm X}^4}
+c_5\frac{\Theta^8}{{\bm X}^5}
+c_6\frac{\varepsilon_{IJKL}\Theta^{I\alpha}\Theta^{J\beta}\Theta^{K\gamma}\Theta^{L\delta}
{\bm X}_{\alpha\beta}{\bm X}_{\gamma\delta}}{{\bm X}^5}~,
\label{5.6}
\ee
where $c_i$ are some coefficients. It is useful to rewrite
(\ref{5.6}) in terms of the symmetric part
$X_{\alpha\beta}$ of ${\bm X}_{\alpha\beta}$ given in (\ref{XX}). The result is 
\be
H=\frac{d_1}{ X}
+d_2 \frac{\Theta^2}{ X^2}
+d_3 \frac{\Theta^4}{ X^3}
+d_4\frac{\Theta^6}{ X^4}
+d_5\frac{\Theta^8}{ X^5}
+d_6\frac{\varepsilon_{IJKL}\Theta^{I\alpha}\Theta^{J\beta}\Theta^{K\gamma}\Theta^{L\delta}
 X_{\alpha\beta} X_{\gamma\delta}}{X^5}~,
\label{5.7}
\ee
where $d_i$ are some coefficients which can be, in principle,
expressed in terms of $c_i$. For the function $H$ in the form
(\ref{5.7}) it is easy to check that it solves (\ref{H-conserv})
for
\be
d_2 = d_3 = d_4 = d_5 =0~,
\ee
and $d_1$, $d_6$ are arbitrary real. Thus, if we denote $d_6=
d_{\cN=4}$ and $d_1 = \tilde d_{\cN=4}$, the solution for $H$ is
\bea
H({\bm X},\Theta)&=& \frac{\tilde d_{\cN=4}}{X}
+d_{\cN=4}\frac{\varepsilon_{IJKL}\Theta^{I\alpha}\Theta^{J\beta}\Theta^{K\gamma}\Theta^{L\delta}
 X_{\alpha\beta} X_{\gamma\delta}}{X^5}
\non\\
&=&\tilde d_{\cN=4}\left(
\frac1{\bm X} +\frac18 \frac{\Theta^4}{{\bm X}^3}+ \frac{3}{128}\frac{\Theta^8}{{\bm X}^5} \right)
+ d_{\cN=4} \frac{\varepsilon_{IJKL}\Theta^{I\alpha}\Theta^{J\beta}\Theta^{K\gamma}\Theta^{L\delta}
{\bm X}_{\alpha\beta}{\bm X}_{\gamma\delta}}{{\bm X}^5}~.~~~~~~~~~
\label{H-N4-result}
\eea
Here, in the second line of (\ref{H-N4-result}), we expressed the
function $H$ in terms of 
${\bm X}_{\alpha\beta}$ given in (\ref{XX}) and transforming covariantly
under the superconformal group.

Using the identities (\ref{222a}) and (\ref{222b}) it is possible
to show that for arbitrary $d_{\cN=4}$ and $\tilde d_{\cN=4}$
the expression (\ref{H-N4-result}) obeys the equation
\be
H(-{\bm X}_1^{\rm
T},-\Theta_1)=\frac{{\bm x}_{12}{}^2}{{\bm x}_{23}{}^2}H({\bm
X}_3,\Theta_3)~,
\ee
which must hold as  a consequence of the symmetry property
\be
\langle J(z_1)J(z_2) J(z_3)\rangle =
\langle J(z_3) J(z_2) J(z_1) \rangle~.
\ee
Thus, the $\cN=4$ supercurrent three-point correlation function
(\ref{JJJ}) with $H$ given by (\ref{H-N4-result})
obeys all the constraints dictated by the superconformal symmetry and
the conservation equation with  {\it two} arbitrary real coefficients
$d_{\cN=4}$ and $\tilde d_{\cN=4}$. The structure of this three-point correlator 
has some similarities with that for the 4D $\cN=2$ supercurrent 
computed in \cite{KT}.

As was demonstrated in \cite{BKS},  the three-point functions of 
the supercurrent in $\cN=1,2,3$
superconformal theories involve only {\it one} free parameter.
In this regard, our $\cN=4$ result given by eqs. (\ref{JJJ}) and (\ref{H-N4-result})
may look rather puzzling, 
since every $\cN=4$ superconformal field theory is a special 
 $\cN=3$ superconformal field theory. The resolution of this puzzle 
 is as follows. We showed in \cite{BKS} that the $\cN=4$ supercurrent 
 consists of two $\cN=3$ multiplets, one of which is the $\cN= 3$ supercurrent 
 and the other multiplet includes conserved currents that are not present 
 in general $\cN=3$ superconformal fields theories (the fourth supersymmetry 
 current and the $R$-symmetry currents associated with the coset space 
 $\sSO(4) / \sSO(3)$). In subsection \ref{AppA}, 
by performing the  $\cN=4 \to \cN=3$ reduction of 
the $\cN=4$ supercurrent correlation function (\ref{JJJ}),
we demonstrate that the first term in (\ref{H-N4-result}) does not contribute to the
three-point function of the $\cN=3$ supercurrent. Hence, it also does not
contribute to the three-point correlation function of the energy-momentum tensor upon
further reduction down to the component fields. This means that just like 
in $\cN=1,2,3$  superconformal theories 
the three-point function of the energy-momentum tensor depends just on a single
tensor structure and a single free coefficient $d_{\cN=4}$.

\subsection{Mixed correlators}

For completeness, we also present mixed three-point correlation functions
involving both the supercurrent and flavour current multiplets. 
It is not difficult to see that 
\bea
\langle
L^{ij\,\bar a}(z_1) J(z_2) J(z_3)
\rangle =0~.
\eea
 However, for the correlator with one
supercurrent and two flavour current multiplet insertions we get
\begin{subequations}
\bea
\langle
L^{ij\,\bar a}(z_1) J(z_2) L^{kl\, \bar b}(z_3)
\rangle &=& \delta^{\bar a\bar b}
\frac{{\bf u}_{13}^{ii'} {\bf u}_{13}^{jj'}}{{\bm x}_{13}{}^2
 {\bm x}_{23}{}^2 } H_{i'j'}{}^{kl}({\bm X}_3,\Theta_3)~,\label{LJL}\\
H^{ij\, kl}({\bm X},\Theta) &=&c\left( N^{i(k} \varepsilon^{l)j} + N^{j(k}
\varepsilon^{l)i}  \right)
=c\frac{{\bf U}^{i(k} \varepsilon^{l)j} + {\bf U}^{j(k}
\varepsilon^{l)i} }{{\bm X}}~,
\label{H-mixed}
\eea
\end{subequations}
where $c$ is a constant. The tensor $H^{ij\, kl}$ is
expressed in terms of the matrices ${\bf U}^{ij}$ and
$N^{ij}$ which are given by (\ref{U-N4a}) and (\ref{2.58a}),
respectively. This tensor is found as the
general solution of the equations
\begin{subequations}
\label{427}
\bea
{\cal D}_\alpha^{\tilde m (m } H^{ij)kl}&=&0~,\label{427a}\\
{\cal Q}^{I\alpha}{\cal Q}^J_\alpha H^{ij\, kl} &=& \frac14
\delta^{IJ}{\cal Q}^{K\alpha}{\cal  Q}_{K\alpha} H^{ij\, kl}~,
\label{427b}
\eea
\end{subequations}
which are the corollaries of the analyticity of the flavour
current multiplet (\ref{4.5a}) and the supercurrent conservation law
(\ref{N4-J-conserv}). In deriving the equations (\ref{427}) the
identities (\ref{useful-prop}) have been used.

The equation
(\ref{427a}) immediately follows from the analyticity of the
matrix $N^{ij}$, see (\ref{2.59}). To check the equation
(\ref{427b}) it is convenient to rewrite it in terms of the
$\sSU(2)$ indices
\be
{\cal Q}^{i(\tilde i \alpha}{\cal Q}^{\tilde j) j}_\alpha H^{i'j'\,
kl}=0~.
\ee
Then, it is easy to see that this equation is satisfied as a
consequence of the following property of the matrix (\ref{2.58a})
\be
{\cal D}^{i(\tilde i \alpha}{\cal D}^{\tilde j) j}_\alpha  N^{kl}
={\cal Q}^{i(\tilde i \alpha}{\cal Q}^{\tilde j) j}_\alpha  N^{kl}
=0~.
\ee

Finally, we note that the tensor (\ref{H-mixed}) obeys the
constraint
\be
H_{ij}{}^{kl}({\bm X}_3,\Theta_3)=
\frac{{\bm X}_1}{{\bm X}_3} {\bf u}_{31ii'} {\bf u}_{31jj'}
{\bf u}_{31}^{kk'}{\bf u}_{31}^{ll'} H_{k'l'}{}^{i'j'}(-{\bm X}_1^{\rm
T},-\Theta_1)~,\label{H-symmetry}
\ee
which is a corollary of the following symmetry property of the
correlation function (\ref{LJL})
\be
\langle
L^{ij\,\bar a}(z_1) J(z_2) L^{kl\, \bar b}(z_3)
\rangle  =
\langle
L^{kl\, \bar b}(z_3) J(z_2)   L^{ij\,\bar a}(z_1)
\rangle~.
\ee
The equation (\ref{H-symmetry}) can be easily verified with the
use of the relation (\ref{UUuu}) which links together the thee-point and
two-point unitary matrices.


\section{Free $\cN=4$ hypermultiplets}

In this section we consider a family of trivial $\cN=4$ superconformal 
field theories -- models for free hypermultiplets. In these models, 
the correlation functions of conserved currents can be computed exactly.  
Using  such results will allow us to derive important relations between 
the numerical parameters appearing in certain two- and three-point functions 
in general $\cN=4$ superconformal field theories. 

\subsection{On-shell hypermultiplets}

In 3D $\cN=4$ supersymmetry, 
there are two types of free on-shell hypermultiplets,
left  $q^i$ and  right $q^{\tilde i}$, 
that transform as isospinors
of the different subgroups
$\sSU(2)_{\rm L}$ and $\sSU(2)_{\rm R}$  of the $R$-symmetry group.
They obey the following constraints
\begin{subequations}\label{q-analyt}
\bea
D_\alpha^{\tilde i (i} q^{j)} &=&0~,
\label{q-analyt-a}
\\
D_\alpha^{i(\tilde i } q^{\tilde j)}&=&0~,
\label{q-analyt-b}
\eea
\end{subequations}
which are similar to those introduced by Sohnius \cite{Sohnius}
to describe the $\cN=2$ hypermultiplet in four dimensions.
These primary superfields  possess 
the superconformal transformation laws
\bea
\delta q^i &=& -\xi q^i -\frac12 \sigma(z) q^i + \Lambda^i_j(z)
q^j~,\non\\
\delta q^{\tilde i} &=& -\xi q^{\tilde i} -\frac12 \sigma(z) q^{\tilde i}
+ \Lambda^{\tilde i}_{\tilde j}(z) q^{\tilde j}~.
\eea
The constraints \eqref{q-analyt-a} and  \eqref{q-analyt-b}
uniquely fix the dimension of 
$q^i$ and $q^{\tilde i}$ to be 1/2.
Associated with $q^i$ and $q^{\tilde i}$ are their conjugates
\be
\overline{\,q^i\,} = \bar q_i = \varepsilon_{ij} \bar q^j~,\qquad
\overline{\, q^{\tilde i} \,} = \bar q_{\tilde i} = \varepsilon_{\tilde i\tilde j} \bar q^{\tilde j}~,
\ee
which may be seen to obey the same constrains 
as $q^i$ and $q^{\tilde i}$.

In accordance with the general results in section \ref{section2}, 
the two-point correlation functions of the primary superfields $q^i$ and
$q^{\tilde i}$ with their conjugates are
\begin{subequations}
\label{qq-correlators}
\bea
\langle
q^i(z_1) \bar q^j(z_2)
 \rangle &=& \frac1{4\pi } \frac{{\bf u}^{ij}_{12}}{{\bm
 x}_{12}}
=\frac1{4\pi } n^{ij}_{12}~ ,
\label{qq-correlator-a}\\
\langle
q^{\tilde i}(z_1) \bar q^{\tilde j} (z_2)
 \rangle &=& \frac1{4\pi } \frac{{\bf u}^{\tilde i\tilde j}_{12}}{{\bm x}_{12}}
 =\frac1{4\pi } n^{\tilde i\tilde j}_{12}~,
 \label{qq-correlator-b}
\eea
\end{subequations}
where the matrices ${\bf u}^{ij}_{12}$, ${\bf u}^{\tilde i
\tilde j}_{12}$ and $n^{ij}_{12}$, $n^{\tilde i\tilde j}_{12}$ are defined in (\ref{uu})
and (\ref{380}), respectively. On the one hand, the expressions
for $\langle q^i(z_1) \bar q^j(z_2) \rangle$ and 
$\langle q^{\tilde i}(z_1) \bar q^{\tilde j} (z_2) \rangle$
in terms of ${\bf u}^{ij}_{12}$, ${\bf u}^{\tilde i \tilde j}_{12}$
and ${\bm x}_{12}$ guarantee that they comply with 
the requirement of superconformal invariance, eq. \eqref{2.77}.
On the other hand,
expressing these correlators in terms of $n_{12}^{ij}$ and $n^{\tilde i\tilde j}_{12}$
allows one to check easily that these two-point
functions obey the analyticity constraints 
\eqref{q-analyt-a} and  \eqref{q-analyt-b}, owing to (\ref{2.54}). 

There exist several off-shell realisations for the hypermultiplet, see Appendix A for a review. 
In any off-shell realisation for the left hypermultiplet, the two-point function 
\eqref{qq-correlator-a} may differ only by contact terms
from that corresponding to the off-shell formulation.  
In particular, in the harmonic superspace approach one deals with 
the  $q^+$-hypermultiplet for which it holds that  
\bea
\langle q^+(z_1, u_1)\,\breve{q}^+(z_2,u_2) \rangle
=  u^+_{1i} \,\langle q^i(z_1)\,\bar q_j(z_2)\rangle \,u^{+j}_{2}
~+~ \mbox{contact terms}~.
\label{5.555}
\eea
Similar comments apply to the right hypermultiplet correlator \eqref{qq-correlator-b}.
For our purposes in this section, it suffices to work with the two-point functions 
\eqref{qq-correlator-a} and \eqref{qq-correlator-b}. A careful treatment of the singularities
of the  two- and three-point functions at coincident points 
is beyond the scope of this paper.

It should be pointed out that 
switching off the Grassmann variables in (\ref{qq-correlators})
leads to 
to the correctly normalised correlators of free complex scalars,
\begin{subequations}
\label{6.5}
\bea
\langle q^i(z_1) \bar q_j(z_2)  \rangle |_{\theta=0}
&=& \langle \vf^i(x_1) \bar \vf_j(x_2) \rangle
 = \frac1{4\pi}\delta^i_j \frac1{\sqrt{(x_1-x_2)^2}}~,\\
\langle
q^{\tilde i}(z_1) \bar q_{\tilde j}(z_2)  \rangle |_{\theta=0}
&=& \langle \vf^{\tilde i}(x_1) \bar \vf_{\tilde j}(x_2) \rangle
 = \frac1{4\pi}\delta^{\tilde i}_{\tilde j} \frac1{\sqrt{(x_1-x_2)^2}}~,
\eea
\end{subequations}
where $\vf^i(x) = q^i(z)|_{\theta=0}$, $\vf^{\tilde i}(x) = q^{\tilde i}(z)|_{\theta=0}$.


\subsection{Two-point correlators}

Let us consider a free model of $m$ left hypermultiplets $q^i$
and $n$ right hypermultiplets $q^{\tilde i}$.  
We assume that $q^i$ transforms in an irreducible representation
of a simple flavour
group $G_{\rm L}$ with generators $\Sigma^{\bar a}$.
Similarly, $q^{\tilde i}$ is assumed to 
transform in an irreducible 
representation of another simple flavour group 
$ G_{\rm R}$
with generators $\Sigma^{\tilde a}$.
Viewing $q^i $ and $q^{\tilde i}$ as column vectors and their 
Hermitian conjugates $\bar q_i $ and $\bar q_{\tilde i}$ 
as row vectors,
the supercurrent $J$ is
\bea
J&=&  \bar q_{\tilde i} q^{\tilde i}- \bar q_i  q^i~,
\label{example-J}
\eea
and the flavour current multiplets $L^{ij\,\bar a}$, $L^{\tilde i\tilde j \, \tilde
a}$ are given by 
\bea
L_{ij}^{\bar a} &= &- \ri \bar q_{(i} \Sigma^{\bar a} q_{j)}~,\qquad
L_{\tilde i\tilde j}^{\tilde a} = - \ri \bar q_{(\tilde i} \Sigma^{\tilde a}
q_{\tilde j)}~.
\label{example-L}
\eea
We assume that the generators of the flavour groups are normalised
such that
\be
\tr(\Sigma^{\bar a} \Sigma^{\bar b}) = k_{\rm L} \delta^{\bar a\bar b}~,
\qquad
\tr(\Sigma^{\tilde a} \Sigma^{\tilde b}) = k_{\rm R} \delta^{\tilde a\tilde
b}~.
\ee
The normalisation constants $ k_{\rm L}$ and $ k_{\rm R}$ depend on the 
representations of the flavour groups $G_{\rm L}$ and $G_{\rm R}$ chosen. 
One can check that, due to the free equations of motion
(\ref{q-analyt}), the current multiplets (\ref{example-J}) and
(\ref{example-L}) obey the conservation laws (\ref{N4-J-conserv})
and (\ref{4.5}), respectively.

The notable feature of of the  supercurrent \eqref{example-J}
is that $J$ is asymmetric with respect to the left and right hypermultiplets. 
The supergravity origin of this property will be discussed in section 
\ref{section8}. 

We compute the two-point correlation functions of the supercurrent and
flavour current multiplets for the free hypermultiplets. 
Since there is no correlation between  the
superfields $q^i$ and $q^{\tilde i}$, the  two-point
function for the supercurrent is given by
\be
\langle J(z_1) J(z_2) \rangle =
 \langle q^{\tilde i}(z_1)\bar q_{\tilde i} (z_1) q^{\tilde j}(z_2)\bar q_{\tilde j}(z_2) \rangle
+ \langle q^i(z_1)\bar q_i (z_1) q^j(z_2)\bar q_j(z_2) \rangle~.
\ee
Performing the Wick contractions and making use of \eqref{qq-correlators}, we find
\be
\langle J(z_1) J(z_2) \rangle =
\frac{m}{(4\pi)^2} \frac{{\bf u}_{12}^{ij}{\bf u}_{12ij}}{{\bm x}_{12}{}^2}
+\frac{n}{(4\pi)^2} \frac{{\bf u}_{12}^{\tilde i\tilde j}{\bf u}_{12\tilde i\tilde j}}{{\bm x}_{12}{}^2} =
\frac1{8\pi^2 }\frac{m+n}{{\bm x}_{12}{}^2}~.
\ee
In a similar way we find two-point correlation functions of
flavour current multiplets
\begin{subequations}
\bea
\langle L^{\bar a}_{ij}(z_1) L^{\bar b}_{kl}(z_2) \rangle
 &=& \frac{ k_{\rm L}}{32\pi^2}
\frac{({\bf u}_{12il}{\bf u}_{12 jk}+{\bf u}_{12 jl}{\bf u}_{12 ik})}{{\bm
x}_{12}{}^2} \delta^{\bar a\bar b} ~,\\
\langle L^{\tilde a}_{\tilde i\tilde j}(z_1) L^{\tilde b}_{\tilde k\tilde l}(z_2) \rangle
 &=& \frac{{k}_{\rm R} }{32\pi^2}
\frac{({\bf u}_{12\tilde i\tilde l}{\bf u}_{12\tilde j\tilde k}
+{\bf u}_{12\tilde j\tilde l}{\bf u}_{12\tilde i\tilde k})}{{\bm x}_{12}{}^2}  \delta^{\tilde a\tilde b} ~.
\eea
\end{subequations}
Comparing these correlation functions with (\ref{L-N4-2pt}) and
(\ref{JJ}) we find the following values for the  coefficients $a_{\cN=4}$ and $c_{\cN=4}$:
\bea
\label{a-coef}
a_{\cN=4} &=& \frac{k_{\rm L}}{16\pi^2}~,\\
c_{\cN=4} &=& \frac{m+n}{8\pi^2}~.
\label{c-coef}
\eea

\subsection{Three-point correlators} 
For the three-point function of the flavour current multiplets
$L_{ij}^{\bar a}$, which are  defined  by (\ref{example-L}), we have
\be
\langle L_{ij}^{\bar a}(z_1) L_{kl}^{\bar b}(z_2) L_{mn}^{\bar
c}(z_3) \rangle =\ri
\langle
\bar q_{(i}(z_1)\Sigma^{\bar a}  q_{j)}(z_1)
\bar q_{(k}(z_2)\Sigma^{\bar b}  q_{l)}(z_2)
\bar q_{(m}(z_3)\Sigma^{\bar c}  q_{n)}(z_3)
\rangle ~.
\ee
Performing the Wick contractions and using the explicit form of
the hypermultiplet two-point function (\ref{qq-correlator-a}) we
find
\be
\langle L_{ij}^{\bar a}(z_1) L_{kl}^{\bar b}(z_2) L_{mn}^{\bar
c}(z_3) \rangle
=-\frac{f^{\bar a\bar b\bar c} k_{\rm L}}{128\pi^3} \frac{
{\bf u}_{12i(k}{\bf u}_{23l)(m}{\bf u}_{31n)j}
+{\bf u}_{12j(k}{\bf u}_{23l)(m}{\bf u}_{31n)i} }{{\bm x}_{12}{\bm x}_{13}{\bm x}_{23}}~.
\label{6.15}
\ee
Using the identity ${\bm x}_{12}{}^2 = {\bm X}_3{}^2 {\bm x}_{13}{}^2 {\bm
x}_{23}{}^2$ the denominator in (\ref{6.15}) can be written as
\be
\frac1{{\bm x}_{12}{\bm x}_{13}{\bm x}_{23} }
=\frac1{{\bm x}_{13}{}^2 {\bm x}_{23}{}^2 {\bm X}_3}~.
\label{xxX}
\ee
With the help of (\ref{UUuu}) the correlation function (\ref{6.15}) gets
exactly the form (\ref{3pt-flavour-N4}) with
\be
b_{\cN=4} = -\frac{{\sqrt2}}{128\pi^3} k_{\rm L}~.
\label{b-coef-N4}
\ee
Comparing this coefficient with (\ref{a-coef}) we observe that
\be
\frac{b_{\cN=4}}{a_{\cN=4}} = -\frac{\sqrt2}{8\pi}~.
\label{ab}
\ee
Although this relation between the coefficients of the two-point and
three-point correlation functions is obtained for the free
hypermultiplets, we propose that it is universal for any $\cN=4$
superconformal field theory. Indeed, the
relation (\ref{ab}) can be considered as a manifestation of a
Ward identity relating the two- and three-point correlation functions
of the flavour current multiplets. Since both of these correlation functions depend
on a single tensor structure the relation between their coefficients can be
found by considering a particular theory.
The explicit form of the relevant $\cN=4$ Ward identity will be derived in the next section.

The three-point correlation function for the flavour current
multiplets $L_{\tilde i \tilde j}^{\tilde a}$ can be analysed in a
similar way, with the same relation (\ref{ab}) between the
coefficients.

Now we turn to computing the three-point correlator for the supercurrent
(\ref{example-J}). In the right-hand side of  
\bea
\langle J(z_1) J(z_2) J(z_3) \rangle &=&
\langle q^{\tilde i}(z_1)\bar q_{\tilde i}(z_1) q^{\tilde j}(z_2)\bar q_{\tilde j}(z_2) q^{\tilde k}(z_3) \bar
q_{\tilde k}(z_3)\rangle
\non\\&&
-\langle q^i(z_1)\bar q_i(z_1) q^j(z_2)\bar q_j(z_2) q^k(z_3) \bar
q_k(z_3) \rangle
\eea
we perform the Wick contractions and make use of (\ref{qq-correlators}) to 
get
\bea
\langle J(z_1) J(z_2) J(z_3) \rangle &=&
\frac m{(4\pi)^3} \frac{
{\bf u}_{12}{}^{\tilde i}{}_{\tilde j}{\bf u}_{23}{}^{\tilde j}{}_{\tilde k}{\bf u}_{31}{}^{\tilde k}{}_{\tilde i}
 +{\bf u}_{13}{}^{\tilde i}{}_{\tilde k}{\bf u}_{21}{}^{\tilde j}{}_{\tilde i}{\bf u}_{32}{}^{\tilde k}{}_{\tilde j}
}{{\bm x}_{13} {\bm x}_{12} {\bm x}_{23}}
\non\\&&
-\frac n{(4\pi)^3} \frac{
{\bf u}_{12}{}^{ i}{}_{ j}{\bf u}_{23}{}^{j}{}_{ k}{\bf u}_{31}{}^{ k}{}_{i}
 +{\bf u}_{13}{}^{ i}{}_{ k}{\bf u}_{21}{}^{ j}{}_{ i}{\bf u}_{32}{}^{ k}{}_{ j}
}{{\bm x}_{13} {\bm x}_{12} {\bm x}_{23}}
\non\\
&=&
\frac{2m}{(4\pi)^3}\frac{{\bf U}_3{}^{\tilde i}{}_{\tilde i}}{{\bm x}_{12}{\bm x}_{23}{\bm x}_{13}}
-\frac{2n}{(4\pi)^3}\frac{{\bf U}_3{}^i{}_i}{{\bm x}_{12}{\bm x}_{23}{\bm
x}_{13}}~.
\label{6.21}
\eea
Here, in the last line, we have applied the relations (\ref{UUuu}).
Next, using the identity (\ref{xxX}), we express (\ref{6.21}) in
terms of  $N^{ij}$ and $N^{\tilde i \tilde j}$ introduced
in (\ref{U-N-N4})
\be
\langle J(z_1) J(z_2) J(z_3) \rangle
=\frac1{32\pi^3}\frac1{{\bm x}_{13}{}^2 {\bm x}_{23}{}^2}(m\, N_3{}^{\tilde
i}{}_{\tilde i} - n\, N_3{}^i{}_i )~.
\ee
Taking into account the explicit form of the matrices $N^{ij}$ and $N^{\tilde i \tilde j}$
given in (\ref{U-N-N4}), we conclude that the correlator has the form (\ref{H-N4-result}), 
\begin{subequations}\label{dd-free}
\bea
\langle J(z_1) J(z_2) J(z_3) \rangle
=\frac1{{\bm x}_{13}{}^2 {\bm x}_{23}{}^2}
\left(  \frac{\tilde d_{\cN=4} }{X_3}+ \frac{ d_{\cN=4} }{X_3{}^5}
\varepsilon_{IJKL}\Theta^{I\alpha}\Theta^{J\beta}\Theta^{K\gamma}\Theta^{L\delta}
 X_{\alpha\beta} X_{\gamma\delta}
\right)~,~~~~~~
\eea
where
\bea
d_{\cN=4} = \frac{m+n}{128 \pi^3}~,\qquad
\tilde d_{\cN=4} = \frac{m-n}{16\pi^3}~.
\label{dd-free-b}
\eea
\end{subequations}

As discussed at the end of subsection \ref{sectJJJ}, it is the 
$d_{\cN=4}$-term in \eqref{dd-free} which contributes 
to the three-point function of the energy-momentum tensor
upon reduction to the component fields.
In accordance with  (\ref{dd-free-b}),
the coefficient $d_{\cN=4}$ receives additive contributions from
the  $q^i$ and $q^{\tilde i}$ hypermultiplets.
The other
coefficient $\tilde d_{\cN=4}$ is non-zero when $m \neq n$.
It is known that the mirror map  $\mathfrak M$ \cite{Zupnik99,Zupnik2009}
turns every left hypermultiplet
$q^i$ into a right one,  $q^{\tilde i}$, and vice versa,   see Appendix A.
Invariance under the mirror map implies that 
the theory has
the same number of the  $q^i$ and $q^{\tilde i}$ hypermultiplets.
Thus, we conclude that the non-vanishing
value of $\tilde d_{\cN=4}$ indicates 
that the superconformal theory under consideration is not invariant under 
the mirror map. 

The ratio of the coefficient $d_{\cN=4}$
in the three-point function (\ref{dd-free}) with $c_{\cN=4}$, 
which determines the two-point correlator (\ref{c-coef}), is
\be
\frac{d_{\cN=4}}{c_{\cN=4}} = \frac1{16\pi}~.
\label{cdN4}
\ee
Although we have found this relation for  the special model of free
$\cN=4$ hypermultiplets, we expect that \eqref{cdN4} is universal
for all $\cN=4$ superconformal models, as a consequence of a Ward
identity. Indeed, there is a Ward identity relating the
two- and three-point functions of the energy-momentum tensor 
(see, e.g., \cite{OP}).  
Since in $\cN=4$ superconformal field theories each of them
is determined by a single tensor structure, 
  the relation between their coefficients can be found by considering a particular theory.
Of course, it is possible to derive a superfield Ward identity expressing 
the $\cN=4 $ superconformal symmetry. Since its main application is to give
another derivation of  \eqref{cdN4}, we will not  indulge in this technical 
issue in the present paper. 


\section{Ward identities for flavour current multiplets}

The Ward identities play an important role in quantum field theory as
they relate different Green functions.
In this section  we derive Ward identities for flavour current
multiplets in $\cN$-extended superconformal field theories,
 with $1\leq \cN \leq 4$. 
 Such Ward identities 
relate the two- and three-point correlation functions of the flavour current 
multiplets and, in principle,
allow one to relate the parameters in these correlators.
The common feature of the four supersymmetry types $1\leq \cN \leq 4$
 is that for each of these cases 
the Yang-Mills multiplet possesses 
 an unconstrained prepotential formulation. 

To derive the Ward identities we will use a standard 
field theoretic construction that can be described as follows.
Consider a superconformal field theory that possesses a flavour current multiplet $L$
(with all indices suppressed). We gauge the flavour symmetry by coupling the theory to
a background vector multiplet described by an unconstrained prepotential $V$
which will be the source for $L$. An $n$-point function for $L$ is obtained 
by computing $n$ functional derivatives of the generating functional $Z[V]$ 
with respect to $V$ and then switching $V$ off, 
\bea
\ri^n \langle L(1) \dots L(n) \rangle =
\frac{\d^n   Z[V]}{ \d V(1) \dots \d V(n) }  \bigg|_{V=0}~,
\eea
where the operator insertions on the left are taken at distinct points. 
The Ward identities follow from the condition of the gauge invariance of $Z[V]$.

\subsection{$\cN=1$ superconformal theories}

In $\cN=1$ superconformal field theory, the flavour current multiplet
is described by a primary real spinor superfield $L^{\bar a}_\alpha$ of dimension $3/2$
subject to the conservation condition
\be
D^\alpha L^{\bar a}_\alpha =0~,
\ee
with  $\bar a$ being the flavour index (see \cite{BKS} for more details).
We now gauge the flavour symmetry by coupling the theory to
a background vector multiplet described by a spinor prepotential 
$V^{\bar a}_\alpha$, which is real but otherwise unconstrained 
(see \cite{GGRS} for the details).
The gauge transformation law of $V^{\bar a}_\alpha$ is 
\be
\delta_\lambda V_\alpha^{\bar a} = D_\alpha \lambda^{\bar a}
 -f^{\bar a\bar b\bar c} V_\alpha^{\bar b} \lambda^{\bar c}~,
\ee
with the superfield gauge parameters $\lambda^{\bar a}$ being real but otherwise unconstrained.
The gauge prepotential  $V^{\bar a}_\alpha$ is  the source
for the flavour current multiplet in the sense that 
\be
\ri\langle L_{\alpha}^{\bar a}(z) \rangle_V = \frac{\delta Z[V]}{\delta V^{\alpha \bar
a}(z)}~,
\ee
where $ Z[V]$ is the generating functional.
As usual,  $\langle \dots \rangle_V$ denotes a correlation function
in the presence of the background field. 
The gauge invariance of 
$Z[V]$ implies 
that
\be
\int \rd^{3|2}z\, (D_\alpha \lambda^{\bar a}
 -f^{\bar a\bar b\bar c} V_{\alpha}^{\bar b} \lambda^{\bar c})\frac{ \delta
Z[V]}{\delta V_\alpha^{\bar a}(z)} =0~.
\ee
Since the gauge parameters $\lambda^{\bar a}$ are arbitrary superfields, 
we conclude that 
\be
\left(D^\alpha \frac\delta{\delta V^{\alpha \bar a}}
 -f^{\bar a\bar b\bar c} V^{\bar b\,\alpha}\frac\delta{\delta V^{\alpha \bar c}}
 \right)Z[V]=0~.
\ee
Varying this identity twice 
and switching off the source $V^{\bar
a}_{\alpha}$, we end up with the Ward identity for $\cN=1$ flavour
current multiplets
\bea
D^\alpha \langle
L_\alpha^{\bar a}(z) L_\beta^{\bar b}(z_1) L_\gamma^{\bar c}(z_2)
\rangle
+\ri f^{\bar a\bar b\bar d}\delta^{3|2}(z-z_1)
\langle L_\beta^{\bar d}(z_1) L_\gamma^{\bar c}(z_2) \rangle&&\non\\
+\ri f^{\bar a\bar c\bar d} \delta^{3|2}(z-z_2)
\langle L_\beta^{\bar b}(z_1)L_\gamma^{\bar d}(z_2) \rangle&=&0~.
\eea
Here $\delta^{3|2}(z-z')$ is the $\cN=1$ superspace delta-function.

\subsection{$\cN=2$ superconformal theories}

The $\cN=2$ flavour current multiplet is described by a primary real 
scalar superfield $L^{\bar a}$ of dimension 1 subject to the conservation
equation
\be
(D^{\alpha(I} D_\alpha^{J)}- \frac12 \delta^{IJ}
D^{\alpha K} D_\alpha^{K} )L^{\bar a} =0~,
\label{178}
\ee
see \cite{BKS} for more details.

In this subsection 
it is useful to deal with 
complex Grassmann coordinates $\theta^\alpha$ and $\bar\theta^\alpha$
for $\cN=2$ superspace that are related to the real ones, $\q^\a_I$,
as follows:
\be
\theta^\alpha=\frac1{\sqrt2}(\theta^{\alpha}_1+\ri\theta^{\alpha}_2)~,\qquad
\bar\theta^\alpha =
\frac1{\sqrt2}(\theta^{\alpha}_1-\ri\theta^{\alpha}_2)~.
\label{4.24}
\ee
The corresponding spinor covariant derivatives are
\be
D_\alpha = \frac1{\sqrt2}(D^1_\alpha - \ri D^2_\alpha)~,\qquad
\bar D_\alpha  =-\frac1{\sqrt2}(D^1_\alpha + \ri
D^2_\alpha)~.
\ee
In this basis, the conservation equations (\ref{178}) turn into the conditions
\be
D^2 L^{\bar a} =0~,\qquad \bar D^2 L^{\bar a}=0~,
\ee
which mean that $L^{\bar a}$ is a real linear superfield. 

We gauge the flavour symmetry by coupling the theory to
a background vector multiplet described by a prepotential 
$V^{\bar a}$, which is real but otherwise unconstrained \cite{HKLR,ZP}.
The gauge transformation of the prepotential is 
\be
\delta_\lambda V^{\bar a} = \frac\ri2 (\bar \lambda^{\bar a} - \lambda^{\bar a})
+\frac12 f^{\bar a\bar b\bar c } V^{\bar b} (\lambda^{\bar c} + \bar \lambda^{\bar c})
+
\ldots
~,
\label{V-gauge_}
\ee
where the gauge parameter $\lambda^{\bar a}$ 
is an arbitrary chiral scalar superfield. 
The ellipsis in (\ref{V-gauge_}) stands for those terms which are at least 
quadratic
in $V^{\bar a}$, and therefore are irrelevant for
the Ward identity relating  the two- and three-point
correlation functions. 
Below, we will systematically neglect the
$\cO(V^2)$-terms in the gauge transformation of $V^{\bar a}$.

The gauge prepotential $V^{\bar a}$ is the  source
for the  flavour current multiplet $L^{\bar a}$ which is obtained 
from the generating functional $Z[V]$ by 
\be
\ri\langle L^{\bar a}(z)\rangle_V =\frac{\delta Z[V]}{\delta V^{\bar
a}(z)}~.
\ee
The gauge invariance of the generating functional 
is expressed as
\be
\int \rd^{3|4}z \left(
\frac\ri2 (\bar \lambda^{\bar a} - \lambda^{\bar a})
+\frac12 f^{\bar a\bar b\bar c } V^{\bar b} (\lambda^{\bar c} + \bar \lambda^{\bar c})
+\ldots
\right)\frac{\delta Z[V]}{\delta V^{\bar a}(z)}=0~.
\ee
Since the gauge parameters
$\lambda^{\bar a}$ are arbitrary chiral superfields, 
we end up with the following identity for the
generating functional $Z$:
\bea
\bar D^2\left(\frac\delta{\delta V^{\bar a}(z)}-\ri f^{\bar a\bar b\bar c}
V^{\bar b}(z)\frac\delta{\delta V^{\bar c}(z)} +\ldots \right)Z[V]&=&0~
\eea
and its conjugate. 
Varying this equation twice and switching off the gauge superfield $V^{\bar
a}$, we obtain the Ward identity relating the  two- and three-point
correlation functions of $\cN=2$ flavour current multiplets
\bea\label{WI0}
\bar D^2\langle L^{\bar a}(z) L^{\bar e}(z_1) L^{\bar d}(z_2) \rangle
-4f^{\bar a \bar e \bar c}\delta_+(z,z_1)
 \langle L^{\bar c}(z_1) L^{\bar d}(z_2) \rangle
 &&\non\\
-4f^{\bar a\bar d\bar c}\delta_+(z,z_2)
 \langle L^{\bar e}(z_1) L^{\bar c}(z_2) \rangle &=&0~.
 \label{WI2}
\eea
Here $\delta_+(z,z')$ 
is the chiral delta-function; it  is expressed in terms of the full superspace
delta-function $\delta^{3|4}(z-z')$ in the standard way
\be
\delta_+(z,z')  = -\frac14 \bar D^2 \delta^{3|4}(z-z')~.
\ee

\subsection{$\cN=3$ superconformal theories}
\label{sect-N3-Ward}

It is  known that the conventional
3D $\cN=3$ Minkowski superspace ${\mathbb M}{}^{3|6} $
is not suitable to realise off-shell $\cN=3$ supersymmetric theories. 
The  adequate superspace setting for them \cite{ZH} is 
${\mathbb M}{}^{3|6} \times {\mathbb C}P^1$, which is  
an extension of ${\mathbb M}{}^{3|6} $ 
by the compact coset space
$\sSU(2) /\sU(1)$ associated with the $R$-symmetry 
group.\footnote{For every positive integer $\cN$,
the 3D $\cN$-extended superconformal group $\sOSp(\cN|4;{\mathbb R})$ is a transformation group of  the  compactified Minkowski superspace
$\overline{\mathbb M}{}^{3|2\cN}$ in which Minkowski superspace
${\mathbb M}{}^{3|2\cN}$ is embedded as a dense open domain \cite{KPT-MvU}.
In the $\cN=3$ case, $\sOSp(3|4;{\mathbb R})$ is also defined to act transitively
on $\overline{\mathbb M}{}^{3|6} \times {\mathbb C}P^1$, 
as shown in  \cite{KPT-MvU}.}
 The most general  $\cN=3$ supersymmetric gauge theories 
 in three dimensions can be described 
 using either the harmonic superspace techniques \cite{ZH,Zupnik99}
 or the projective ones \cite{KLT-M11}. These formulations 
are 3D analogues of the 4D $\cN=2$ harmonic  \cite{GIKOS,GIOS} 
and projective \cite{KLR,LR} superspace approaches 
(see also \cite{K-Lectures} for a review of the projective superspace formalism). 
The 3D $\cN=3$ projective superspace setting has been used to construct 
the most general off-shell $\cN=3$ superconformal 
$\s$-models \cite{KPT-MvU}  and supergravity-matter couplings \cite{KLT-M11}.
The 3D $\cN=3$ harmonic superspace has been shown to be  
efficient for studying the quantum aspects of $\cN=3$ superconformal theories
\cite{BILPSZ}. It also  provides an elegant description of the ABJM theory
\cite{Buchbinder:2008vi}. In this subsection we will use the harmonic superspace 
to derive Ward identities for $\cN=3$ flavour current multiplets. 

We will use
$\sSU(2)$ harmonic variables $u^+_i$ and $u^-_i$ 
constrained by 
\be
u^{+i} u^-_j - u^{-i} u^+_j = \delta^i_j~, \qquad \overline{ u^{+i} } = u^-_i~.
\label{u-harmonics}
\ee
Associated with these
variables there are the following vector fields
\be
\partial^{++} = u^+_i\frac\partial{\partial u^-_i}~,\quad
\partial^{--} = u^-_i\frac\partial{\partial u^+_i}~,\quad
\partial^0=
u^+_i \frac\partial{\partial u^+_i}
-u^-_i \frac\partial{\partial u^-_i}~,
\label{harm-derivatives}
\ee
which form the $\sSU(2)$ algebra
\bea
[ \pa^0 , \pa^{++} ]= 2 \pa^{++}~, \quad [ \pa^0 , \pa^{--} ]= -2 \pa^{--}~,
\quad [\partial^{++},\partial^{--}] = \pa^0~.
\eea
Using these harmonic variables 
allows one to introduce a new basis for 
the Grassmann
variables $\theta^{ij}_\alpha$ and the spinor covariant 
derivatives $D^{ij}_\alpha$:
\begin{subequations}
\bea
\theta^{ij}_\alpha & ~\longrightarrow ~& 
(\theta^{++}_\alpha, \theta^{--}_\alpha, \theta^0_\alpha)
 = (u^+_i u^+_j  \q^{ij}_\a, \, u^-_i u^-_j  \q^{ij}_\a, \, u^+_i u^-_j  \q^{ij}_\a) 
 ~,\\
D^{ij}_\alpha & ~\longrightarrow ~& (D^{++}_\alpha, D^{--}_\alpha, D^0_\alpha)
 = (u^+_i u^+_j D^{ij}_\a, \, u^-_i u^-_j D^{ij}_\a, \, u^+_i u^-_j D^{ij}_\a)
 ~.
\eea
\end{subequations}

As discussed in section \ref{section3}, 
the $\cN=3$ flavour current multiplet is described by a
primary superfield $L^{ij}=L^{ji}$ subject to the conservation law
(\ref{L-conserv}). Associated with this superfield are the following
harmonic projections:
\be
L^{++} = u^+_i u^+_j L^{ij}~,\quad
L^{--} = u^-_i u^-_j L^{ij}~,\quad
L^0 = u^+_i u^-_j L^{ij}~.
\label{L-projections}
\ee
It is sufficient to study only one of these projections, say
$L^{++}$, since the others can be obtained by acting on $L^{++}$
with $\partial^{--}$. By construction, $L^{++}$ is annihilated by $\pa^{++}$, 
\be
\partial^{++} L^{++} =0 ~.
\label{7.22}
\ee
It is important that the equation (\ref{L-conserv}) has the
following corollary
\be
D^{++}_\alpha L^{++} =0~,
\label{7.23}
\ee
which is usually referred to as the analyticity condition.

The main feature of the harmonic superspace 
is that it allows one to introduce new off-shell multiplets 
that are annihilated by $D^{++}_\alpha $. Such superfields 
are defined on a supersymmetric subspace of
${\mathbb M}{}^{3|6} \times {\mathbb C}P^1$ known as the analytic subspace.
It is parametrised by coordinates 
\be
\zeta = (x_A^{a} , \theta^{++}_\alpha, \theta^0_\alpha,
u^\pm_i)~,
\label{analyt-coord}
\ee
where
\be
x_A^a = x^a + \ri \gamma^a_{\alpha\beta}
\theta^{++\alpha}\theta^{--\beta}~.
\ee
In the analytic coordinate basis for ${\mathbb M}{}^{3|6} \times {\mathbb C}P^1$
consisting of the variables $\z$ and $\q^{--}_\a$, 
the spinor covariant derivative
$D^{++}_\alpha$ becomes short,
\be
D^{++}_\alpha = \frac\partial{\partial \theta^{--\alpha}}~,
\ee
while the  harmonic derivative $\partial^{++}$ acquires additional
terms
\be
{\mathscr D}^{++} = \partial^{++}
+2\ri \gamma^a_{\alpha\beta} \theta^{++\alpha} \theta^{0\alpha}
\frac\partial{\partial x^a_A}
+\theta^{++\alpha} \frac\partial{\partial\theta^{0\alpha}}
+2\theta^{0\alpha}\frac\partial{\partial\theta^{--\alpha}}~.
\ee
Therefore, in the analytic basis,
the
equation (\ref{7.23}) 
tells us that 
$L^{++}= L^{++}(\zeta)$ while (\ref{7.22}) becomes a non-trivial
constraint
\be
{\mathscr D}^{++} L^{++} =0~.
\ee

We are prepared to derive Ward identities in a superconformal field 
theory possessing flavour current multiplets  $L^{++\, \bar a}$. 
For this we gauge the flavour symmetry by coupling the theory to
a background vector multiplet described by a prepotential 
$V^{++\,\bar a}$ which is an analytic real superfield.
Its gauge transformation reads
\be
\delta_\lambda V^{++\, \bar a} ={\mathscr D}^{++}\lambda^{\bar a } -f^{\bar a\bar b\bar
c}V^{++\,\bar b} \lambda^{\bar c}~,
\label{V-gauge}
\ee
where the gauge parameters $\lambda^{\bar a}$ are unconstrained analytic superfields. 
The gauge prepotential $V^{++ \bar a}$ 
is the  source
for the flavour current multiplet $L^{++ \bar a}$ which is obtained 
from the generating functional $Z[V]$ by 
\be
\ri\langle L^{++\,\bar a}(\zeta) \rangle_V = \frac{\delta Z[V]}{\delta V^{++\,\bar
a}(\zeta)}~.
\ee
The gauge invariance of $Z[V]$ implies the equation
\be
\int \rd\zeta^{(-4)} ({\mathscr D}^{++}\lambda^{\bar a}-f^{\bar a\bar b\bar c}V^{++\,\bar b} \lambda^{\bar c})
\frac{\delta Z[V]}{\delta V^{++\,\bar a}(\zeta)}=0~,
\label{7.31}
\ee
where $\rd\zeta^{(-4)}$ is the invariant measure on the analytic
subspace (\ref{analyt-coord}). Since  the gauge parameters $\lambda^{\bar
a}$ in (\ref{7.31}) are arbitrary, we conclude that 
\be
\left(
{\mathscr D}^{++}\frac\delta{\delta V^{++\,\bar a}(\zeta)} - f^{\bar a\bar b\bar c} V^{++\,\bar b}(\zeta)
\frac\delta{\delta V^{++\,\bar c}(\zeta)} \right)Z[V]=0~.
\label{DW=0}
\ee
Finally, varying this relation twice and switching off the gauge superfield $V^{++}$
we end up with the Ward identity for the correlation functions of flavour current
multiplets $L^{++\,\bar a}$
\bea
{\mathscr D}^{++}_{(\zeta)} \langle  L^{++\,\bar a}(\zeta)
 L^{++\,\bar b}(\zeta_1) L^{++\,\bar c}(\zeta_2)
\rangle
+\ri f^{\bar a\bar b\bar d}\delta_A^{(4,0)}(\zeta,\zeta_1)
 \langle  L^{++\,\bar d}(\zeta_1) L^{++\,\bar c}(\zeta_2) \rangle &&
 \non\\
+\ri f^{\bar a\bar c\bar d} \delta_A^{(4,0)}(\zeta,\zeta_2)
\langle L^{++\,\bar b}(\zeta_1) L^{++\,\bar d}(\zeta_2) \rangle&=&0~,
\label{L++Ward}
\eea
where $\delta_A^{(4,0)}(\zeta,\zeta')$ is the delta-function in the analytic
subspace.

\subsection{$\cN=4$ superconformal theories}
\label{sect7.4}

The $R$-symmetry group of the $\cN=4$ super-Poincar\'e algebra
is $\sSU(2)_{\rm L}\times \sSU(2)_{\rm R}$. 
It is the superspace \eqref{biharmonic} which is adequate to formulate 
general off-shell $\cN=4$ supersymmetric theories. 
Hence, one can introduce harmonic variables for
either of the $\sSU(2)$ subgroups, or  for both of them. For
studying Ward identities involving correlation functions of the
left flavour current multiplets $L^{ij}$ it is sufficient to introduce
harmonic variables for the subgroup $\sSU(2)_{\rm L}$ which acts on the indices
$i,j$. We will use the same harmonic variables $u^\pm_i$ constrained by 
(\ref{u-harmonics}) and the corresponding harmonic
derivatives (\ref{harm-derivatives}). Now we project the $\cN=4$
Grassmann variables $\theta^{i \tilde i }_\alpha$ and
spinor covariant derivatives $D^{i \tilde i }_\alpha$ as
\bea
\theta^{i \tilde i}_\alpha &\longrightarrow& 
(\theta^{\tilde i +}_\alpha, \theta^{\tilde i - }_\alpha) 
= (u^+_i \theta^{i \tilde i }_\alpha, \, u^-_i  \theta^{i \tilde i }_\alpha)~,\\
D^{i \tilde i  }_\alpha &\longrightarrow& (D^{\tilde i +}_\alpha,
D^{\tilde i - }_\alpha) = (u^+_i D^{ i\tilde i }_\alpha  , \,u^-_i D^{i \tilde i}_\alpha)~.
\eea
The flavour current multiplet $L^{ij}$ has the same
harmonic projections as in (\ref{L-projections}). The equation
(\ref{7.22}) remains unchanged in the $\cN=4$ case while 
the analyticity constraint (\ref{7.23}) turns into
\be
D^{\tilde i +}_\alpha L^{++} =0~.
\label{L++analyt}
\ee
This equation follows from (\ref{4.5a}) by contracting the indices
$i,k,l$ with the $u^+$-harmonics.

Let us consider the analytic subspace of the $\cN=4$ harmonic superspace
parametrised by the variables 
\be
\zeta = (x_A^a , \theta^{\tilde i +}_\alpha, u^\pm_i)~,
\label{zetaA}
\ee
where
\be
x_A^a = x^a +\ri \gamma^a_{\alpha\beta} \theta^{\tilde i+
\,\alpha}\theta^{-\,\beta}_{\tilde i}~.
\ee
In the coordinate system $(\z, \theta^{\tilde i - }_\alpha )$,  
the spinor covariant derivative
$D^{\tilde i+}_\alpha$ becomes short,
\be
D^{\tilde i+}_\alpha = \frac\partial{\partial \theta^{-\,\alpha}_{\tilde
i}}~,
\ee
while the covariant harmonic derivatives $\partial^{++}$ and $\partial^{--}$ acquire
additional terms
\begin{subequations}
\bea
{\mathscr D}^{++} = \partial^{++} + \ri\gamma^a_{\alpha\beta}
 \theta^{\tilde i+\,\alpha} \theta^{+\,\beta}_{\tilde i}
  \frac\partial{\partial x^a_A}
 +\theta^{\tilde i+}_\alpha \frac\partial{\partial\theta^{\tilde i -
 }_\alpha}~,
 \label{D++}\\
{\mathscr D}^{--} = \partial^{--} + \ri\gamma^a_{\alpha\beta}
 \theta^{\tilde i-\,\alpha} \theta^{-\,\beta}_{\tilde i}
  \frac\partial{\partial x^a_A}
 +\theta^{\tilde i-}_\alpha \frac\partial{\partial\theta^{\tilde i
 + }_\alpha}~.
\eea
\end{subequations}
The crucial feature of using the analytic coordinates in the
$\cN=4$ harmonic superspace is that the equation
(\ref{L++analyt}) is automatically solved by the analytic
superfield $L^{++} = L^{++}(\zeta)$ while (\ref{7.22}) turns
into a non-trivial constraint
\be
{\mathscr D}^{++} L^{++} =0~.
\label{6.422}
\ee

Once the analytic subspace (\ref{zetaA}) in the $\cN=4$ superspace is
introduced, the further derivation of the Ward identity for
$L^{++}$ goes exactly the same way as in sect.\
\ref{sect-N3-Ward} and the equations (\ref{V-gauge})--(\ref{DW=0})
remain unchanged. Thus, we end up with the Ward identity for
$L^{++}$ exactly in the form (\ref{L++Ward})
\bea
{\mathscr D}^{++}_{(\zeta)} \langle  L^{++\,\bar a}(\zeta)
 L^{++\,\bar b}(\zeta_1) L^{++\,\bar c}(\zeta_2)
\rangle
+\ri f^{\bar a\bar b\bar d}\delta_A^{(4,0)}(\zeta,\zeta_1)
 \langle  L^{++\,\bar d}(\zeta_1) L^{++\,\bar c}(\zeta_2) \rangle &&
 \non\\
+\ri f^{\bar a\bar c\bar d} \delta_A^{(4,0)}(\zeta,\zeta_2)
\langle L^{++\,\bar b}(\zeta_1) L^{++\,\bar d}(\zeta_2) \rangle&=&0~.~~~
\label{N4Ward}
\eea

In a similar way one can find the Ward identity for the right flavour
current multiplet $L^{\tilde i \tilde j}$ by introducing the
harmonic variables for the subgroup $\sSU(2)_{\rm R}$ of the
$R$-symmetry group.

It is instructive to check that the Ward identity (\ref{N4Ward})
is satisfied for the free hypermultiplets. 

Consider the action for a single hypermultiplet 
\bea
S = \int \rd\zeta^{(-4)} \breve{q}^+ {\mathscr D}^{++} q^+~,
\label{hyperact}
\eea
where $q^+$ is constrained by 
\be
D^{\tilde i +}_\alpha q^+ 
=0~,
\ee
and the same constraints hold for its  smile-conjugate $\breve{q}^+$.
 The superfield $q^+$ contains infinitely many
auxiliary component fields at the component level.
These auxiliary fields vanish on the equation of motion
\be
{\mathscr D}^{++} q^+ =0~,
\ee
which implies that the hypermultiplet superfields takes the form
\be
q^+(z,u)  = u^+_i q^i (z)~,\qquad
\breve{q}^+(z,u) = u^{+i} \bar q_i (z)~.
\ee

The two-point function
$\langle q^+(\zeta_1)\breve{q}^+(\zeta_2) \rangle $
corresponding to the action \eqref{hyperact} is
\be
\langle q^+(\zeta_1)\breve{q}^+(\zeta_2)\rangle = -\ri G^{(+,+)}(\zeta_1,\zeta_2)~.
\label{6.488}
\ee
Here we have introduced the Green function
$G^{(+,+)}(\zeta_1,\zeta_2)$ 
as a solution of the equation
\be
{\mathscr D}^{++} G^{(+,+)}(\zeta_1,\zeta_2) =
-\delta_A^{(3,1)}(\zeta_1,\zeta_2)~.
\label{Green-eq}
\ee
Explicitly, it can be represented in the following form (see Appendix \ref{AppC} for
the details)
\be
G^{(+,+)}(\zeta_1,\zeta_2) = -\frac\ri{4\pi } \frac{(u^+_1 u^+_2)}{
\sqrt{\hat x^a_{12} \hat x_{12a}}}~,
\label{Green}
\ee
where
\be
\hat x^a_{12} = x^a_{A\,1} - x^a_{A\, 2}
-\frac\ri{(u^+_1 u^+_2)}
[(u^-_1 u^+_1)\theta^{+\tilde i}_1 \gamma^a \theta^+_{1\tilde i}
-(u^+_1 u^-_2)\theta^{+\tilde i}_2 \gamma^a \theta^+_{2\tilde i}
+2 \theta^{+\tilde i}_1 \gamma^a \theta^+_{2\tilde i}]
\label{hat-x12}
\ee
is a manifestly analytic coordinate difference that is invariant under
$Q$-supersymmetry transformations. We point out that the two-point function 
\eqref{6.488} is related to \eqref{qq-correlator-a} according to 
eq. \eqref{5.555}.

Let us consider a free superconformal theory describing 
a  column vector ${\bm q}^+$ of several  $q^+$ hypermultiplets and 
their smile conjugates $\breve{\bm q}^+$ viewed as a row vector. 
The corresponding action, which is the sum of $n$ free actions  \eqref{hyperact}
is invariant under rigid flavour transformations 
\bea 
\d {\bm q}^+ =\ri \lambda^{\bar a } \S^{\bar a} {\bm q}^+~, \qquad
\d \breve{\bm q}^+ =- \ri \lambda^{\bar a } \breve{\bm q}^+\S^{\bar a} ~,
\eea
with constant real parameters $\l^{\bar a}$ and Hermitian generators
$\S^{\bar a}$ of the flavour group. 
We gauge this symmetry by coupling the hypermultiplets to  an analytic 
gauge prepotential $V^{++} = V^{++\bar a} (\z) \S^{\bar a} $ taking its values 
in the Lie algebra of the flavour group, 
\bea
S = \int \rd\zeta^{(-4)} \breve{\bm q}^+ {\mathscr D}^{++} {\bm q}^+
\quad \longrightarrow \quad 
\int \rd\zeta^{(-4)} \breve{\bm q}^+ \Big( {\mathscr D}^{++} 
+\ri V^{++}  \Big){\bm q}^+~.
\eea
From the action obtained we read off the  flavour current multiplet
\be
L^{++\bar a}  (\z)= \ri \breve{\bm q}^+ \Sigma^{\bar a} {\bm q}^+~.
\label{6.555}
\ee
By construction, $L^{++\bar a}$ respects the analyticity constraint 
\eqref{L++analyt}. It also obeys the condition \eqref{6.422}
on the mass shell.  Thus our new representation \eqref{6.555} for the flavour current 
multiplet is equivalent to \eqref{example-L} we used before. 

Let us use the new representation \eqref{6.555} 
to compute the two- and three-point functions of the flavour current 
multiplets.  
Performing the Wick contractions gives
\bea
\langle L^{++\bar a}(\zeta_1) L^{++\bar b}(\zeta_2) \rangle&=&
k_{\rm L} \delta^{\bar a \bar b} \langle q^+(\zeta_1) \bar q^+(\zeta_2)
\rangle \langle q^+(\zeta_1) \bar q^+(\zeta_2)
\rangle
 ~,\label{7.53}\\
\langle L^{++\bar a}(\zeta_1) L^{++\bar b}(\zeta_2) L^{++\bar c}(\zeta_3)
\rangle&=&-k_{\rm L} f^{\bar a\bar b \bar c}
\langle q^+(\zeta_1) \bar q^+(\zeta_3) \rangle
\langle q^+(\zeta_1) \bar q^+(\zeta_2) \rangle
\langle q^+(\zeta_2) \bar q^+(\zeta_3) \rangle~~~~~~~~
\label{7.54}
\eea
where the propagator is given by \eqref{6.488}.
Now we use the explicit form
of the hypermultiplet Green's function (\ref{Green}) and obtain
a new representation for the correlators of  the flavour current multiplets
\bea
\langle L^{++\bar a}(\zeta_1) L^{++\bar b}(\zeta_2) \rangle&=&
a_{\cN=4} \delta^{\bar a\bar b}\frac{(u^+_1 u^+_2)^2}{\hat x_{12}{}^2}
 ~,\label{7.53+}\\
\langle L^{++\bar a}(\zeta_1) L^{++\bar b}(\zeta_2) L^{++\bar c}(\zeta_3)
\rangle&=&\sqrt2 b_{\cN=4} f^{\bar a\bar b\bar c}
\frac{(u^+_1 u^+_2)(u^+_2 u^+_3)(u^+_3 u^+_1)}{
\sqrt{\hat x_{12}{}^2 \hat x_{23}{}^2 \hat x_{31}{}^2}}~.
\label{7.54+}
\eea
These expressions are manifestly analytic in all arguments.
They are equivalent to (\ref{L-N4-2pt}) and (\ref{3pt-flavour-N4}) modulo contact terms which vanish for non-coincident superspace points. We stress that
for generic values of $a_{\cN=4}$ and $b_{\cN=4}$ the form of the correlation functions (\ref{7.53+}) and (\ref{7.54+}) is universal for any $\cN=4$ superconformal theory although they were derived for free hypermultiplets. 
The values of the coefficients $a_{\cN=4}$ and $b_{\cN=4}$ for the case of free hypermultiplets are given by (\ref{a-coef}) and (\ref{b-coef-N4}), respectively.

Recall that the hypermultiplet Green's function (\ref{Green}) obeys the equation (\ref{Green-eq}). Using this equation we compute the derivative of the expression
(\ref{7.54+})
\bea
&&{\mathscr D}^{++}_{(1)} \langle L^{++\bar a}(\zeta_1) L^{++\bar b}(\zeta_2) L^{++c}(\zeta_3)
\rangle =
4\sqrt2 \ri\pi b_{\cN=4}  f^{\bar a \bar b \bar c}
[\delta_A^{(4,0)}(\zeta_1,\zeta_2)-\delta_A^{(4,0)}(\zeta_1,\zeta_3)]
\frac{(u^+_2 u^+_3)}{\hat x_{12}{}^2}~
\non\\&&=
4\sqrt2 \ri\pi \frac{b_{\cN=4}}{a_{\cN=4}}  f^{\bar a \bar b \bar d}
[\delta_A^{(4,0)}(\zeta_1,\zeta_2)-\delta_A^{(4,0)}(\zeta_1,\zeta_3)]
\langle L^{++\bar c}(\zeta_2) L^{++\bar d}(\zeta_3) \rangle
~.
\eea
Hence, the correlation functions (\ref{7.53+}) and (\ref{7.54+}) obey the Ward identity (\ref{N4Ward}) if the coefficients $a_{\cN=4}$ and $b_{\cN=4}$ are related to each other by the equation (\ref{ab}) which was found previously for the case of free hypermultiplets. 
Here we have demonstrated that it holds for every $\cN=4$ superconformal field theory.

\section{Relations between correlation functions
in superconformal field theories with $1\leq \cN \leq 4$}

The study of correlation functions performed in the present paper
is the continuation of our earlier work \cite{BKS}. 
In \cite{BKS} and in sections \ref{section3}  and \ref{section4} of the present paper, we derived explicit expressions for the two- and three-point
correlation functions of the supercurrent and flavour current
multiplets in three-dimensional $\cN$-extended superconformal field theories
 with $1\leq \cN \leq 4$.
As was discussed above, the coefficients of the two- and three-point function
are not independent but are related by the Ward identities.
The aim of this section is to derive the relations between these coefficients for $1\leq \cN \leq 4$.
Our derivation will be based on the following two observations.

\begin{itemize}

\item
If both the two- and the three-point functions are fixed up to overall coefficients
and are related to each other by the Ward identities, we can find a universal, model independent relation between the coefficients
by considering a particular theory. We have already used this observation to obtain 
the relations \eqref{ab} and \eqref{cdN4}
which are valid in any $\cN=4$ superconformal field theory.\footnote{The coefficient $\tilde{d}_{{\cal N}=4}$ does not
contribute to the three-point function of the energy-momentum tensor and, hence, does not appear in the Ward identities.
Thus, it is not related to the coefficients $c_{\cN=4}$ and $d_{\cN=4}$ in a universal manner.}

\item
Every $\cN=4$ superconformal theory can also be viewed 
as a special $\hat \cN$-extended superconformal theory, 
with $\hat \cN \leq 3$.
Since the relevant two- and three-point functions
in theories with $\cN=1, 2, 3$ supersymmetries are fixed up to 
overall coefficients\footnote{In the case
of the three-point function of flavour current multiplets in $\cN=2$ superconformal 
field theories, there is a second structure proportional to 
the totally symmetric tensor of the flavour symmetry group
\cite{BKS}. However, this structure
does not contribute  to the three-point functions of  conserved currents. 
Hence, it does not
contribute to the Ward identities and
can be ignored for our discussion.} \cite{BKS}
we can find similar universal relations between 
the  coefficients using eqs.~\eqref{ab} and~\eqref{cdN4}.
\end{itemize}

To perform explicit calculations we use the fact that the correlation functions of conserved currents for
different $\cN$ are related to each other by the superspace
reduction. Indeed, as explained in \cite{BKS}, the supercurrents
in $1\leq\cN\leq3$ superconformal theories can be derived from the
supercurrent $J$ in the $\cN=4$ theory by applying covariant
spinor derivative and switching off some of the Grassmann
coordinates. The flavour current multiplets in $1\leq\cN\leq3$
theories can also be derived from the $\cN=4$ flavour current
multiplets by applying the rules of the superspace reduction
discussed in \cite{BKS}.

\subsection{Superspace reduction of the supercurrent correlation functions}
\subsubsection{$\cN=3$ supercurrent}
Let us start with the $\cN=4$ supercurrent $J$ whose
correlation functions are given by (\ref{JJ}), (\ref{JJJ}) and \eqref{H-N4-result}. The $\cN=3$
supercurrent $J_\alpha$ is related to $J$ as follows
\be
J_\alpha = \ri D^4_\alpha J |~,
\ee
where the bar-projection means that we set $\theta_{4}^{\alpha} =0$. Hence,
for the correlation functions of $J_\alpha$ we have
\bea
\langle J_\alpha(z_1) J_\beta(z_2)\rangle &=& -D^4_{(1)\alpha}
D^4_{(2)\beta} \langle J(z_1) J(z_2)\rangle|~,\\
\langle J_\alpha(z_1) J_\beta(z_2) J_\gamma(z_3) \rangle &=&
-\ri D^4_{(1)\alpha}
D^4_{(2)\beta} D^4_{(3)\gamma} \langle J(z_1) J(z_2) J(z_3) \rangle|~.
\label{8.3}
\eea
Computation of the required derivatives of (\ref{JJ}) and (\ref{JJJ}) is a
straightforward but tedious task. The details of this procedure
are given in subsection \ref{AppA}. Here we present the results:
\begin{subequations}
\label{N3-J-correlators}
\bea
\langle
J_\alpha(z_1) J_\beta(z_2)
 \rangle &=& \ri c_{\cN=3} \frac{{\bm x}_{12\alpha\beta}}{{\bm
 x}_{12}{}^4}~,\label{8.4a}\\
\langle
J_\alpha(z_1)J_\beta(z_2) J_\gamma(z_3) \rangle
&=& d_{\cN=3}\frac{{\bm x}_{13\alpha\alpha'}{\bm x}_{23\beta\beta'}}{{\bm x}_{13}{}^4 {\bm x}_{23}{}^4}
H^{\alpha'\beta'}{}_\gamma({\bm X}_{3},\Theta_{3})~,\label{8.4b}\\
H^{\alpha\beta}{}_\gamma({\bm X},\Theta)&=&\frac{1 }{{\bm X}^5}
\Big[
(\delta^\beta_\gamma {\bm X}^{\alpha\rho}
 +\delta^\alpha_\gamma {\bm X}^{\rho\beta}){\bm X}^{\mu\nu}\Theta^I_\mu \Theta^J_\nu\Theta^K_\rho
  \varepsilon_{IJK}\non\\&&
 +{\bm X}^{\beta\alpha}{\bm X}^{\mu\nu}\Theta^I_{\mu}\Theta^J_{\nu}
 \Theta^K_{\gamma}\varepsilon_{IJK}
+2{\bm X}^{\alpha\mu}{\bm X}^{\nu\beta}\Theta^I_{\mu}\Theta^J_{\nu}\Theta^K_{\gamma}
 \varepsilon_{IJK}\Big]~,~~~~~
 \label{8.4c}
\eea
\end{subequations}
where
\begin{subequations}
\bea
c_{\cN=3} &=& 2c_{\cN=4}~,\label{8.5a}\\
d_{\cN=3} &=& 4d_{\cN=4}~.
\label{8.5b}
\eea
\end{subequations}
Eqs.~\eqref{N3-J-correlators} are precisely the expressions for the correlation functions of the $\cN=3$ supercurrent
obtained in~\cite{BKS}.
Using eq.~(\ref{cdN4}) we then obtain the following relation between the
coefficients
\be
\frac{d_{\cN=3}}{c_{\cN=3}} = \frac1{8\pi}~.
\label{8.6}
\ee
We expect that this relation is valid in any $\cN=3$ superconformal theory.

\subsubsection{$\cN=2$ supercurrent}
The $\cN=2$ supercurrent $J_{\alpha\beta}$ is related to the
$\cN=3$ supercurrent $J_\alpha$ as follows
\be
J_{\alpha\beta} = D^3_\alpha J_\beta |~,
\ee
where the bar-projection means that $\theta_3^{\alpha}=0$. Hence,
the correlation functions of $\cN=2$ supercurrents can be found
from (\ref{N3-J-correlators}) by the rules
\bea
\langle J_{\alpha\alpha'}(z_1) J_{\beta\beta'}(z_2)\rangle &=& -D^3_{(1)\alpha}
D^3_{(2)\beta} \langle J_{\alpha'}(z_1) J_{\beta'}(z_2)\rangle|~,\\
\langle J_{\alpha\alpha'}(z_1) J_{\beta\beta'}(z_2) J_{\gamma\gamma'}(z_3) \rangle &=&
- D^3_{(1)\alpha}
D^3_{(2)\beta} D^3_{(3)\gamma} \langle J_{\alpha'}(z_1) J_{\beta'}(z_2) J_{\gamma'}(z_3) \rangle|~.
\label{8.9}
\eea
The details of computations of these derivatives are given in
subsection \ref{AppB}. The resulting expressions are:
\bea
\langle J_{\a \b} (z_1) J^{\a' \b'} (z_2) \rangle &=& c_{\cN=2}
\frac{{\bm x}_{12 \a}{}^{ (\a'} {\bm x}_{12 \b}{}^{ \b')}}{ {\bm x}_{12}{}^6}~,
\label{n2.3}
\\
\langle J_{\a \a'} (z_1)   J_{\b \b'} (z_2) J_{\g \g'} (z_3) \rangle
&=&d_{\cN=2}
\frac{{\bm x}_{13 \a \r} {\bm x}_{13 \a' \r'}   {\bm x}_{23  \b \s}
 {\bm x}_{23 \b' \s'}  }{ {\bm x}_{13}{}^6 {\bm x}_{23}{}^6}
H^{\r \r', \s \s'}{}_{\g \g'} ({\bm X}_{3}, \Theta_{3})~,
\label{n2.4}
\eea
where
\bea
&&H^{\alpha\alpha',\beta\beta',\gamma\gamma'}({\bm X},\Theta)
=
\frac{2\ri}{{\bm X}^3}\left[
\varepsilon^{\alpha(\beta}\varepsilon^{\beta')\alpha'}\Theta_I^\gamma\Theta_J^{\gamma'}
+\varepsilon^{\alpha(\gamma}\varepsilon^{\gamma')\alpha'}\Theta_I^\beta \Theta_J^{\beta'}
+\varepsilon^{\beta(\gamma}\varepsilon^{\gamma')\beta'}\Theta_I^\alpha\Theta_J^{\alpha'}
\right] \varepsilon^{IJ}\non\\&&
+\frac{\ri}{{\bm X}^5}\left[
3{\bm X}^{\alpha\alpha'}{\bm X}^{\gamma\gamma'}\Theta_I^\beta \Theta_J^{\beta'}
+3{\bm X}^{\beta\beta'}{\bm X}^{\gamma\gamma'}\Theta_I^\alpha \Theta_J^{\alpha'}
-5{\bm X}^{\alpha\alpha'}{\bm X}^{\beta\beta'}\Theta_I^\gamma\Theta_J^{\gamma'}
\right]\varepsilon^{IJ}\non\\&&
+\frac{\ri}{{\bm X}^5}\left[
5\varepsilon^{\alpha(\gamma}\varepsilon^{\gamma')\alpha'}{\bm X}^{\beta\beta'}
+5\varepsilon^{\beta(\gamma}\varepsilon^{\gamma')\beta'} {\bm X}^{\alpha\alpha'}
-3\varepsilon^{\alpha(\beta}\varepsilon^{\beta')\alpha'} {\bm X}^{\gamma\gamma'}
\right]{\bm X}^{\delta\delta'}\Theta^I_\delta
\Theta^J_{\delta'}\varepsilon_{IJ}\non\\&&
+\frac52\frac{\ri}{{\bm X}^7}{\bm X}^{\alpha\alpha'}{\bm X}^{\beta\beta'}{\bm X}^{\gamma\gamma'}
 {\bm X}^{\delta\delta'}\Theta^I_\delta \Theta^J_{\delta'}\varepsilon_{IJ}
~.
\label{7.41}
\eea
The coefficients $c_{\cN=2}$ and $d_{\cN=2}$ in (\ref{n2.3}) and
(\ref{n2.4}) are related to $c_{\cN=3}$ and $d_{\cN=3}$ as
\be
c_{\cN=2} = -4 c_{\cN=3}~,\qquad
d_{\cN=2} = -6 d_{\cN=3}~.
\label{cdN2}
\ee
Eq.~\eqref{n2.3}, \eqref{n2.4}, \eqref{7.41} are precisely the correlation functions of the $\cN=2$ supercurrent
obtained in~\cite{BKS}.

Taking into account (\ref{8.6}), we find the ratio of the coefficients
(\ref{cdN2}):
\be
\frac{d_{\cN=2}}{c_{\cN=2}} = \frac3{16\pi}~,
\label{8.14}
\ee
which we expect to be valid in any $\cN=2$ superconformal theory.

\subsubsection{$\cN=1$ supercurrent}
Consider now the reduction of the $\cN=2$ supercurrent
$J_{\alpha\beta}$ to the $\cN=1$ supercurrent
\be
J_{\alpha\beta\gamma} = \ri D^2_\alpha J_{\beta\gamma} |~,
\ee
where the bar-projection means that $\theta_2^{\alpha} =0$.
The corresponding relation for the supercurrent correlation
functions reads
\bea
\langle J_{\alpha \alpha' \alpha''}(z_1) J_{\beta\beta' \beta''}(z_2)
 \rangle &=&
-D^2_{(1)\alpha} D^2_{(2)\beta} \langle
J_{\alpha'\alpha''}(z_1) J_{\beta' \beta''}(z_2) \rangle |~,\\
\langle J_{\alpha\alpha'\alpha''}(z_1)
J_{\beta\beta'\beta''}(z_2) J_{\gamma\gamma'\gamma''}(z_3) \rangle
&=&-\ri D^{ 2}_{(1)\alpha} D^{2}_{(2)\beta} D^{2}_{(3)\gamma }
\langle J_{\alpha'\alpha''}(z_1)
J_{\beta'\beta''}(z_2) J_{\gamma'\gamma''}(z_3) \rangle
|~.~~~~~~~~~~
\eea
The computations of these expressions were performed in
\cite{BKS}. Here we give only the result:
\bea
\langle  J_{\a \b \g} (z_1)  J^{\a' \b' \g'} (z_2)\rangle &=&\ri c_{{\cal N}=1}
\frac{ {\bm x}_{12 \a}{}^{( \a'} {\bm x}_{12 \b}{}^{ \b'}
 {\bm x}_{12 \g}{}^{ \g')} }{ {\bm x}_{12}{}^8}~,
 \label{8.18}
 \\
\langle
J_{\alpha\alpha'\alpha''}(z_1)
J_{\beta\beta'\beta''}(z_2) J_{\gamma\gamma'\gamma''}(z_3)
\rangle &=& \ri d_{\cN=1} \frac{{\bm x}_{13\alpha}{}^\rho {\bm x}_{13\alpha'}{}^{\rho'}
{\bm x}_{13\alpha''}{}^{\rho''} {\bm x}_{23\beta}{}^{\sigma}
{\bm x}_{23\beta'}{}^{\sigma'} {\bm x}_{23\beta''}{}^{\sigma'} }{
{\bm x}_{13}{}^8 {\bm x}_{23}{}^8}
\non\\&&\times
H_{\rho\rho'\rho''\, \sigma\sigma'\sigma''\,
\gamma\gamma'\gamma''}({\bm X}_3,\Theta_3)~,
\label{8.19}
\eea
where the tensor
$H^{ \a \a' \a''\,  \b \b' \b''\, \g \g' \g''} =(\g_m)^{\a' \a''} (\g_n)^{\b' \b''} (\g_k)^{\g' \g''} H^{\a m\, \b n\, \g k}$
has complicated, but explicit form:
\bea
H^{\a m\, \b n\, \g k} (X, \Theta)&=&(\g_p)^{\a \b} \left[  \Theta^{\g}   C^{(mnp), k} +\frac{1}{2} (\gamma_r)^{\g}_{\ \d} \Theta^{\d}
\ve^{k r q} \eta_{q q'} C^{(mnp), q'}
+ (\gamma_r)^{\g}_{\ \d} \Theta^{\d}   D^{(mnp), (kr)}\right]~,\non\\
C^{mnp,k}&=&\frac1{X^3} (\eta^{mn}\eta^{kp}+\eta^{mk}\eta^{np}+\eta^{nk}\eta^{mp})
\non\\&&
+\frac3{X^5} (X^mX^k\eta^{np} + X^n X^k \eta^{mp}+ X^p X^k \eta^{mn})
\non\\&&
-\frac5{X^5} (X^m X^n \eta^{pk}+X^n X^p \eta^{mk} + X^m X^p \eta^{nk})
-\frac5{X^7} X^m X^n X^p X^k~,\non\\
D^{(mnp), (kr)} &= & \ve^{mks} \eta_{s s'} T^{(np), r, s'} + \ve^{nks} \eta_{s s'} T^{(mp), r, s'} +
\ve^{pks} \eta_{s s'} T^{(mn), r, s'}
\nonumber \\
&&+
 \ve^{mrs} \eta_{s s'} T^{(np), k, s'} + \ve^{nrs} \eta_{s s'} T^{(mp), k, s'} +
\ve^{prs} \eta_{s s'} T^{(mn), k, s'}~,\non\\
T^{(np), r, s}&=&\frac{1}{2} \left[\frac{\eta^{nr} X^p X^s  +\eta^{pr} X^n X^s -\eta^{np} X^r X^s  }{X^5} +
\frac{3 X^n X^p X^r X^s}{X^7}\right]~.
\eea
It is important that the coefficients $c_{\cN=1}$ and $d_{\cN=1}$
in (\ref{8.18}) and (\ref{8.19}) are expressed in terms of
$c_{\cN=2}$ and $d_{\cN=2}$ as
\be
c_{\cN=1} = 6 c_{\cN=2}~,\qquad
d_{\cN=1} = -5 d_{\cN=2}~.
\ee
From (\ref{8.14}) we find the ratio of these coefficients:
\be
\frac{d_{\cN=1}}{c_{\cN=1}} = -\frac5{32\pi}~.
\ee


\subsection{Correlation functions of flavour current multiplets}

We now turn to deriving relations between the coefficients in the two- and three-point
correlators of flavour current multiplets. 

\subsubsection{$\cN=3$ flavour current multiplets}

The two- and three-point correlation functions of $\cN=4$ flavour current multiplets
are found in the form (\ref{L-N4-2pt}) and (\ref{3pt-flavour-N4}).
They contain free coefficients $a_{\cN=4}$ and $b_{\cN=4}$ which
are related to each other by (\ref{ab}). Owing to the identities
(\ref{4.10}), the same relation must hold for the coefficients
among two-point and three-point functions in $\cN=3$
superconformal theories
\be
\frac{b_{\cN=3}}{a_{\cN=3}} = -\frac{\sqrt2}{8\pi}~.
\label{ab-N3}
\ee

\subsubsection{$\cN=2$ flavour current multiplets}
Let us consider the reduction of the $\cN=3$ $\langle LLL\rangle $
correlator
to the $\cN=2$ superspace. Recall that the $\cN=2$ flavour
current multiplet is described by a primary real dimension-1 superfield $L$
subject to the  constraint
\be
(D^{\a I}D_{\a}^J - \frac{1}{2}\d^{IJ}D^{K \a}D _{\a}^K )L =0~,
\ee
which defines the $\cN=2$ linear multiplet. 
Such a  superfield can be obtained by bar-projecting 
one of the three components of 
the $\cN=3$ flavour current multiplet $L^I$, 
\be
L = L^3 |~,
\ee
where the bar-projection assumes that $\theta_3^{\alpha}=0$.
Hence, the correlation functions of the $\cN=2$ flavour current
multiplets can be obtained by evaluating the bar-projections
\begin{subequations}
\bea
\langle L^{\bar a}(z_1) L^{\bar b}(z_2) \rangle &=&
 \langle L^{3\,\bar a}(z_1) L^{3\,\bar b}(z_2) \rangle |~,\\
\langle L^{\bar a}(z_1) L^{\bar b}(z_2) L^{\bar c}(z_3) \rangle &=&
 \langle L^{3\,\bar a}(z_1) L^{3\,\bar b}(z_2) L^{3\,\bar c}(z_3) \rangle
 |~.
\eea
\end{subequations}
Now, given the explicit form of the correlation functions of
$\cN=3$ flavour current multiplets, eqs. (\ref{2ptL}) and (\ref{3ptL_}), 
we derive 
\begin{subequations}
\label{8.28}
\bea
\langle L^{\bar a}(z_1) L^{\bar b}(z_2) \rangle &=& a_{\cN=3}
\frac{\delta^{\bar a\bar b}}{{\bm x}_{12}{}^2}~,\\
\langle L^{\bar a}(z_1) L^{\bar b}(z_2) L^{\bar c}(z_3) \rangle &=&
-\frac12 b_{\cN=3} \frac{f^{\bar a\bar b \bar c}}{{\bm x}_{13}{}^2 {\bm x}_{23}{}^2}
\frac{\ri \Theta_3^{\hat I\alpha}
X_{3\alpha\beta}\Theta_3^{\hat J\beta}\varepsilon_{\hat I\hat
J}}{X_3{}^3}~,
\eea
\end{subequations}
where $\hat I, \hat J$ are the $\sSO(2)$ indices.
Recall that the $\cN=2$ flavour current correlation functions were
found in \cite{BKS} in the form
\begin{subequations}
\bea
\langle L^{\bar a}(z_1) L^{\bar b}(z_2) \rangle &=& a_{\cN=2}
\frac{\delta^{\bar a\bar b}}{{\bm x}_{12}{}^2}~,\\
\langle L^{\bar a}(z_1) L^{\bar b}(z_2) L^{\bar c}(z_3) \rangle
&=&\frac1{{\bm x}_{13}{}^2 {\bm x}_{23}{}^2}
\left[f^{\bar a\bar b\bar c}
b_{\cN=2}\frac{\ri\varepsilon_{\hat I\hat J}\Theta_3^{\hat I\alpha} X_{3\alpha\beta} \Theta_3^{\hat J\beta}  }{X_3{}^3}
+d^{\bar a\bar b\bar c}
 \frac{\tilde b_{\cN=2}}{X_3}\right]~.~~~
 \label{e1}
\eea
\end{subequations}
Comparing these expressions with (\ref{8.28}) we conclude that
\be
a_{\cN=2} = a_{\cN=3}~,\qquad
b_{\cN=2} = -\frac12 b_{\cN=3}~,\qquad
\tilde b_{\cN=2}=0~.
\ee
As a consequence of (\ref{ab-N3}) we find the ratio of
coefficients $b_{\cN=2}$ and $a_{\cN=2}$
\be
\frac{b_{\cN=2}}{a_{\cN=2}} = \frac{\sqrt2}{16\pi}~.
\label{8.31}
\ee

Let us point out that $\tilde b_{\cN=2}$ is found to be zero because the last term in
(\ref{e1}) cannot be lifted to $\cN=3$ supersymmetry, but in generic 
$\cN=2$ supersymmetric theories it is not
necessarily zero. It can be shown that this term does not contribute to the three-point function of conserved
currents~\cite{BKS} and, hence, is irrelevant for our present discussion.

\subsubsection{$\cN=1$ flavour current multiplets}
Finally, let us discuss the reduction of the $\cN=2$ flavour
current multiplets correlation functions down to $\cN=1$. The
$\cN=1$ flavour current is described by a primary dimension-3/2
superfield $L_\alpha$ obeying the conservation law $D^\alpha L_\alpha
=0$. It can be obtained from the $\cN=2$ flavour current multiplet
$L$ by the rule
\be
L_{\a} = \ri D^2_{\a}L|~,
\ee
where the bar-projection assumes that $\theta_2^{\alpha}=0$.
The corresponding relations among the correlation functions of
$\cN=2$ and $\cN=1$ flavour current multiplets were found in \cite{BKS}
\bea
\langle L^{\bar a}_\alpha(z_1) L^{\bar b}_\beta (z_2) \rangle
&=& - D^2_{(1)\alpha} D^2_{(2)\beta}
 \langle L^{\bar a}(z_1) L^{\bar b} (z_2) \rangle |
 =2\ri a_{\cN=2} \delta^{\bar a\bar b} \frac{{\bm x}_{12\alpha\beta}}{{\bm
 x}_{12}{}^4}~,\label{8.33}\\
\langle L^{\bar a}_\alpha(z_1) L^{\bar b}_\beta (z_2) L^{\bar c}_{\gamma}(z_3) \rangle
&=& -\ri D^2_{(1)\alpha} D^2_{(2)\beta} D^2_{(3)\gamma}
 \langle L^{\bar a}(z_1) L^{\bar b} (z_2) L^{\bar c}(z_3) \rangle |
\non\\&=&
2\ri b_{\cN=2}\frac{{\bm x}_{13\alpha\alpha'} {\bm x}_{23\beta\beta'}}{{\bm x}_{13}{}^4 {\bm x}_{23}{}^4}
\frac{X_3^{\alpha\beta} \Theta_3^\gamma
-\varepsilon^{\alpha\gamma}X_3^{\beta\rho}\Theta_{3\rho}
-\varepsilon^{\beta\gamma}X_3^{\alpha\rho}\Theta_{3\rho}}{X_3{}^3}~.~~~~~~
\label{8.34}
\eea
The same expressions for these correlation functions were found in \cite{BKS}
by using the superconformal invariance and conservation
conditions. The free coefficients of these correlation functions
are related to the ones in (\ref{8.33}) and (\ref{8.34})
\be
a_{\cN=1} = 2 a_{\cN=2}~,\qquad
b_{\cN=1} = 2 b_{\cN=2}~.
\ee
Hence, these coefficients have the same ratio as in (\ref{8.31})
\be
\frac{b_{\cN=1}}{a_{\cN=1}} = \frac{\sqrt2}{16\pi}~.
\ee


\section{Concluding comments}\label{section8}

In this paper, we have studied some implications of $\cN=4$ superconformal 
symmetry in three dimensions. 
A rather unexpected result of our analysis
is that the three-point function of the supercurrent in 
${\cal N}=4$ superconformal field theories 
is allowed to possess {\it two}  independent tensor structures, which is 
a consequence of the superconformal symmetry and the conservation equation.
It may look surprising since
any ${\cal N}=4$  superconformal field theory can also be thought of as a special case of 
one with $\cN<4$ 
and, as we showed in~\cite{BKS}, similar three-point functions in superconformal field theories with ${\cal N} <4$
contain only one tensor structure.
An apparent disagreement has a simple resolution. 
From the viewpoint of 
 ${\cal N}=3$ (or even less extended)
supersymmetry,
the ${\cal N}=4$ supercurrent consists of two $\cN=3$ multiplets, one of which 
is the $\cN=3$ supercurrent  and the other contains
additional currents, like the $R$-symmetry currents.  Such ${\cal N} \to {\cal N}-1$ decompositions  can be found in the introduction of~\cite{BKS}.
As we explained in section 4 (see also subsection C.1), 
only one tensor structure in the three-point function of the  ${\cal N}=4$ supercurrent 
contributes to the three-point function of the  ${\cal N}=3$ (and, hence, ${\cal N}<3$) supercurrent. Thus, just like in general ${\cal N}\leq 3$  superconformal theories,  
 the three-point correlator of the  energy-momentum tensor in ${\cal N}=4$ superconformal theories is determined by a single tensor structure. 
As concerned the second tensor structure,  in section 5 we pointed out  that it is 
present in those $\cN=4$ superconformal field theories which are not invariant  
under the mirror map (see also below). 
In the case of free $\cN=4$ hypermultiplet models,
 it is proportional to  the difference between the number of 
left and right supermultiplets with respect to 
the $R$-symmetry group $\sSU (2)_{\rm L} \times \sSU (2)_{\rm R}$. 

Another important result of the paper consists in the relations between the coefficients of the  two- and  three-point  correlation functions
of the supercurrent and flavour current multiplets in {\it all} $1 \leq {\cal N} \leq 4$ superconformal theories. These relations are derived
in section 7 and the analysis is based on two observations. 
First, if both the two- and  three-point functions of either 
the supercurrent or the flavour current multiplets are fixed up to a single coefficient and are related to each other by the Ward identities,
we can derive the {\it universal}  ratio of the coefficients by simply considering any specific theory. Second, as already mentioned, any $\cN$-extended supersymmetric 
theory is a special case of a $(\cN-1)$-extended theory.
In particular, any ${\cal N}=4$ superconformal theory can be considered 
as an ${\cal N}=1$, ${\cal N}=2$ or ${\cal N}=3$ superconformal theory. 
As a result, we can derive all  {\it universal} relations between the coefficients of the two- and  three-point functions by considering 
one relatively simple specific example, namely, 
the ${\cal N}=4$ superconformal theory of free hypermultiplets.

The hypermultiplet supercurrent \eqref{example-J} is asymmetric with respect to the left and right hypermultiplets. More generally, given an $\cN=4$ superconformal theory
that is invariant with respect to the mirror map, its supercurrent must change sign under  
the mirror map $\mathfrak M$. A simple illustrating example is provided by the model describing  
an equal number of left and right hypermultiplets. The corresponding supercurrent
\bea
J=  \bar q_{\tilde i} q^{\tilde i}- \bar q_i  q^i
\eea
is odd under the mirror map $ q^{\tilde i} \longleftrightarrow q^i$.
This property has its origin in $\cN=4$ conformal supergravity. 
To explain this important point, we have to recall three results from supergravity.
Firstly, as shown in \cite{KLT-M11}, the $\cN=4$ super-Cotton tensor $X(z) $
changes its sign under the mirror map.\footnote{The algebra of covariant derivatives 
for $\cN=4$ conformal supergravity is known to be 
invariant  under the mirror map \cite{KLT-M11}. 
The super-Cotton tensor is obtained from the completely antisymmetric curvature
tensor $X^{IJKL}$, which is invariant under the mirror map, by the rule
$X^{IJKL} = \ve^{IJKL} X$.  Since the Levi-Civita tensor $\ve^{IJKL}$ changes its sign under the mirror map, eq. \eqref{A.28}, the same is true of the super-Cotton tensor.}
Secondly, the off-shell action 
$S_{\rm CSG}$ for $\cN=4$  conformal  
supergravity \cite{BKNT-M2} proves to be   invariant under the mirror map, and its 
variation can be represented as 
\bea
\d S_{\rm CSG} \propto \int \rd^{3|8} z \,E \, \d H X~, \qquad 
E = {\rm Ber}(E_M{}^A)~. 
\eea
Here $ \rd^{3|8} z \,E $ is the integration measure of $\cN=4$ curved superspace, 
and $H (z)$ denotes the conformal supergravity prepotential
(see also \cite{BKNT-M, KNT-M}). Therefore, the prepotential $H$ changes its sign 
under the mirror map. Thirdly, given a system of matter multiplets coupled 
to conformal supergravity, an infinitesimal disturbance of $H$ changes 
the matter action $S_{\rm matter} $ as follows 
\bea
\d S_{\rm matter} = \int \rd^{3|8} z \,E \, \d H J~,
\eea
where $J(z)$ is the matter supercurrent. If $S_{\rm matter} $ is invariant 
with respect to $\mathfrak M$, then $J$ is indeed odd under the mirror map. 

As shown in subsection \ref{sectJJJ}, 
the most general expression for the three-point function of the $\cN=4$ supercurrent is
\bea
\langle J(z_1) J(z_2) J(z_3) \rangle
=\frac1{{\bm x}_{13}{}^2 {\bm x}_{23}{}^2}
\left(  \frac{\tilde d_{\cN=4} }{X_3}+ \frac{ d_{\cN=4} }{X_3{}^5}
\varepsilon_{IJKL}\Theta^{I\alpha}\Theta^{J\beta}\Theta^{K\gamma}\Theta^{L\delta}
 X_{\alpha\beta} X_{\gamma\delta}
\right)~.~~~
\label{8.5}
\eea
The parameter $\tilde d_{\cN=4} $ must vanish, $\tilde d_{\cN=4} =0$,  
in every theory invariant under the mirror map. The second term in \eqref{8.5}
is odd under the mirror map due to the property 
\bea
\mathfrak M : ~\varepsilon_{IJKL}\Theta^{I\alpha}\Theta^{J\beta}\Theta^{K\gamma}
\Theta^{L\delta} X_{\alpha\beta} X_{\gamma\delta}~
 \longrightarrow~ - \varepsilon_{IJKL}\Theta^{I\alpha}\Theta^{J\beta}\Theta^{K\gamma}\Theta^{L\delta}
 X_{\alpha\beta} X_{\gamma\delta}~.
 \eea
All other building blocks in \eqref{8.5} are invariant under $\mathfrak M$. 

It would be interesting to extend the results of the present paper to the cases of superconformal theories with $  {\cal N} >4$. 
The $  {\cal N} >4$ 
supercurrent is described by a primary superfield $J^{IJKL}$ of dimension 1 subject to the conservation law~\cite{BKNT-M, KNT-M}
\be 
D_{\alpha}^I J^{JKLP}= D_{\alpha}^{[I} J^{JKLP]} - \frac{4}{{\cal N}-3} D_{\alpha}^Q J^{Q[JKL}  \delta^{P] I}\,, \quad I=1, \dots, {\cal N}\,. 
\label{conc1}
\ee
The construction of the correlation functions involving the supercurrent $J^{IJKL}$ has its own complications due to a large number of the
$R$-symmetry indices. We postpone this problem for later study.
\\

\noindent
{\bf Acknowledgements:}\\
This work is supported in part by the
Australian Research Council, project
 DP140103925. The work of E.I.B. is also supported by the ARC Future Fellowship FT120100466.


\appendix

\section{Comments on off-shell hypermultiplets}

The superfield constraints  \eqref{q-analyt-a} and  \eqref{q-analyt-b} 
are on-shell. There exist
off-shell models for left and right hypermultiplets 
such that their equations of motion are equivalent to the 
 constraints  \eqref{q-analyt-a} and  \eqref{q-analyt-b}. 
Before discussing such off-shell hypermultiplets, some general comments are in order. 
Off-shell descriptions exists for many 3D $\cN=4$ supersymmetric field theories
such as general $\cN=4$ nonlinear $\s$-models \cite{KPT-MvU}. 
However, it turns out that the conventional
$\cN=4$ Minkowski superspace ${\mathbb M}{}^{3|8} $
is not suitable to realise the most interesting off-shell couplings.
An adequate superspace setting for them is an extension of ${\mathbb M}{}^{3|8} $ by
auxiliary bosonic dimensions parametrising a compact
manifold, in the spirit of the superspace \cite{Rosly}
${\mathbb M}{}^{4|8} \times {\mathbb C}P^1$
which is at the heart of the 4D $\cN=2$ harmonic \cite{GIKOS,GIOS} 
and projective \cite{KLR,LR} superspace approaches.\footnote{The 
relationship between the 4D $\cN=2$ harmonic and projective superspace 
formulations
is spelled out in \cite{K-double}.} 
All known off-shell $\cN=4$ supersymmetric field theories 
in three dimensions can be realised in
the following superspace\footnote{For every positive integer $\cN$,
the 3D $\cN$-extended superconformal group $\sOSp(\cN|4;{\mathbb R})$ is a transformation group of
 the so-called compactified Minkowski superspace
$\overline{\mathbb M}{}^{3|2\cN}$ in which
${\mathbb M}{}^{3|2\cN}$ is embedded as a dense open domain \cite{KPT-MvU}.
In the $\cN=4$ case, $\sOSp(4|4;{\mathbb R})$ is also defined to act transitively
on $\overline{\mathbb M}{}^{3|8} \times {\mathbb C}P^1_{\rm L}
\times {\mathbb C}P^1_{\rm R} $, as shown in  \cite{KPT-MvU}.
}
\cite{Zupnik98,Zupnik99}
\bea
{\mathbb M}{}^{3|8} \times {\mathbb C}P^1_{\rm L}
\times {\mathbb C}P^1_{\rm R} 
= {\mathbb M}{}^{3|8} \times  \left[ {\sSU(2)}/{\sU(1)} \right]_{\rm L}
\times \left[ {\sSU(2)} / {\sU(1)} \right]_{\rm R} ~,
\label{biharmonic}
\eea
which may be called harmonic or projective depending 
on the type of  $\cN=4$ off-shell multiplets one is interested in. 
All such multiplets are functions over 
either ${\mathbb C}P^1_{\rm L}$ or $ {\mathbb C}P^1_{\rm R} $. 
For definiteness, let us consider left multiplets associated with 
 ${\mathbb C}P^1_{\rm L}$. Our presentation below is similar to 
 \cite{K-Lectures}.
 
Let $ v_{\rm L}  \equiv (v^i) \in {\mathbb C}^2 \setminus \{0\}$ 
be homogeneous coordinates for  ${\mathbb C}P^1_{\rm L}$, 
and $v_{\rm L}{}^\dagger= (\overline{v^i})  :=( \bar v_i) $ be their conjugates
(in what follows, the subscripts `L' and `R' will always be omitted if no confusion
may occur). 
Any superfield living in ${\mathbb M}{}^{3|8} \times {\mathbb C}P^1$
may be identified with a function $ \f(z,v, \overline{v})$
that only scales under arbitrary re-scalings of $v$: 
\bea
\f(z,c\, v, {\bar c}\, \overline{v})= c^{n_+} \, {\bar c}^{n_-} \,\f(z,v, \overline{v})~,
\qquad c\in {\mathbb C}^* \equiv {\mathbb C} \setminus \{0\}
\eea
for some parameters $n_\pm$ such that ${ n_+ - n_-}$ is an integer. 
Since $v^\dagger v = \bar v_i v^i \neq 0$, 
we can always choose ${ n_-=0}$ 
by redefining $ \f(z,v, \bar v) \to \f(z, v, \bar v)/  (v^\dagger v  ) ^{n_-}$.
Any superfield with the homogeneity property 
\bea
 \f^{(n)} (z,c\, v, {\bar c}\, \overline{v})= c^{n}  \f^{(n)}(z,v, \overline{v})~,
\qquad c\in {\mathbb C}^*
\label{HomCon}
\eea
is said to have {weight $n$}.
Let us introduce fermionic operators
\bea 
{\mathfrak D}^{\tilde i}_{ \a} \equiv D_\a^{\tilde i (1)}
:=v_{ i} { D}^{i \tilde i}_{ \a} ~, 
\label{5.77}
\eea
where $v_i:= \ve_{ij}\,v^j$.
 In accordance with \eqref{2.622}, these operators strictly anticommute 
 with each other, 
 \bea
 \{ {\mathfrak D}^{\tilde i}_{ \a}, {\mathfrak D}^{\tilde j}_{ \b} \} =0~,
 \eea
which allows us to introduce left {\it isochiral multiplets} (following the terminology of
\cite{RS}) constrained by 
\bea
{\mathfrak D}^{\tilde i}_{ \a} \f^{(n)}(z,v, \overline{v}) =0~.
\label{5.99}
\eea
These constraints are consistent with the homogeneity condition
\eqref{HomCon}.

Given an isochiral superfield $\f^{(n)}(z, v^i, {\bar v}_j)$, its complex conjugate 
\bea
{\bar \f}^{(n)} (z, {\bar v}_i, {v}^j) 
:=
\overline{ \f^{(n)}(z, v^i, {\bar v}_j) }
\eea
is no longer isochiral.
However, by analogy with the 4D $\cN=2$ case \cite{Rosly,GIKOS}
one can define a modified conjugation that maps every isochiral superfield
$\f^{(n)}(z, v, {\bar v})$ into an isochiral one  $\breve{\f}^{(n)}(z, v, {\bar v})$ 
of the same weight defined  as follows:  
\bea
 \f^{(n)}(v^i, {\bar v}_j) \longrightarrow  {\bar \f}^{(n)} ({\bar v}_i, {v}^j) 
  \longrightarrow  {\bar \f}^{(n)} \Big({\bar v}_i \to -v_i, {v}^j \to {\bar v}^j\Big) =:\breve{\f}^{(n)}(v^i, {\bar v}_j)~.
~~~
\label{smile-iso}
\eea
The weight-$n$ isochiral superfield $\breve{\f}^{(n)}(z,v, {\bar v})$ is said to be the 
smile-conjugate of  $\f^{(n)}(z,v, {\bar v})$. One can check that 
\bea
\breve{ \breve{\f}}^{(n)}(z,v, {\bar v}) =(-1)^n {\f}^{(n)}z,(v, {\bar v})~.
\eea
Therefore, if the weight $n$ is even, real isochiral superfields can be defined, 
 $\breve{\f}^{(2m)} = {\f}^{(2m)}$.

Within the 3D $\cN=4$ projective superspace approach \cite{KPT-MvU}, 
 off-shell multiplets are described in terms of weight-$n$ isochiral
superfields $Q^{(n)}(z,v)$,
\bea
 {\mathfrak D}^{\tilde i}_{ \a} Q^{(n)}=0~, \qquad
  Q^{(n)} (z,c\, v )= c^{n}  \, Q^{(n)}(z,v)~,
\quad c\in {\mathbb C}^* ~,
 \eea
which are { {\it holomorphic } over an {\it open domain} of
$ {\mathbb C}P^1_{\rm L}$},
\bea
\frac{\pa}{\pa {\bar v}_i} \, Q^{(n)} =0~.
\eea
Such isochiral superfields are called left {\it projective multiplets} of weight $n$.
The action principle in projective superspace involves a contour integral, 
and not an integral over ${\mathbb C}P^1$. This is why there is no need for projective
multiplets to be smooth over ${\mathbb C}P^1$.
This approach is useful  to construct the most general $\cN=4$ supersymmetric $\s$-models, 
both in Minkowski superspace \cite{KPT-MvU} and in supergravity \cite{KLT-M11}. 
The structure of superconformal projective multiplets is well understood \cite{KPT-MvU}.

Somewhat different isochiral superfields are used in the framework 
of the 3D $\cN=4$ harmonic superspace approach \cite{Zupnik98,Zupnik99}.
The equivalence $v^i \sim c\,v^i$,  
which is intrinsic to ${\mathbb C}P^1$, 
allows one to switch to the description in terms of normalised  isotwistors:
\bea
u^{+i} := \frac{v^i}{\sqrt{v^\dagger v}}~, \qquad u^-_i :=  \frac{{\bar v}_i}{\sqrt{v^\dagger v}}
=\overline{u^{+i}} 
\quad \Longrightarrow \quad 
 \big(u_i{}^- , u_i{}^+ \big) \in \sSU(2)~.
\eea
The variables $u^\pm_i$ are called {\it harmonics}. They  are defined modulo the equivalence relation 
 $ u^\pm_i \sim 
 \re^{\pm {\rm i}\a}u^\pm_i$,  with $\a \in {\mathbb R}$.
It is clear that  the harmonics parametrize the coset space
$\sSU(2)/\sU(1) \cong S^2$.
Given an isochiral superfield  $\f^{(n)}(z,v, \overline{v})$
we can associate with it the following superfield 
 \bea
 \vf^{(n)} (z,u^+,u^-) := \f^{(n)}\left(z, \frac{v}{ \sqrt{v^\dagger v} }, \frac{\bar v}{ \sqrt{ v^\dagger v} }\right)  =
 \frac{1}{ (\sqrt{v^\dagger v})^n } \f^{(n)}(z,v, \overline{v})
 \eea
obeying the  homogeneity condition
\bea
\vf^{(n)}(z, {\rm e}^{ {\rm i}\a}\, u^+, {\rm e}^{ -{\rm i}\a}\,u^-)
= {\rm e}^{ {\rm i}n\a} \,\vf^{(n)}(z,u^+, u^-)~.
\eea
This property tells us that  $\vf^{(n)}(z, u^\pm) $ has $\sU(1)$ charge $n$. 
Thus the weight of $\f^{(n)}(z,v, \overline{v})$ is replaced with 
the $\sU(1)$ charge of $\vf^{(n)}(z, u^\pm) $.
It is obvious that we have the one-to-one correspondence
$\f^{(n)}(z,v, \overline{v}) \longleftrightarrow  \vf^{(n)}(z, u^\pm) $. 
The fermionic operators \eqref{5.77} turn into
\bea
{D}^{\tilde i +}_{ \a}:=  \frac{1}{ \sqrt{v^\dagger v}}
{\mathfrak D}^{\tilde i}_{ \a} = u^+_{ i} { D}^{i \tilde i}_{ \a}~,
\eea
and therefore the isochirality condition \eqref{5.99} takes the form 
\bea
{D}^{\tilde i +}_{ \a} \vf^{(n)}(z, u^\pm) =0~.
\label{A.166}
\eea

{In harmonic superspace, every isochiral superfield
$\vf^{(n)}(z, u^\pm) $ is required to be a 
{\it smooth} charge-$n$ function over $\sSU(2)$ or, equivalently, a {\it smooth} tensor field over 
the two-sphere ${ S}^2 $}. 
Such a superfield is called left {\it analytic}.
It can be represented, say for $n\geq 0$, by a convergent Fourier series  
\bea
\vf^{(n)}(z, u^\pm) = \sum_{p=0}^{\infty} 
\vf^{ (i_1 \dots i_{n+p}  j_1 \dots j_p )} (z)\,
u^+_{i_1} \dots u^+_{i_{n+p}} u^-_{j_1} \dots  u^-_{ j_p }  ~,
\eea
in which the coefficients $\vf^{ i_1 \dots i_{n+2p}   } (z)= \vf^{( i_1 \dots i_{n+2p}   )} (z)$
are ordinary $\cN=4$ superfields obeying first-order differential constraints that follow 
from (\ref{A.166}).
The beauty of this approach is that  the power of harmonic analysis can be used.

We are now prepared to discuss off-shell hypermultiplets. 
In harmonic superspace, the most suitable off-shell description 
of a single hypermultiplet makes use of an analytic superfield 
$q^+ (z, u^\pm) \equiv q^{(1)} (z,u^\pm)$
and its smile-conjugate $\breve{q}^+ (z,u^\pm)$. 
The free hypermultiplet equation of motion, 
which corresponds to the action \eqref{hyperact}, is 
\bea
\pa^{++} q^+ =0 \quad \Longrightarrow \quad 
q^+ (z, u^\pm) = q^i (z) u^+_i~, 
\eea
where $q^i(z)$ obeys the constraint \eqref{q-analyt-a}. 

In projective superspace, the most suitable off-shell description 
of a single hypermultiplet makes use of an arctic multiplet
$\U^{(1)} (z,v) $ and its its smile-conjugate $\breve{\U}^{(1)} (z,v) $.
By definition, $\U^{(1)} (z,v) $ is a weight-1 projective multiplet 
which is  holomorphic over the so-called north chart ${\mathbb C} $ of 
${\mathbb C}P^1 ={\mathbb C} \cup \{\infty \}$. Here the point 
$\infty \in {\mathbb C}P^1$ is identified with  
the ``{north pole}'' $v^{i}_{\rm north} \sim (0,1)$.
In the north chart, 
it is useful to introduce a complex 
({\it inhomogeneous}) coordinate $\z$ defined by 
\bea 
 v^i = v^{1} \,(1, \z) ~,\qquad \z:=\frac{v^{2}}{v^{1}} ~,\qquad\quad 
{ i=1 ,2}~.
\label{Zeta}
\eea
The arctic multiplet $\U^{(1)} (z,v) $  looks like
\bea
\U^{(1)} (z, v) &=&  v^{1}
\sum_{k=0}^{\infty} \U_k (z) \z^k 
~, 
\label{arctic1}
\eea
and  its smile-conjugate {\it antarctic} multiplet $\breve{\U}^{(1)} (z,v) $, is
 \bea
\breve{\U}^{(1)} (z,v) &=& v^{2} 
\sum_{k=0}^{\infty}  {\bar \U}_k (z)\,
\frac{(-1)^k}{\z^k}~.
\label{antarctic1}
\eea
The dynamics of free polar hypermultiplet is described by the action
\bea
S=\frac{1}{2\p} \oint_{\g}  { v_i {\rm d} v^i }
\int {\rm d}^3x \, D^{(-4)} \cL^{(2)} (z,v) \Big|_{\q =0}~, 
\qquad \cL^{(2)} = \breve{\U}^{(1)} \U^{(1)}~,
\label{A.222}
\eea
where we have defined
\bea
D^{(-4)} &:=&
\frac{1}{48}D^{(-2){\tilde i}{\tilde j}}D^{(-2)}_{{\tilde i}{\tilde j}}~,\quad
D^{(-2)}_{{\tilde i}{\tilde j}}:=D^{(-1)\g}_{{\tilde i}}D_{{\tilde j}\g}^{(-1)}~,
\quad 
D^{(-1) {\tilde i}  }_\a:=\frac{1}{(v,u)}u_{i}D_\a^{i{\tilde i}}~.~~~
\eea
The fourth-order operator $D^{(-4)} $ in  \eqref{A.222}
involves a constant isotwistor
$u_i$ constrained by the only condition $(v,u):=v^iu_i \neq 0$ 
which must hold along the closed integration contour $\g$.  
The action \eqref{A.222} proves to be independent of $u_i$. 
It can be shown that the equation of motion, which follows from the action 
 \eqref{A.222}, is 
\bea
\U^{(1)} (z, v) &=&  v^{1} \Big( \U_0 (z)  + \U_1(z) \z \Big) 
\equiv q^i(z) v_i~,
\eea
where $q^i(z)$ obeys the constraint \eqref{q-analyt-a}. 
Thus, the $q^+$ hypermultiplet and the polar hypermultiplet provide two 
different off-shell realisations for the hypermultiplet.  
Both actions \eqref{hyperact} and \eqref{A.222} are superconformal.

There is a family of isochiral multiplets that are holomorphic over ${\mathbb C}P^1$, 
and therefore they are suitable for both the harmonic and projective superspace 
settings. These are the so-called $\cO(n)$ multiplets, where $n =1,2, \dots$,  
\bea
H^{(n)} (z,v) = H^{i_1 \dots i_n} (z) v_{i_1} \dots v_{i_n} ~, 
\qquad D_\a^{\tilde j (j} H^{i_1 \dots i_n )} =0~. 
\eea
Such a multiplet is (i) on-shell for $n=1$ and  describes a free hypermultiplet; 
and (ii) off-shell for $n>1$. When $n$ is even, one can define real multiplets with respect
to the smile-conjugation. The flavour current multiplet $L^{ij }$ is described 
by a real  $\cO(2)$ multiplet $L^{(2)} $.
It may be shown that real $\cO(2n)$ multiplets with $n>1$ can be used 
to describe neutral hypermultiplets. However, the corresponding free hypermultiplet
actions are not superconformal.

The mirror map \cite{Zupnik99,Zupnik2009} is defined as 
\bea
{\frak M} : ~ \sSU (2)_{\rm L}~ \longleftrightarrow ~ \sSU (2)_{\rm R}~.
\eea
It  changes the tensor types of superfields 
as 
${\rm D}^{(p/2)}_{\rm L} \otimes {\rm D}^{(q/2)}_{\rm R} 
\to {\rm D}^{(q/2)}_{\rm L} \otimes {\rm D}^{(p/2)}_{\rm R}$, 
where ${\rm D}^{(p/2)}$ denotes the spin-$p/2$ 
representation of $\sSU(2)$. The mirror map interchanges the on-shell left $q^i$ 
and right $q^{\tilde i}$ hypermultiplets, 
\bea
{\mathfrak M} \cdot q^i = q^{\tilde i}~, \qquad {\mathfrak M} \cdot q^{\tilde i} = q^{i}~.
\eea
It also interchanges the left $L^{ij} $ and right $L^{\tilde i \tilde j} $ flavour current multiplets. 
Since the latter multiplets are (anti) self-dual, eq. \eqref{ASD}, 
the mirror map must act on the Levi-Civita tensor $\ve^{IJKL} $ as 
\bea
\mathfrak M \cdot  \ve^{IJKL}  = - \ve^{IJKL} ~.
\label{A.28}
\eea


\section{$\cN=4$ hypermultiplet propagator}
\label{AppC}
The free equation of motion for  $q^+$ hypermultiplet is ${\mathscr D}^{++} q^+
=0$, where ${\mathscr D}^{++}$ is defined in (\ref{D++}). By definition, the
Green function of the free hypermultiplet
$G^{(+,+)}(\zeta_1,\zeta_2)$ obeys the equation
\be
{\mathscr D}^{++} G^{(+,+)}(\zeta_1,\zeta_2) =
-\delta_A^{(3,1)}(\zeta_1,\zeta_2)~,
\label{G++eq}
\ee
where $\delta_A^{(3,1)}(\zeta_1,\zeta_2)$ is the analytic delta
functions. The solution to this equation is very similar to the
four-dimensional $q$-hypermultiplet Green's function \cite{GIKOS,GIOS1}
\be
G^{(+,+)}(\zeta_1,\zeta_2) = \frac1{\square} (D^+_1)^4 (D^+_2)^4
\frac{\delta^3(x_1-x_2)\delta^8(\theta_1-\theta_2)}{(u^+_1
u^+_2)^3}~,
\label{G++app}
\ee
where $(D^+)^4=\frac1{16}(D^{+\tilde 1 \alpha}D^{+\tilde 1}_\alpha)
(D^{+\tilde 2 \beta}D^{+\tilde 2}_\beta)$.
To check that (\ref{G++app}) obeys (\ref{G++eq}) one has to take into account
that ${\mathscr D}^{++}$ commutes with $D^{+\tilde i}_\alpha$ and
hits only the harmonic distribution in (\ref{G++app}) producing
the harmonic delta-function (see \cite{GIOS} for a review of properties
of harmonic distributions)
\be
\partial^{++} \frac1{(u^+_1 u^+_2)^3}  = \frac12 (\partial^{--})^2
\delta^{(3,-3)} (u_1,u_2)~.
\ee
This harmonic delta function is part of the analytic delta
function
\be
\delta_A^{(3,1)}(\zeta_1,\zeta_2) = (D^+_2)^4 \delta^3(x_1-x_2)
\delta^8(\theta_1 - \theta_2) \delta^{(3,-3)} (u_1,u_2)~.
\ee
As a result we have
\be
{\mathscr D}^{++}_1 G^{(+,+)}(\zeta_1,\zeta_2) = \frac12\frac1{\square}
(D^+_1)^4 ({\mathscr D}^{--}_1)^2 \delta_A^{(3,1)}(\zeta_1,\zeta_2)
=-\delta_A^{(3,1)}(\zeta_1,\zeta_2)~.
\ee
Here we applied the identity
\be
(D^+)^4 ({\mathscr D}^{--})^2 \phi_A = -2 \square \phi_A~,
\ee
which holds for arbitrary analytic superfield $\phi_A$.

We point out that the operator $1/\square$ in (\ref{G++app}) acts
only on the bosonic delta-function $\delta^3(x_1-x_2)$ and gives
the scalar field Green's function $G(x_1,x_2)$ which we represent as
the integral over the proper time $s$
\be
\frac1\square \delta^3(x_1-x_2)=-G(x_1,x_2) = -\ri\int_0^\infty
{\rm d}s \, U(x_1,x_2|s)~,\label{B7_}
\ee
where $U(x_1,x_2|s)$ is the heat kernel of the three-dimensional
d'Alembert operator
\be
U(x_1,x_2|s)=\frac\ri{(4\pi\ri s)^{3/2}}e^{\ri\frac{(x_1-x_2)^2}{4s}}~.
\ee
The integration over the proper time in (\ref{B7_}) can be done
explicitly, and the result has slightly different forms for the
point inside and outside the lightcone
\be
G(x_1,x_2) = \left\{
\begin{array}{ll}
\frac1{4\pi}\frac1{\sqrt{-(x_1-x_2)^2}}\qquad &(x_1-x_2)^2<0\\
\frac\ri{4\pi}\frac1{\sqrt{(x_1-x_2)^2}} \qquad& (x_1-x_2)^2>0~.
\end{array}
\right.
\ee
These two cases can be unified in a single formula such that
\be
\frac1{\square} \delta^3(x_1-x_2) =
-\frac\ri{4\pi}\frac1{\sqrt{(x_1-x_2)^2}}
\ee
is valid for $(x_1-x_2)^2\ne0$. Then, we rewrite (\ref{G++app}) as
\be
G^{(+,+)}(\zeta_1,\zeta_2) =- \frac\ri{4\pi}
(D^+_1)^4 (D^+_2)^4\left( \frac1{\sqrt{(x_1-x_2)^2}}
 \frac{\delta^8(\theta_1-\theta_2)}{(u^+_1 u^+_2)^3} \right)~.
 \label{B8}
\ee
It is important to realize that the supersymmetrized coordinate
difference (\ref{hat-x12}) at coincident Grassmann coordinates is simply
\be
\hat x_{12}^a |_{\theta_1 = \theta_2 } = (x_1-x_2)^a~.
\ee
Thus, in (\ref{B8}) we can apply the identity
\be
\frac{\delta^8(\theta_1 - \theta_2)}{\sqrt{(x_1-x_2)^2}}
=\frac{\delta^8(\theta_1 - \theta_2)}{\sqrt{\hat x_{12}{}^2}}
\ee
and use the analyticity of (\ref{hat-x12}) in both superspace
arguments to represent (\ref{B8}) as follows
\be
G^{(+,+)}(\zeta_1,\zeta_2) = -\frac\ri{4\pi}\frac1{\sqrt{\hat x_{12}{}^2}}
(D^+_1)^4 (D^+_2)^4 \frac{\delta^8(\theta_1-\theta_2)}{(u^+_1
u^+_2)^3}~.
\ee
Finally, we employ the identity
\be
(D^+_1)^4 (D^+_2)^4 \delta^8(\theta_1-\theta_2) = (u^+_1 u^+_2)^4
\ee
to get the following final expression for the hypermultiplet
Green's function
\be
G^{(+,+)}(\zeta_1,\zeta_2) =-\frac\ri{4\pi} \frac{(u^+_1 u^+_2)}{\sqrt{\hat
x_{12}{}^2}}~.
\ee
This representation of the hypermultiplet Green's function was used
in sect.\ \ref{sect7.4} in studying Ward identities of $\cN=4$
flavour current multiplets. Note that similar representation of
the four-dimensional hypermultiplet propagator was found
in \cite{Sokatchev} (see also \cite{GIOS}).


\section{Superspace reduction of correlation functions}\label{AppA-0}

The procedure of superspace reduction of supercurrent correlation
functions is straightforward, but quite tedious. It was applied in
\cite{BKS} to find the relations among three-point correlation
functions of the $\cN=2$ and $\cN=1$ supercurrents. Here we will
follow the same procedure to perform the $\cN=4\to \cN=3\to
\cN=2$ reductions of the supercurrent correlators.

\subsection{$\cN=4 \to \cN=3$ reduction of the correlation functions for the supercurrent}
\label{AppA}

The $\cN=4$ supercurrent is described by the primary scalar
superfield $J$ of dimension 1. When reduced to the $\cN=3$
superspace, it has two independent $\cN=3$ superfield components:
a scalar $S$ and a spinor $J_\alpha$  \cite{BKS}
\begin{subequations}
\bea
S&=& J|~,\label{A1a}\\
J_\alpha &=& \ri D^4_\alpha J|~,
\eea
\end{subequations}
where the bar-projection means $\theta_{4\alpha}=0$. The $\cN=4$
supercurrent conservation condition (\ref{N4-J-conserv}) turns
to the following constraints for the $\cN=3$ superfields $S$ and
$J_\alpha$
\begin{subequations}
\label{A2_}
\bea
(D^{\hat I\alpha}D^{\hat J}_\alpha - \frac13 \delta^{\hat I\hat J}
 D^{\hat K\alpha}D^{\hat K}_\alpha) S &=&0~,\\
 D^{\hat I\alpha} J_\alpha &=&0~.
\eea
\end{subequations}
Here $\hat I,\hat J,\hat K =1,2,3$ are the indices of $\sSO(3)$
group.

The superfield $J_\alpha$ is the $\cN=3$ supercurrent. In
components, it contains the energy-momentum tensor, conserved
currents of $\cN=3$ supersymmetry and conserved currents of the
$\sSO(3)$ subgroup of the $\sSO(4)$ $R$-symmetry of $\cN=4$
theory. The $\cN=3$ scalar contains among its components the
current of the fourth supersymmetry and the currents of the
remaining $\sSO(4)/\sSO(3)$ $R$-symmetry. Therefore, when we
consider an $\cN=4$ superconformal theory in the $\cN=3$
superspace, the conserved quantities are described by the
following four types of three-point correlation functions
\be
\langle S S S \rangle ~,\quad
\langle S S J_\alpha \rangle~,\quad
\langle S J_\alpha J_\beta \rangle~,\quad
\langle J_\alpha J_\beta J_\gamma \rangle~.
\label{A3_}
\ee
In this Appendix we derive these correlators
from the three-point function of  the $\cN=4$ supercurrent
which was obtained in sect.\ \ref{sectJJJ} in the form
\begin{subequations}
\label{AppJJJ}
\bea
\langle J(z_1)J(z_2) J(z_3)\rangle &=&
\frac1{{\bm x}_{13}{}^2{\bm x}_{23}{}^2} H(X_3,\Theta_3)~,\\
H(X_3,\Theta_3)&=& \frac{\tilde d_{\cN=4}}{X_3}
+d_{\cN=4}\frac{\varepsilon_{IJKL}\Theta_3^{I\alpha}\Theta_3^{J\beta}\Theta_3^{K\gamma}\Theta_3^{L\delta}
 X_{3\alpha\beta} X_{3\gamma\delta}}{X_3{}^5}~.
\label{AppH}
\eea
\end{subequations}
The distinguishing feature of this correlation function as compared to the
ones in the $\cN=1,2,3$ superconformal theories is that it has two
completely different terms with two independent parameters $\tilde
d_{\cN=4}$ and $d_{\cN=4}$. As we will show further, the two terms
in (\ref{AppH}) contribute to different correlators
(\ref{A3_}).

\subsubsection{Correlator $\langle SSS\rangle$}
Since the superfield $S$ is just the lowest component of $J$, see
(\ref{A1a}), its three-point correlator appears simply by
switching off the Grassmann coordinate $\theta_{4\alpha}$ at each
superspace point
\be
\langle S(z_1)S(z_2)S(z_3)\rangle = \langle J(z_1)J(z_2)
J(z_3)\rangle| = \frac1{{\bm x}_{13}{}^2{\bm x}_{23}{}^2}
\frac{\tilde d_{\cN=4}}{X_3}~.
\label{A6_}
\ee
Note that the last term in (\ref{AppH}) vanishes in this reduction
and only the first term with the coefficient $\tilde d_{\cN=4}$
survives.

\subsubsection{Correlator $\langle J_\alpha SS\rangle$}

To compute this correlation function we have to hit (\ref{AppJJJ})
by one spinor covariant derivative
\be
\langle J_\alpha(z_1) S(z_2)S(z_3)\rangle = \ri D^4_{(1)\alpha}
\langle J(z_1)J(z_2) J(z_3)\rangle |=
\ri D^4_{(1)\alpha}\frac1{{\bm x}_{13}{}^2{\bm x}_{23}{}^2}
H(X_3,\Theta_3)|~.
\ee
Note that the spinor covariant derivative acts on the two-point
function (\ref{super-interv-X}) by the rule
\be
D^I_{(1)\alpha} {\bm x}^{\mu\nu}_{12} = -2\ri \delta_\alpha^\mu
\theta_{12}^{I\nu}~.
\label{A7_}
\ee
Hence, all terms in which the derivative $D^4_{(1)\alpha}$ hits
the bosonic two-point and three-point structures vanish under the
bar-projection and only the last term in (\ref{AppH}) contributes
\bea
\langle J_\alpha(z_1) S(z_2)S(z_3)\rangle&=&\ri d_{\cN=4}
\frac1{{\bm x}_{13}{}^2{\bm x}_{23}{}^2}
D^4_{(1)\alpha}
\frac{\varepsilon_{IJKL}\Theta_3^{I\mu}\Theta_3^{J\nu}\Theta_3^{K\rho}\Theta_3^{L\sigma}
 X_{3\mu\nu} X_{3\rho\sigma}}{X_3{}^5}\Big|\non\\
 &=&-\ri d_{\cN=4}\frac{{\bm x}_{13\alpha\beta}}{{\bm x}_{13}{}^4{\bm x}_{23}{}^2}
 {\cal D}^{4\beta}_{(3)}
 \frac{\varepsilon_{IJKL}\Theta_3^{I\mu}\Theta_3^{J\nu}\Theta_3^{K\rho}\Theta_3^{L\sigma}
 X_{3\mu\nu} X_{3\rho\sigma}}{X_3{}^5}\Big|\non\\
&=&-4\ri d_{\cN=4}\frac{{\bm x}_{13\alpha\beta}}{{\bm x}_{13}{}^4{\bm x}_{23}{}^2}
\frac{
 X_3^{\mu\nu} X_3^{\rho\sigma}
 \Theta^{\hat I}_{3\mu} \Theta^{\hat J}_{3\nu}
 \Theta^{\hat K}_{3\gamma} \varepsilon_{\hat I \hat J \hat K}}{X_3{}^5}~.
\eea
Here, in the second line, we applied the identity
(\ref{useful-prop-a}). Note that, in contrast to (\ref{A6_}),
this correlation function depends on the coefficient $d_{\cN=4}$
rather than $\tilde d_{\cN=4}$.

\subsubsection{Correlator $\langle J_\alpha J_\beta S \rangle$}

To compute this correlation function we have to hit (\ref{AppJJJ})
by two spinor covariant derivatives
\be
\langle J_\alpha(z_1) S(z_2) J_\beta(z_3) \rangle =
D^4_{(3)\beta}D^4_{(1)\alpha}
\langle J(z_1) J(z_2) J(z_3) \rangle  |
=D^4_{(3)\beta}D^4_{(1)\alpha} \frac1{{\bm x}_{13}{}^2{\bm x}_{23}{}^2}
H(X_3,\Theta_3)|~.
\label{A10_}
\ee
As is seen from (\ref{A7_}), when two covariant spinor
derivatives hit the correlation function (\ref{AppJJJ}), only the
following two terms survive under the bar-projection
\begin{subequations}
\bea
\langle J_\alpha(z_1) S(z_2) J_\beta(z_3) \rangle &= &A+B~,\\
A&=&\frac1{{\bm x}_{23}{}^2}\left(
D^4_{(3)\beta} D^4_{(1)\alpha}\frac1{{\bm x}_{13}{}^2} \right)
H(X_3,\Theta_3)|~,\label{A11b_}\\
B&=&\frac1{{\bm x}_{13}{}^2 {\bm x}_{23}{}^2}
D^4_{(3)\beta}D^4_{(1)\alpha} H(X_3,\Theta_3)|~.\label{A11c_}
\eea
\end{subequations}
In the part $A$ we easily compute the derivatives owing to (\ref{A7_})
\be
D^4_{(3)\beta} D^4_{(1)\alpha} \frac1{{\bm x}_{13}{}^2} \Big|
=2\ri \frac{{\bm x}_{13\alpha\beta}}{{\bm x}_{13}{}^4}~.
\label{A12_}
\ee
Thus, for (\ref{A11b_}) we have
\be
A= 2\ri \frac{{\bm x}_{13\alpha\beta}}{{\bm x}_{13}{}^4 {\bm x}_{23}{}^2}
\frac{\tilde d_{\cN=4}}{X_3}~.
\label{A13_}
\ee

In the part $B$ given by (\ref{A11c_}) two derivatives hit the
function $H$. For one of them we apply the identity
(\ref{useful-prop-a}) to represent it in the form
\be
D^4_{(3)\beta} D^4_{(1)\alpha} H|=
 D^4_{(3)\beta} {\bm x}^{-1}_{13\gamma\alpha}
  u^{4I}_{13} {\cal D}^{I\gamma} H|
={\bm x}^{-1}_{13\gamma\alpha}
 [D^4_{(3)\beta}{\cal D}^{4\gamma}H|
  + (D^4_{(3)\beta}u_{13}^{4\hat I}){\cal D}^{\hat I\gamma}H|]~.
 \label{A14_}
\ee
The two terms in the right-hand side of (\ref{A14_}) give the
following two contributions to the correlation function
\begin{subequations}
\bea
B_1 &=& \frac{{\bm x}_{13\alpha}{}^\gamma}{{\bm x}_{13}{}^4 {\bm
x}_{23}{}^4} D^4_{(3)\beta} {\cal D}^4_{(3)\gamma} \frac{\tilde
d_{\cN=4}}{X_3}\Big|~,\label{A15a_}\\
B_2 &=& \frac{{\bm x}_{13\alpha}{}^\gamma}{{\bm x}_{13}{}^4 {\bm x}_{23}{}^4}
(D^4_{(3)\gamma}u_{13}^{4\hat I}){\cal D}^{\hat I}_{(3)\gamma} \frac{\tilde
d_{\cN=4}}{X_3}\Big|~.\label{A15b_}
\eea
\end{subequations}
Here we have taken into account that the last term in (\ref{AppH}) does not
contribute to (\ref{A14_}).

In the right-hand side of (\ref{A15a_}) we use the explicit form
(\ref{generalized-DQ}) of the derivative ${\cal D}^4_\gamma$ to
represent this expression as
\bea
B_1 &=& \ri \tilde d_{\cN=4} \frac{{\bm x}_{13\alpha}{}^\gamma}{{\bm x}_{13}{}^4 {\bm x}_{23}{}^4}
(D^4_{(3)\beta}\Theta^{4\delta}_{(3)})\partial_{\gamma\delta}
\frac1{X_3}\Big| \non\\
&=&-\ri \tilde d_{\cN=4} \frac{{\bm x}_{13\alpha}{}^\gamma}{{\bm x}_{13}{}^4 {\bm x}_{23}{}^4}
 ({\bm x}_{13}^{-1}{}^\delta{}_\beta - {\bm x}_{23}^{-1}{}^\delta{}_\beta)
 \frac{X_{3\gamma\delta}}{X_3{}^3}\non\\
&=&-\ri \tilde d_{\cN=4} \frac{{\bm x}_{13\alpha\gamma}}{{\bm x}_{13}{}^4 {\bm x}_{23}{}^4}
 \left(-{\bm X}_{3\beta\delta} + \ri \varepsilon_{\beta\delta}
 \frac{\theta_{13}{}^2}{{\bm x}_{13}{}^2} +2\ri{\bm x}^{-1}_{13\beta\mu}
  \theta_{13}^{\hat I \mu}\theta_{32}^{\hat I \nu}{\bm x}_{32\nu\delta}^{-1}\right)
   \frac{X_{3}^{\gamma\delta}}{X_3{}^3}~.
   \label{A16_}
\eea
Here, in the last line, we applied the identity
\be
{\bm x}^{-1}_{13\alpha\beta} - {\bm x}^{-1}_{23\alpha\beta}
=-{\bm X}_{3\beta\alpha}
 +\ri \varepsilon_{\beta\alpha} \frac{\theta_{13}{}^2}{{\bm x}_{13}{}^2}
 +2\ri{\bm x}^{-1}_{13\beta\mu} \theta_{13}^{\hat I\mu} \theta_{32}^{\hat I\nu}
  {\bm x}^{-1}_{32\nu\alpha}~.
\ee
In this identity, only the first term in the right-hand side
is given by the three-point structure while the other two terms
are non-covariant in the sense that they are
represented by the combination of two-point superconformal
structures and cannot be expressed solely in terms of three-point ones.
These non-covariant terms should cancel against
the contributions form (\ref{A15b_}). Indeed, using the definition
(\ref{two-point-u}) we compute the derivatives in (\ref{A15b_})
\bea
B_2&=& 2\tilde d_{\cN=4} \frac{{\bm x}_{13\alpha}{}^\gamma}{{\bm x}_{13}{}^4 {\bm x}_{13}{}^2}
 {\bm x}^{-1}_{13\beta\rho }\theta_{13}^{\hat I\rho }\Theta^{\hat
 I\delta}_{3}\partial_{\gamma\delta}\frac1{X_3}\non\\
 &=&\tilde d_{\cN=4} \frac{{\bm x}_{13\alpha}{}^{\gamma}
 \theta_{13}{}^2}{{\bm x}_{13}{}^6 {\bm x}_{23}{}^2}
 \frac{X_{3\gamma\beta} }{X_3{}^3}
 +2\tilde d_{\cN=4} \frac{{\bm x}_{13\alpha\gamma}}{{\bm x}_{13}{}^4 {\bm x}_{23}{}^2}
  {\bm x}^{-1}_{13\beta\rho} {\bm x}^{-1}_{\delta\sigma}
   \theta_{13}^{\hat I\rho} \theta_{23}^{\hat I\sigma}
    \frac{X_3^{\gamma\delta}}{X_3{}^3}~.
    \label{A18_}
\eea
Thus, in the sum of (\ref{A16_}) and (\ref{A18_}) only one term
remains which we represent in the following form using (\ref{XX})
\be
B_1+B_2 = \ri \tilde d_{\cN=4} \frac{{\bm x}_{13\alpha\gamma}}{{\bm x}_{13}{}^4 {\bm x}_{23}{}^2}
 \frac{{\bm X}_{3\beta\delta} X_3^{\gamma\delta}}{X_3{}^3}
 =\ri \tilde d_{\cN=4} \frac{{\bm x}_{13\alpha\gamma}}{{\bm x}_{13}{}^4 {\bm x}_{23}{}^2}
 \left(-\frac{\delta_\beta^\gamma}{X_3}
 -\frac\ri2 \frac{X^\gamma_{3\beta}\Theta_3{}^2}{X_3{}^3}
 \right)~.
 \label{A19_}
\ee

Finally, we put together the contributions (\ref{A13_}) and (\ref{A19_})
and get the resulting expression for the correlation function
(\ref{A10_}) in the form
\begin{subequations}
\bea
\langle J_\alpha(z_1) S(z_2) J_\beta(z_3) \rangle &=&\ri \tilde
d_{\cN=4} \frac{{\bm x}_{13\alpha\gamma}}{{\bm x}_{13}{}^4 {\bm
x}_{23}{}^2} H^\gamma_\beta(X_3,\Theta_3)~,\label{A20a_}\\
H^\gamma_\beta(X,\Theta)&=&\ri \frac{\delta^\gamma_\beta}{X}
+\frac12 \frac{X^\gamma_\beta \Theta^2}{X^3}~.
\label{A20b_}
\eea
\end{subequations}
One can verify that the tensor (\ref{A20b_}) obeys the equations
\be
{\cal D}^{\hat I}_\alpha H^\alpha_\beta =0~,\qquad
({\cal D}^{\hat I\alpha}{\cal D}^{\hat J}_\alpha
 -\frac13 \delta^{\hat I\hat J}{\cal D}^{\hat K \alpha}{\cal D}^{\hat K}_\alpha)
  H^\gamma_\beta =0~,
\ee
which are the corollaries of (\ref{A2_}).

It is interesting to note that the correlation function
(\ref{A20a_}) depends only on the parameter $\tilde d_{\cN=4}$
similar to (\ref{A6_}).

\subsubsection{Correlator $\langle J_\alpha J_\beta J_\gamma \rangle$}
To compute the correlation function with three
$\cN=3$ supercurrents $J_\alpha$ we have to hit (\ref{AppJJJ}) by
three spinor covariant derivatives
\be
\langle J_\alpha(z_1) J_\beta(z_2) J_\gamma(z_3) \rangle =
-\ri D^4_{(1)\alpha}
D^4_{(2)\beta} D^4_{(3)\gamma} \langle J(z_1) J(z_2) J(z_3)
\rangle|~.
\label{A2}
\ee
First of all, we point out that the first term in (\ref{AppH})
does not contribute to (\ref{A2}). Indeed, due to the identity
(\ref{A7_}), when three derivatives hit this term we always get the
contribution which vanishes under the bar-projection
\be
-\ri \tilde d_{\cN=4} D^4_{(1)\alpha}
D^4_{(2)\beta} D^4_{(3)\gamma} \frac1{{\bm x}_{13}{}^2 {\bm x}_{23}{}^2
X_3}\Big| =0~.
\label{A4}
\ee
Hence, we need to consider only the second term in (\ref{AppH}).

Taking into account (\ref{A4}) we represent (\ref{A2}) as a
sum of two contributions
\begin{subequations}
\bea
&&\langle J_\alpha(z_1) J_\beta(z_2) J_\gamma(z_3) \rangle =
A+B~,\\&&
A=
\frac{\ri}{{\bm x}_{13}{}^2} \left( D^4_{(3)\gamma} D^4_{(2)\beta} \frac1{{\bm
x}_{23}{}^2} \right)
 D^4_{(1)\alpha} \tilde H(X_3,\Theta_3)\non\\&&\qquad\qquad
-\frac{\ri}{{\bm x}_{23}{}^2} \left( D^4_{(3)\gamma} D^4_{(1)\alpha}\frac1{{\bm
x}_{13}{}^2}\right)
D^4_{(2)\beta} \tilde H(X_3,\Theta_3)\Big|~,\label{A5B}
\\&&
B=\frac{\ri}{{\bm x}_{13}{}^2 {\bm x}_{23}{}^2}
 D^4_{(3)\gamma} D^4_{(2)\beta} D^4_{(1)\alpha} \tilde H(X_3,\Theta_3) \Big|~,
\label{A5C}
\eea
\end{subequations}
where $\tilde H$ is the second term in (\ref{AppH})
\be
\tilde H(X,\Theta)=d_{\cN=4}\frac{\varepsilon_{IJKL}\Theta^{I\alpha}\Theta^{J\beta}\Theta^{K\gamma}\Theta^{L\delta}
 X_{\alpha\beta} X_{\gamma\delta}}{X^5}
=\frac{4d_{\cN=4}}{X^5} (\Theta^{\hat I}_\alpha\Theta^{\hat J}_\mu
 \Theta^{\hat K}_\nu \varepsilon_{\hat I\hat J\hat
 K})\Theta^4_\delta X^{\alpha\mu} X^{\nu\delta} ~.
\label{tildeH}
\ee

In the right-hand side of (\ref{A5B}) we apply the following
relations
\be
D^4_{(3)\gamma}D^4_{(2)\beta}\frac1{{\bm x}_{23}{}^2} \Big| =
2\ri\frac{{\bm x}_{23\beta\gamma}}{{\bm x}_{23}{}^4}~,\qquad
D^4_{(3)\gamma}D^4_{(1)\alpha}\frac1{{\bm x}_{13}{}^2} \Big| =
2\ri\frac{{\bm x}_{13\alpha\gamma}}{{\bm x}_{13}{}^4}~.
\label{A7}
\ee
Next, using the identities (\ref{useful-prop}) we have
\be
D^4_{(1)\alpha} \tilde H(X_3,\Theta_3)| = -\frac{{\bm x}_{13\alpha\rho}}{{\bm x}_{13}{}^2}
 {\cal D}^{4\rho}_{(3)} \tilde H(X_3,\Theta_3) |~,\quad
D^4_{(2)\beta}\tilde H(X_3,\Theta_3)|  = \frac{{\bm x}_{23\beta\rho}}{{\bm
x}_{23}{}^2} {\cal D}^{4\rho}_{(3)} \tilde H(X_3,\Theta_3)|~,
\label{A8}
\ee
where the derivative of (\ref{tildeH}) reads
\be
{\cal D}^{4\rho } \tilde H(X,\Theta)| = 4d_{\cN=4}\frac{X^{\rho\beta} X^{\gamma\delta}
 \varepsilon_{\hat I\hat J\hat K}\Theta^{\hat I}_\beta \Theta^{\hat J}_\gamma
 \Theta^{\hat K}_\delta}{X^5}~.
\label{A9}
\ee
Substituting now (\ref{A7})--(\ref{A9}) into (\ref{A5B}) we get
the corresponding contribution to the correlation function
\begin{subequations}
\bea
A &= & \frac{{\bm x}_{13\alpha\alpha'} {\bm x}_{23\beta\beta'}}{{\bm x}_{13}{}^4 {\bm x}_{23}{}^4}
H_{(A)}^{\alpha'\beta'}{}_\gamma~,\\
H_{(A)}^{\alpha'\beta'}{}_\gamma &=& 8 d_{\cN=4}
\frac{\varepsilon_{\hat I\hat J\hat K}\Theta^{\hat I}_{\gamma'}\Theta^{\hat J}_\mu\Theta^{\hat K}_\nu X^{\mu\nu}
(X^{\beta'\gamma'}\delta_\gamma^{\alpha'} +
X^{\alpha'\gamma'}\delta_\gamma^{\beta'}) }{X^5}~.
\label{H-A_}
\eea
\end{subequations}

Using (\ref{useful-prop}) we get
the following representation for the part (\ref{A5C})
\be
B=\frac{{\bm x}_{13\alpha\alpha'} {\bm x}_{23\beta\beta'}}{{\bm x}_{13}{}^4 {\bm x}_{23}{}^4}
H_{(B)}^{\alpha'\beta'}{}_{\gamma}(X_3,\Theta_3)
~,
\label{C21}
\ee
where
\be
H_{(B)}^{\alpha\beta}{}_{\gamma }
 =
\bigg(-2 \Theta^{\hat I}_\gamma \frac\partial{\partial\Theta^{\hat I}_\beta}
 \frac\partial{\partial\Theta^{4}_\alpha}
+ X_{\mu\gamma}\frac\partial{\partial X_{\beta\mu}}
 \frac\partial{\partial\Theta^{4}_\alpha}
+X_{\mu\gamma}\frac\partial{\partial X_{\alpha\mu}}
 \frac\partial{\partial\Theta^{4}_\beta}
-X_{\mu\gamma}\frac\partial{\partial X_{\alpha\beta}}
\frac\partial{\partial\Theta^{4}_\mu}
 \bigg)\tilde H |~.
\label{222+}
\ee
Compute now the derivatives of the function (\ref{tildeH}) which
are necessary for (\ref{222+}):
\bea
&&-2 \Theta^{\hat I}_\gamma \frac\partial{\partial\Theta^{\hat I}_\beta}
 \frac\partial{\partial\Theta^{4}_\alpha} \tilde H| \non\\
&=&\frac{16 d_{\cN=4} }{X^5} X^{\alpha\mu} X^{\beta\nu}
 \Theta^{\hat I}_\gamma \Theta^{\hat J}_\mu \Theta^{\hat K}_\nu
  \varepsilon_{\hat I\hat J\hat K}
+\frac{8 d_{\cN=4}}{X^5} X^{\alpha\beta} X^{\mu\nu}
 \Theta^{\hat I}_\gamma \Theta^{\hat I}_\mu \Theta^{\hat K}_\nu
 \varepsilon_{\hat I\hat J\hat K}~,\label{H14}\\
&&\bigg(
 X_{\mu\gamma}\frac\partial{\partial X_{\beta\mu}}
 \frac\partial{\partial\Theta^{4}_\alpha}
+X_{\mu\gamma}\frac\partial{\partial X_{\alpha\mu}}
 \frac\partial{\partial\Theta^{4}_\beta}
-X_{\mu\gamma}\frac\partial{\partial X_{\alpha\beta}}
\frac\partial{\partial\Theta^{4}_\mu}
 \bigg)\tilde H |
 \non\\
&=&-\frac{4 d_{\cN=4}}{X^5} X^{\alpha\beta}X^{\mu\nu}
 \Theta^{\hat I}_\gamma \Theta^{\hat J}_\alpha \Theta^{\hat K}_{\beta}
  \varepsilon_{\hat I\hat J\hat K}
-\frac{8d_{\cN=4}}{X^5} X^{\alpha\mu} X^{\beta\nu}
 \Theta^{\hat I}_\gamma \Theta^{\hat J}_\alpha \Theta^{\hat
 K}_\beta \varepsilon_{\hat I\hat J\hat K}
 \non\\
&&-\frac{4d_{\cN=4}}{X^5} X^{\mu\nu} (X^{\alpha\rho}\delta^\beta_\gamma
 +X^{\beta\rho}\delta^\alpha_\gamma)
 \Theta^{\hat I}_{\rho}\Theta^{\hat J}_\mu \Theta^{\hat K}_\nu
  \varepsilon_{\hat I\hat J\hat K}~.
  \label{H15}
\eea
Finally, we collect the results of computations (\ref{H-A_}),
(\ref{H14}) and (\ref{H15}) in a single expression
\begin{subequations}
\label{A16}
\bea
\langle J_\alpha(z_1) J_\beta(z_2) J_\gamma(z_3) \rangle &=&
\frac{{\bm x}_{13\alpha\alpha'} {\bm x}_{23\beta\beta'}}{{\bm x}_{13}{}^4 {\bm x}_{23}{}^4}
H^{\alpha'\beta'}{}_\gamma(X_3,\Theta_3)~, \label{A16a}\\
H^{\alpha\beta}{}_\gamma( X,\Theta)&=&\frac{4 d_{\cN=4} }{ X^5}
\Big[
(\delta^\beta_\gamma X^{\alpha\rho}
 +\delta^\alpha_\gamma  X^{\rho\beta}) X^{\mu\nu}\Theta^{\hat I}_\mu \Theta^{\hat J}_\nu\Theta^{\hat K}_\rho
  \varepsilon_{\hat I\hat J\hat K}\non\\&&
 + X^{\beta\alpha} X^{\mu\nu}\Theta^{\hat I}_{\mu}\Theta^{\hat J}_{\nu}
 \Theta^{\hat K}_{\gamma}\varepsilon_{\hat I\hat J\hat K}
+2 X^{\alpha\mu}X^{\nu\beta}\Theta^{\hat I}_{\mu}\Theta^{\hat J}_{\nu}\Theta^{\hat K}_{\gamma}
 \varepsilon_{\hat I\hat J\hat K}\Big]~.~~~~
\label{A16b}
\eea
\end{subequations}
This expression for $H^{\alpha\beta}{}_\gamma$ coincides with
(\ref{8.4c}) upon the replacement $X_{\alpha\beta}\to {\bm
X}_{\alpha\beta}$. Although $X_{\alpha\beta}$ and ${\bm
X}_{\alpha\beta}$ differ in a $\Theta$-dependent term, see
(\ref{XX}), one can check that these additional terms do not
contribute to (\ref{A16b}). To match the expressions (\ref{8.4b}) and (\ref{A16})
one has to make also the identification of their parameters
(\ref{8.5b}).


\subsubsection{Reduction of the two-point function}

The superspace reduction of the two-point function of $\cN=4$
supercurrents (\ref{JJ}) to the $\cN=3$ superspace is much simpler
than the same procedure for the three-point functions. Indeed, the
correlator of the superfield $S$ immediately follow from (\ref{JJ})
\be
\langle S(z_1) S(z_2) \rangle  = \langle J(z_1) J(z_2) \rangle |
=\frac{c_{\cN=4}}{{\bm x}_{12}{}^2}~,
\ee
while for the correlation function of the $\cN=3$ supercurrent
$J_\alpha$ we have
\be
\langle J_\alpha(z_1) J_\beta(z_2) \rangle =
D^4_{(2)\beta} D^4_{(1)\alpha} \langle J(z_1) J(z_2) \rangle |
=c_{\cN=4} D^4_{(2)\beta} D^4_{(1)\alpha} \frac1{{\bm
x}_{12}{}^2}\Big|~.
\ee
Applying the identity (\ref{A12_}) we find
\be
\langle J_\alpha(z_1) J_\beta(z_2) \rangle =
2\ri c_{\cN=4} \frac{{\bm x}_{12\alpha\beta}}{{\bm x}_{12}{}^4}~.
\ee
Comparing this two-point function with (\ref{8.4a}) allows us to
get the relation (\ref{8.5a}) among the coefficients $c_{\cN=3}$
and $c_{\cN=4}$.


\subsection{$\cN=3 \to \cN=2$ reduction of the supercurrent correlation function}
\label{AppB}

Recall that the $\cN=3$ supercurrent $J_\alpha$ contains the
following two independent $\cN=2$ supermultiplets \cite{BKS}:
\begin{subequations}
\bea
R_\a &:=& J_\a|~, \qquad \qquad D^{\hat I \a} R_\a = 0 ~; \label{1.6a}\\
J_{\a\b} &:= &  D^3_{(\a} J_{\b)} |~, \qquad D^{\hat I \a} J_{\a\b} =0~,\qquad
\hat I =1,2~.
\eea
\end{subequations}
Here $J_{\a\b}$ is the $\cN=2$ supercurrent, while
$R_\a$ contains the third supersymmetry current and two
$R$-symmetry currents corresponding to $\sSO(3)/\sSO(2)$.
In this appendix we consider all two- and three-point correlation
functions of $R_\alpha$ and $J_{\alpha\beta}$ which follow from the
corresponding correlators of $\cN=3$ supercurrent $J_\alpha$.

\subsubsection{Two-point correlators}
Consider the two-point correlation function of the $\cN=3$
supercurrent (\ref{8.4a}). Obviously, the two-point correlator of
$R_\alpha$ superfield has the same form in the $\cN=2$ superspace
\be
\langle
R_\alpha(z_1) R_\beta(z_2)
 \rangle=
\langle J_\alpha(z_1) J_\beta(z_2)
 \rangle | = \ri c_{\cN=3} \frac{{\bm x}_{12\alpha\beta}}{{\bm
 x}_{12}{}^4}~,
\ee
where the bar-projection assumes $\theta_{3\alpha}=0$.

To find the two-point function of $\cN=2$ supercurrent we need
to hit (\ref{8.4a}) by two spinor covariant derivatives
\be
\langle J_{\alpha\alpha'}(z_1) J_{\beta\beta'}(z_2) \rangle
= - D^3_{(1)\alpha} D^3_{(2)\beta} \langle J_{\alpha'}(z_1)
 J_{\beta'}(z_2) \rangle|
  = \ri c_{\cN=3} D^3_{(2)\beta} D^3_{(1)\alpha}
   \frac{{\bm x}_{12\alpha'\beta'}}{{\bm x}_{12}{}^4}\Big|~.
\ee
It is straightforward to compute these derivatives using the
definition of the two-point structure (\ref{super-interv-X})
\be
\langle J_{\a \b} (z_1) J^{\a' \b'} (z_2) \rangle =-4 c_{\cN=3}
\frac{{\bm x}_{12 \a}{}^{ (\a'} {\bm x}_{12 \b}{}^{ \b')}}{ {\bm
x}_{12}{}^6}~.
\ee
Comparing this expression with (\ref{n2.3}) we find the
relation among the coefficients $c_{\cN=2}$ and $c_{\cN=3}$ given
in (\ref{cdN2}).


\subsubsection{Three-point correlators involving $R_\alpha$}

There are three correlation functions involving $R_\alpha$:
$$\langle R_\alpha(z_1)R_\beta(z_2) R_\gamma(z_3) \rangle~ ,\quad
\langle J_{\alpha\delta}(z_1)R_\beta(z_2) R_\gamma(z_3) \rangle~
,\quad
\langle J_{\alpha\delta}(z_1) J_{\beta\rho}(z_2) R_\gamma(z_3) \rangle~.
$$
It is easy to see that
\be
\langle R_\alpha(z_1)R_\beta(z_2) R_\gamma(z_3) \rangle=
\langle J_\alpha(z_1)J_\beta(z_2) J_\gamma(z_3) \rangle|=0~.
\ee
Indeed, the tensor $H_{\alpha\beta\gamma}$ which defines the
$\cN=3$ supercurrent correlator (\ref{8.4c}) vanishes under the
bar-projection owing to $\Theta^I_\alpha \Theta^J_\beta \Theta^K_\gamma
\varepsilon_{IJK}|=0$. Similarly, it is possible to show that
\be
\langle J_{\alpha\delta}(z_1) J_{\beta\rho}(z_2) R_\gamma(z_3) \rangle
=-D^3_{(1)\delta} D^3_{(2)\rho} \langle J_{\alpha}(z_1) J_{\beta}(z_2) J_\gamma(z_3)
\rangle| =0~.
\ee
Thus, we need to consider only $\langle J_{\alpha\delta}(z_1)R_\beta(z_2) R_\gamma(z_3)
\rangle$ which is non-trivial
\bea
\langle J_{\alpha\delta}(z_1) R_\beta(z_2) R_\gamma(z_3) \rangle
&=&D^3_{(1)\delta} \langle J_\alpha(z_1) J_\beta(z_2) J_\gamma(z_3)
\rangle| \non\\
&=&-d_{\cN=3} \frac{{\bm x}_{13\alpha\alpha'}{\bm x}_{13\delta\delta'}
 {\bm x}_{23\beta\beta'}}{{\bm x}_{13}{}^6 {\bm x}_{23}{}^4}
 {\cal D}^{3\delta'}H^{\alpha'\beta'}{}_{\gamma}|~.
 \label{A43__}
\eea
Here we have applied the identity (\ref{useful-prop-a}) and the
representation (\ref{8.4b}) for the $\cN=3$ supercurrent
three-point function. It is easy to evaluate the derivative of the
tensor (\ref{8.4c}) since under the bar projection only those
terms survive in which ${\cal D}^{3\delta}$ acts on the
generalized Grassmann variable $\Theta^I_\alpha$ but not on ${\bm
X}_{\mu\nu}$. As a result, we get the following representation for
the correlator (\ref{A43__})
\begin{subequations}
\bea
\langle J_{\alpha\delta}(z_1) R_\beta(z_2) R_\gamma(z_3)
\rangle&=&d_{\cN=3} \frac{{\bm x}_{13\alpha\alpha'} {\bm x}_{13\delta\delta'}
 {\bm x}_{23\beta\beta'}}{{\bm x}_{13}{}^6 {\bm x}_{23}{}^4}
 H^{\delta'\alpha'\beta'}{}_\gamma(X_3,\Theta_3)~,\\
H^{\delta\alpha\beta}{}_\gamma(X,\Theta)&=&\frac1{X^5}\bigg[
\delta_\gamma^\beta X^{\alpha\delta} X^{\mu\nu}(\Theta\Theta)_{\mu\nu}
+2\delta_\gamma^\beta X^{\alpha\mu} X^{\delta\nu}
 (\Theta\Theta)_{\mu\nu} \non\\&&
+2X^{\alpha\delta} X^{\nu\beta}(\Theta\Theta)_{\nu\gamma}
+2\delta_\gamma^{(\alpha} X^{\delta)\beta} X^{\mu\nu} (\Theta\Theta)_{\mu\nu}
 \non\\&&
+4\delta_\gamma^{(\delta} X^{\alpha)\mu} X^{\nu\beta}(\Theta\Theta)_{\mu\nu}
+4X^{\beta(\alpha}X^{\delta)\nu} (\Theta\Theta)_{\nu\gamma}
\bigg]~,
\eea
\end{subequations}
where we use the notation $(\Theta\Theta)_{\mu\nu}=\Theta^I_\mu \Theta^J_\nu
\varepsilon_{IJ}$.


\subsubsection{Three-point correlator of the $\cN=2$ supercurrent}

Consider now the three-point function of the $\cN=2$ supercurrent
\be
\langle
J_{\alpha\alpha'}(z_1)J_{\beta\beta'}(z_2)J_{\gamma\gamma'}(z_3)
\rangle = - D^3_{(1)\alpha} D^{3}_{(2)\beta} D^3_{(3)\gamma}
\langle
 J_{\alpha'}(z_1)J_{\beta'}(z_2) J_{\gamma'}(z_3)
\rangle|~.
\label{A17}
\ee
The $\cN=3$ supercurrent three-point correlation function is found in the form
(\ref{A16}) involving the tensor $H^{\alpha\beta}{}_\gamma$. For the following
calculations it will be convenient to use the form of this tensor
with the pair of spinor indices $\alpha\beta$ converted into a
vector one $m$
\bea
\label{307}
H^m_\gamma &=&-\frac12 \gamma^m_{\alpha\beta}
H^{\alpha\beta}{}_\gamma = -\ri\frac 6{X^3} \Theta^3_\gamma
(\Theta\Theta)^m +\ri \frac{18}{X^5} X^m X^p \Theta^3_\gamma
(\Theta\Theta)_p
\non\\&&
-\ri\frac2{X^3} \varepsilon^{pmq} (\gamma_q)^\mu_\gamma
\Theta^3_\mu (\Theta\Theta)_p
+\ri\frac8{X^5} X^m X_r \varepsilon^{prq}(\gamma_q)^\mu_\gamma
  \Theta^3_\mu (\Theta\Theta)_p \non\\&&
+\ri\frac4{X^5} \varepsilon^{mrq}(\gamma_q)^\mu_\gamma X_r X^p
 \Theta^3_\mu (\Theta\Theta)_p
+\ri\frac2{X^5} X_r X^q \varepsilon^{mrq}(\gamma_q)^\mu_\gamma
\Theta^3_\mu(\Theta\Theta)_p~,
\eea
where we employ the short-notation $(\Theta\Theta)_m$ for
\be
(\Theta \Theta)_m = -\frac{\ri}{2} (\g_m)^{\a \b} \Theta_{\a}^I
\Theta_{\b}^J \varepsilon_{IJ}~.
\label{ThetaTheta}
\ee

We substitute (\ref{A16a}) into (\ref{A17}) and represent it as a
sum of two parts with specific distribution of covariant  spinor derivatives
on the factors
\begin{subequations}
\bea
&&
D^3_{(3)\gamma} D^3_{(2)\beta} D^3_{(1)\alpha} \frac{{\bm x}_{13\alpha'\alpha''}{\bm x}_{23\beta'\beta''}}{{\bm x}_{13}{}^4{\bm x}_{23}{}^4}
H^{\alpha''\beta''}{}_{\gamma'}({X}_{3},\Theta_{3})|=A+B~,\label{A20a}\\
A&=&\frac{{\bm x}_{13\alpha'\alpha''}}{{\bm x}_{13}{}^4}
\left(D^3_{(3)\gamma}D^3_{(2)\beta}\frac{{\bm x}_{23\beta'\beta''}}{{\bm x}_{23}{}^4}\right)
 D^3_{(1)\alpha}H^{\alpha''\beta''}{}_{\gamma'}
\non\\&&
-\frac{{\bm x}_{23\beta'\beta''}}{{\bm x}_{23}{}^4}
 \left( D^3_{(3)\gamma}D^3_{(1)\alpha} \frac{{\bm x}_{13\alpha'\alpha''}}{{\bm x}_{13}{}^4} \right)
  D^3_{(2)\beta}H^{\alpha''\beta''}{}_{\gamma'}| ~,\label{partA}\\
B&=&\frac{{\bm x}_{13\alpha'\alpha''}{\bm x}_{23\beta'\beta''}}{{\bm x}_{13}{}^4{\bm x}_{23}{}^4}
 D^3_{(3)\gamma} D^3_{(2)\beta} D^3_{(1)\alpha}
 H^{\alpha''\beta''}{}_{\gamma'}|~.
 \label{partB}
\eea
\end{subequations}
One can check that in (\ref{A20a}) the terms in which the covariant spinor
derivatives are distributed in other ways vanish under the
bar-projection. Now consider the computations of contribution
(\ref{partA}) and (\ref{partB}) separately.

In the part $A$ given by (\ref{partA}) we need the following
relations
\bea
D^3_{(3)\gamma} D^3_{(1)\alpha}
\frac{{\bm x}_{13\alpha'\alpha''}}{{\bm x}_{13}{}^4}|&=&\frac{2\ri}{{\bm x}_{13}{}^6}
({\bm x}_{13\alpha\alpha''}{\bm x}_{13\alpha'\gamma} +{\bm x}_{13\alpha\gamma}
{\bm x}_{13\alpha'\alpha''})~,\non\\
D^3_{(3)\gamma} D^3_{(2)\beta}
\frac{{\bm x}_{23\beta'\beta''}}{{\bm x}_{23}{}^4}|&=&
 \frac{2\ri}{{\bm x}_{23}{}^6} ({\bm x}_{23\beta\beta''}{\bm x}_{23\beta'\gamma}+
 {\bm x}_{23\beta\gamma}{\bm x}_{23\beta'\beta''})~,
 \label{A21}
\eea
which follow from the definition (\ref{super-interv-X}).
With the use of identities (\ref{useful-prop}) we have
\bea
D^3_{(1)\alpha}H^{\alpha'\beta'\gamma}(X_3,\Theta_3)|&=&
 -\frac{{\bm x}_{13\alpha\rho}}{{\bm x}_{13}{}^2} {\cal D}_{(3)}^{3\rho}H^{\alpha'\beta'\gamma}(
 X_3,\Theta_3)| ~,\non\\
D^3_{(2)\beta}H^{\alpha'\beta'\gamma}(X_3,\Theta_3)|&=&
 \frac{{\bm x}_{23\beta\rho}}{{\bm x}_{23}{}^2}{\cal D}^{3\rho}_{(3)}H^{\alpha'\beta'\gamma}(X_3,\Theta_3)| ~.
\label{A22}
\eea
Taking into account (\ref{A21}) and (\ref{A22}) we represent the
part $A$ in the form
\be
A=\frac{{\bm x}_{13 \alpha \rho} {\bm x}_{13 \alpha' \rho'}   {\bm x}_{23  \beta \sigma}
 {\bm x}_{23 \beta' \sigma'}  }{ {\bm x}_{13}{}^6 {\bm x}_{23}{}^6}
H_{(A)}^{\rho \rho'\, \sigma \sigma'}{}_{\gamma \gamma'} (X_{3}, \Theta_{3})~,
\label{A23}
\ee
where
\be
H_{(A)}^{\rho\rho'\,\sigma\sigma'}{}_{\gamma\gamma' } =-4\ri(
\delta_\gamma^{\sigma}{\cal D}^{3\rho}H^{\rho'\sigma'}{}_{\gamma'}
+\delta_\gamma^{\sigma'}{\cal D}^{3\rho}H^{\rho'\sigma}{}_{\gamma'}
+\delta_\gamma^{\rho} {\cal D}^{3\sigma} H^{\rho'\sigma'}{}_{\gamma'}
+\delta_\gamma^{\rho'} {\cal D}^{3\sigma} H^{\rho\sigma'}{}_{\gamma'})|~.
 \label{311}
\ee

In the expression (\ref{partB}) we use the identities
(\ref{useful-prop}) to represent it in the form
\bea
D^3_{(3)\gamma} D^3_{(2)\beta} D^3_{(1)\alpha}
 H^{\alpha'\beta'\gamma}(X_3,\Theta_3)|&=&
  \ri{\bm x}^{-1}_{13\rho\alpha}{\bm x}^{-1}_{23\sigma\beta}
 D^3_{(3)\gamma}  u_{13}^{3J} u_{23}^{3K} {\cal Q}^{K\sigma}
  {\cal D}^{J\rho} H^{\alpha'\beta'\gamma}(X_3,\Theta_3)|\non\\&
=&\ri({\bm x}^{-1}_{13})^\rho{}_\alpha({\bm x}^{-1}_{23})^\sigma{}_\beta
 D^3_{(3)\gamma}
[{\cal Q}^3_\sigma {\cal D}^3_\rho
 + u_{13}^{31}{\cal Q}_\sigma^3 {\cal D}_\rho^1
  + u_{13}^{32}{\cal Q}^3_\sigma {\cal D}^2_\rho
  \non\\&&
  + u_{23}^{31}{\cal Q}^1_\sigma {\cal D}^3_\rho
  + u_{23}^{32}{\cal Q}^2_\sigma {\cal D}^3_\rho]H^{\alpha'\beta'\gamma}(X_3,\Theta_3)|
  ~.
  \label{A25}
\eea
Consider various terms in this expression separately.

For the first term in the square brackets in (\ref{A25}) we use
explicit forms of generalised spinor covariant derivative and the
supercharge (\ref{generalized-DQ}) to rewrite it as
\bea
&&
 D^3_{(3)\gamma}
{\cal Q}^3_\sigma {\cal D}^3_\rho H^{\alpha'\beta'\gamma}| \non\\
&=& [({\bm x}^{-1}_{13})^\mu{}_\gamma - ({\bm x}^{-1}_{23})^\mu{}_\gamma]
\left( \frac\partial{\partial X^{\sigma\mu}}\frac\partial{\partial\Theta^{3\rho}}
 +\frac\partial{\partial X^{\sigma\mu}}\frac\partial{\partial\Theta^{3\rho}}
 -\frac\partial{\partial X^{\rho\sigma}} \frac\partial{\partial\Theta^{3\mu}}
 \right)H^{\alpha'\beta'\gamma} |~.
 \label{A26}
\eea
Here we used the fact that in the bar-projection only those terms
survive in which the derivative $D^3_{(3)\gamma}$ acts on
$\Theta^{3\mu}$ and produces the factor
$[({\bm x}^{-1}_{13})^\mu{}_\gamma - ({\bm
x}^{-1}_{23})^\mu{}_\gamma]$. As is pointed out in \cite{BKS},
this factor cannot be expressed solely in terms of ${\bm X}_{3\alpha\beta}$ and
$\Theta^I_{3\alpha}$, but it involves the two-point structures as
well
\be
({\bm x}_{13}^{-1})_{\alpha\beta} - ({\bm x}_{23}^{-1})_{\alpha\beta}
=-{\bm X}_{3\alpha\beta} +\ri\frac{\varepsilon_{\alpha\beta}}{{\bm x}_{23}{}^2}
\theta_{23}^2
+2\ri ({\bm x}^{-1}_{13})_{\alpha\mu} \theta^\mu_{13}
 \theta^{\nu}_{32} ({\bm x}_{32}^{-1})_{\nu\beta}~.
\label{C17}
\ee
The last two terms here are non-covariant in the sense that they
are expressed in terms of two-point superconformal invariants
rather than the three-point ones. Then, taking into account
(\ref{C17}) we rewrite (\ref{A26}) as
\bea
 D^3_{(3)\gamma}
{\cal Q}^3_\sigma {\cal D}^3_\rho H^{\alpha'\beta'\gamma}| &=&
- {\bm X}_3{}^\mu{}_\gamma
\left( \frac\partial{\partial X^{\sigma\mu}}\frac\partial{\partial\Theta^{3\rho}}
 +\frac\partial{\partial X^{\sigma\mu}}\frac\partial{\partial\Theta^{3\rho}}
 -\frac\partial{\partial X^{\rho\sigma}} \frac\partial{\partial\Theta^{3\mu}}
 \right)H^{\alpha'\beta'\gamma}|\non\\&&
 +\mbox{non-covariant terms},
 \label{A28}
\eea
where the `non-covariant terms' are those which correspond to the
last two terms in (\ref{C17}). Here we do not write down these terms
explicitly as they cancel against the contributions coming
from the remaining terms in the square brackets in (\ref{A25})
\footnote{This cancellation has been explicitly demonstrated in Appendix C.1 of \cite{BKS}
for the case of superspace reduction of the $\cN=2$ supercurrent correlation function
down to $\cN=1$. In the present case the cancellation of the non-covariant terms
can be checked in the same way.}
\bea&&
 D^3_{(3)\gamma}
[ u_{13}^{31}{\cal Q}_\sigma^3 {\cal D}_\rho^1
  + u_{13}^{32}{\cal Q}^3_\sigma {\cal D}^2_\rho
  + u_{23}^{31}{\cal Q}^1_\sigma {\cal D}^3_\rho
  + u_{23}^{32}{\cal Q}^2_\sigma {\cal D}^3_\rho]H^{\alpha'\beta'\gamma}|
  \non\\
  &=&2(\Theta^1_{\gamma}\frac\partial{\partial\Theta^{1\sigma}}\frac\partial{\partial\Theta^{3\rho}}
+\Theta^2_{\gamma}\frac\partial{\partial\Theta^{2\sigma}}\frac\partial{\partial\Theta^{3\rho}})H^{\alpha'\beta'\gamma}|
-\mbox{non-covariant terms}.
\label{A29}
\eea
Thus, when we take the sum of (\ref{A28}) and (\ref{A29}) these
`non-covariant terms' cancel and we get the contribution to the
$\cN=2$ supercurrent correlation functions in the form
\be
B=\frac{{\bm x}_{13 \alpha \rho} {\bm x}_{13 \alpha' \rho'}   {\bm x}_{23  \beta \sigma}
 {\bm x}_{23 \beta' \sigma'}  }{ {\bm x}_{13}{}^6 {\bm x}_{23}{}^6}
H_{(B)}^{\rho \rho'\, \sigma \sigma'}{}_{\gamma \gamma'} (X_{3}, \Theta_{3})~,
\label{A30}
\ee
where
\bea
H_{(B)}^{\rho\rho'\,\sigma\sigma'}{}_{\gamma\gamma'}(X,\Theta)&=&\ri\Big(
2\Theta^1_{\gamma} \frac\partial{\partial\Theta^1_\rho} \frac\partial{\partial\Theta^3_\sigma}
+2\Theta^2_{\gamma} \frac\partial{\partial\Theta^2_\rho} \frac\partial{\partial\Theta^3_\sigma}
-X_{\mu\gamma}\frac\partial{\partial X_{\sigma\mu}}\frac\partial{\partial\Theta^3_\rho}
\non\\&&
-X_{\mu\gamma}\frac\partial{\partial X_{\rho\mu}}\frac\partial{\partial\Theta^3_\sigma}
+X_{\mu\gamma}\frac\partial{\partial X_{\rho\sigma}}\frac\partial{\partial\Theta^3_\mu}
\Big)H^{\rho'\sigma'}{}_{\gamma'}(X,\Theta)|~.
\eea

Summarizing (\ref{A23}) and (\ref{A30}) we find the $\cN=2$
supercurrent three-point correlation function
\be
\langle
J_{\alpha\alpha'}(z_1)J_{\beta\beta'}(z_2)J_{\gamma\gamma'}(z_3)
\rangle =
\frac{{\bm x}_{13 \alpha \rho} {\bm x}_{13 \alpha' \rho'}   {\bm x}_{23  \beta \sigma}
 {\bm x}_{23 \beta' \sigma'}  }{ {\bm x}_{13}{}^6 {\bm x}_{23}{}^6}
H^{\rho \rho'\, \sigma \sigma'}{}_{\gamma \gamma'} (X_{3},
\Theta_{3})~,
\ee
where the tensor $H^{\rho \rho'\, \sigma \sigma'}{}_{\gamma \gamma'}$
is expressed in terms of derivatives of the tensor
$H^{\alpha\beta}{}_\gamma$ in the $\cN=3$ theory given by
(\ref{A16b})
\begin{subequations}
\label{A33}
\bea
H^{\rho\rho'\,\sigma\sigma'}{}_{\gamma\gamma' } &=&
H_{(A)}^{\rho\rho'\,\sigma\sigma'}{}_{\gamma\gamma' }
+H_{(B_1)}^{\rho\rho'\,\sigma\sigma'}{}_{\gamma\gamma' }
+H_{(B_2)}^{\rho\rho'\,\sigma\sigma'}{}_{\gamma\gamma' }
+H_{(B_3)}^{\rho\rho'\,\sigma\sigma'}{}_{\gamma\gamma' }~,\\
H_{(A)}^{\rho\rho'\,\sigma\sigma'}{}_{\gamma\gamma' }&=&
2\ri
\bigg(\delta_\gamma^{\sigma} \frac\partial{\partial\Theta^3_\rho}
 H^{\rho'\sigma'}{}_{\gamma'}
+
 \delta_\gamma^{\sigma'} \frac\partial{\partial\Theta^3_\rho} H^{\rho'\sigma}{}_{\gamma'}
 \non\\&&
+ \delta_\gamma^{\rho} \frac\partial{\partial\Theta^3_\sigma}
 H^{\rho'\sigma'}{}_{\gamma'}
+ \delta_\gamma^{\rho'} \frac\partial{\partial\Theta^3_\sigma}
 H^{\rho\sigma'}{}_{\gamma'}\bigg) |~,\label{H-A}\\
H_{(B_1)}^{\rho\rho'\,\sigma\sigma'}{}_{\gamma\gamma' }&=&2\ri\left(
\Theta^1_{\gamma} \frac\partial{\partial\Theta^1_\rho} \frac\partial{\partial\Theta^3_\sigma}
+\Theta^2_{\gamma} \frac\partial{\partial\Theta^2_\rho} \frac\partial{\partial\Theta^3_\sigma}
\right)H^{\rho'\sigma'}{}_{\gamma'}|~,\label{B-1}\\
H_{(B_2)}^{\rho\rho'\,\sigma\sigma'}{}_{\gamma\gamma' }&=&
-\ri\left(X_{\mu\gamma}\frac\partial{\partial X_{\sigma\mu}}\frac\partial{\partial\Theta^3_\rho}
+X_{\mu\gamma}\frac\partial{\partial X_{\rho\mu}}\frac\partial{\partial\Theta^3_\sigma}
\right)H^{\rho'\sigma'}{}_{\gamma'} |~,\label{B-2}\\
H_{(B_3)}^{\rho\rho'\,\sigma\sigma'}{}_{\gamma\gamma' }&=&
\ri X_{\mu\gamma}\frac\partial{\partial X_{\rho\sigma}}\frac\partial{\partial\Theta^3_\mu}
H^{\rho'\sigma'}{}_{\gamma'}  | ~. \label{B-3}
\eea
\end{subequations}
As a result, the problem is reduced to computing the derivatives
of the tensor (\ref{A16b}).

Let us convert the spinor indices of $H^{\rho\rho'\,\sigma\sigma'}{}_{\gamma\gamma' }$
into the vector ones
\be
H^{mnk} = -\frac18
\gamma^m_{\rho\rho'}\gamma^n_{\sigma\sigma'}(\gamma^k)^{\gamma\gamma'}H^{\rho\rho'\,\sigma\sigma'}{}_{\gamma\gamma'}~.
\label{318}
\ee
It is known that the tensor $H^{mnk}$ which defines the
three-point correlation function of $\cN=2$ supercurrent
can be represented in the form \cite{BKS}
\be
H^{mnk} = (\Theta\Theta)_p C^{mnp,k}~,
\label{A35}
\ee
where $(\Theta\Theta)_p$ is given in (\ref{ThetaTheta}) and
$C^{mnp,k}$ is symmetric and traceless in the indices $mnp$,
\be
C^{mnp,k} = C^{(mnp),k} ~, \qquad
\eta_{mn}C^{mnp,k} =0~.
\label{A36}
\ee
Hence, the tensor $H^{\rho\rho'\,\sigma\sigma'}{}_{\gamma\gamma'}$
has the following symmetry property
\be
H^{\rho\rho'\,\sigma\sigma'}{}_{\gamma\gamma'}
= H^{(\rho\rho'\sigma\sigma')}{}_{\gamma\gamma'}~.
\ee
As a consequence, the relation (\ref{318}) can equivalently be
rewritten as
\be
H^{mnk} = -\frac18
\gamma^m_{\rho\sigma}\gamma^n_{\rho'\sigma'}(\gamma^k)^{\gamma\gamma'}H^{\rho\rho'\,\sigma\sigma'}{}_{\gamma\gamma'}~,
\ee
and it appears to be more convenient to use the tensor (\ref{307}) for further computations.
Indeed, the expression (\ref{H-A}) can be rewritten as a
derivative of (\ref{307})
\bea
H_{(A)}^{mnp}&=&2\ri(\gamma^n)_{\rho\sigma} (\gamma^k)^{\gamma\sigma}
 \frac\partial{\partial\Theta^3_\rho}
 H^m_\gamma=C_{(A)}^{mnp,k}(\Theta\Theta)_p~,\non\\
C_{(A)}^{mnp,k}&=&-24\left(
\frac1{X^3}\eta^{kn}\eta^{mp}- \frac3{X^5}\eta^{nk}X^m X^p
\right)\non\\&&
-\frac{24}{X^5}(X^m X^k \eta^{np}+X^k X^p \eta^{mn})
+\frac{24}{X^5}(\eta^{kp}X^m X^n+\eta^{km}X^n X^p)~.
\label{A40}
\eea
In contrast to (\ref{A36}), the tensor $C_{(A)}^{mnp,k}$ is not
symmetric and traceless in the indices $mnp$. However, when all
contributions (\ref{A33}) are taken into account, the resulting
tensor $C^{mnp,k}$ should obey the symmetry (\ref{A36}). Hence, it
is sufficient for the following to consider only symmetric in
indices $mnp$ part of (\ref{A40})
\bea
C_{(A)}^{(mnp),k}&=&-\frac{8}{X^3}(\eta^{nk}\eta^{mp}+\eta^{mk}\eta^{np}+\eta^{kp}\eta^{mn})
\non\\&&
+\frac{40}{X^5}(\eta^{kn}X^m X^p + \eta^{km}X^n X^p + \eta^{kp}X^m
X^n)\non\\&&
-\frac{16}{X^5}(X^k X^m \eta^{nr}+ X^k X^n \eta^{mp}+X^k X^p \eta^{mn} )
~.\label{C-A}
\eea
In the same way using the tensor (\ref{307}) we compute the
expression (\ref{B-1})--(\ref{B-2}),
\begin{subequations}
\bea
H_{(B_1)}^{mnk}&=&
\frac\ri2\gamma^n_{\rho\sigma}(\gamma^k)^{\gamma\gamma'}
\Big(
\Theta^1_{\gamma} \frac\partial{\partial\Theta^1_\rho} \frac\partial{\partial\Theta^3_\sigma}
+\Theta^2_{\gamma} \frac\partial{\partial\Theta^2_\rho} \frac\partial{\partial\Theta^3_\sigma}
\Big)H^{m}_{\gamma'}
=C_{(B_1)}^{mnp,k}(\Theta\Theta)_p~,\\
C_{(B_1)}^{mnp,k}&=&-\frac{12}{X^3}\eta^{mn}\eta^{kp}
+\frac{36}{X^5}X^m X^n \eta^{kp}\non\\&&
-\frac{12}{X^5}(X^k X^m \eta^{np}+X^k X^n \eta^{mp})
+\frac{12}{X^5}(\eta^{km}X^n X^p + \eta^{kn}X^m X^p)~,\label{A41b}
\eea
\end{subequations}
\begin{subequations}
\bea
H_{(B_2)}^{mnk}&=&-\frac\ri2 \gamma^n_{\rho\sigma} (\gamma^k)^{\gamma\gamma'}
 X_{3\mu\gamma}\frac\partial{\partial X_{3\sigma\mu}}\frac\partial{\partial\Theta^3_\rho}
H^m_{\gamma'}=C_{(B_2)}^{mnp,k}~,\\
C_{(B_2)}^{mnp,k}&=&\frac{6}{X^3} \eta^{nk}\eta^{mp}
+\frac{12}{X^3} (\eta^{mk}\eta^{np} + \eta^{mn}\eta^{kp} )
\non\\&&
-\frac{42}{X^5} \eta^{nk}X^m X^p
-\frac{78}{X^5} \eta^{km}X^n X^p
-\frac{30}{X^5} \eta^{kp}X^m X^n
\non\\&&
+\frac{30}{X^5} \eta^{mn}X^k X^p
+\frac{6}{X^5} X^m X^k \eta^{np}
+\frac{24}{X^5} X^n X^k \eta^{mp}~,\label{A41d}
\eea
\end{subequations}
\begin{subequations}
\bea
H_{(B_3)}^{mnk}&=&
-\frac\ri2 (\gamma^k)^{\gamma\gamma'} X_{3\mu\gamma}\frac\partial{\partial X_{3n}}\frac\partial{\partial\Theta^3_\mu}
H^{m}_{\gamma'}=C_{(B_3)}^{mnp,k}(\Theta\Theta)_p~,\\
C_{(B_3)}^{mnp,k}&=&-\frac{12}{X^3}\eta^{mk}\eta^{np}
+\frac{12}{X^5} \eta^{nk}X^m X^p
+\frac{48}{X^5} \eta^{mk}X^n X^p
\non\\&&
-\frac{18}{X^5}(\eta^{mp}X^k X^n + \eta^{mn}X^k X^p)
-\frac{6}{X^5} \eta^{np} X^k X^m
+\frac{30}{X^7}X^m X^n X^k X^p~,~~~~~~~~~~ \label{A41f}
\eea
\end{subequations}
We need only symmetric parts in the indices $mnp$ of the tensors
(\ref{A41b}), (\ref{A41d}) and (\ref{A41f})
\allowdisplaybreaks
\begin{subequations}
\label{A42}
\bea
C_{(B_1)}^{(mnp)k}&=&-\frac{4}{X^3}(\eta^{mn}\eta^{kp}+\eta^{kn}\eta^{mp}+\eta^{km}\eta^{np})
\non\\&&
+\frac{20}{X^5}(\eta^{kp}X^m X^n + \eta^{km}X^n X^p + \eta^{kn}X^m X^p)
\non\\&&
-\frac{8}{X^5}(X^k X^m \eta^{np}+X^kX^n \eta^{mp}+X^k
X^p\eta^{mn})~,\\
C_{(B_2)}^{(mnp)k}&=&\frac{10}{X^3}(\eta^{km}\eta^{np}+\eta^{kn}\eta^{mp}+\eta^{kp}\eta^{mn})
\non\\&&
-\frac{50}{X^5}(\eta^{kn}X^m X^p+\eta^{km}X^n X^p + \eta^{kp}X^m
X^p)\non\\&&
+\frac{20}{X^5}(X^k X^p \eta^{mn}+X^k X^n \eta^{mp}+X^k X^m
\eta^{np})~,\\
C_{(B_3)}^{(mnp)k}&=&-\frac{4}{X^3}(\eta^{km}\eta^{np}+\eta^{kn}\eta^{mp}+\eta^{kp}\eta^{mn})
\non\\&&
+\frac{20}{X^5}(\eta^{kn}X^m X^p+\eta^{km}X^n X^p + \eta^{kp}X^m
X^p)\non\\&&
-\frac{14}{X^5}(X^k X^p \eta^{mn}+X^k X^n \eta^{mp}+X^k X^m
\eta^{np})+\frac{30}{X^7}X^m X^n X^k X^p~.
\eea
\end{subequations}

The sum of (\ref{A40}) and (\ref{A42}) is
\bea
C^{mnp,k}&=&C_A^{(mnp)k}+C_{B_1}^{(mnp)k}+C_{B_2}^{(mnp)k}+C_{B_3}^{(mnp)k}\non\\
&=&-6d_{\cN=3}\bigg[
\frac1{X^3}
(\eta^{km}\eta^{np}+\eta^{kn}\eta^{mp}+\eta^{kp}\eta^{mn})\non\\&&
-\frac5{X^5}(\eta^{kp}X^m X^n + \eta^{kn}X^m X^p)
\non\\&&-\frac5{X^5}(\eta^{kp}X^m X^n+\eta^{kn}X^m X^p + \eta^{km}X^n X^p)
\non\\&&+\frac3{X^5}(X^k X^m\eta^{np}+X^kX^n \eta^{mp}+X^k X^p \eta^{mn})
-\frac5{X^7}X^m X^n X^k X^p
\bigg].
\eea
Substituting this tensor back into (\ref{A35}) we find the $\cN=2$
supercurrent correlation function in the form (\ref{n2.4})
where the parameter $d_{\cN=2}$ is related to $d_{\cN=3}$ as
\be
d_{\cN=2} = -6 d_{\cN=3}~.
\ee


\begin{footnotesize}

\end{footnotesize}


\begin{thebibliography}{66}

\bibitem{BKS}
  E.~I.~Buchbinder, S.~M.~Kuzenko, I.~B.~Samsonov,
  ``Superconformal field theory in three dimensions:
  Correlation functions of conserved currents,'' 
  JHEP {\bf 1506}, 138 (2015)
  [arXiv: 1503.04961 [hep-th]].


\bibitem{A-GF}
L.~Alvarez-Gaum\'e and D.~Z.~Freedman,
``Geometrical structure and ultraviolet finiteness in
the supersymmetric sigma model,''
Commun.\ Math.\ Phys.\  {\bf 80}, 443 (1981).


\bibitem{Bagger}
  J.~A. Bagger,
  ``Supersymmetric sigma models,''
Lectures given at the {1984 NATO Advanced Study Institute
on Supersymmetry}, (Bonn, Germany, August 1984); reprinted in
{\it Supergravities in Diverse Dimensions}, A. Salam and E. Sezgin (Eds.)
North-Holland/World Scientific, 1989, Vol. 1, pp. 569--611.

\bibitem{KPT-MvU}
S.~M.~Kuzenko, J.-H.~Park, G.~Tartaglino-Mazzucchelli and R.~Unge,
``Off-shell superconformal nonlinear sigma-models in three dimensions,''
JHEP {\bf 1101}, 146 (2011)
  [arXiv:1011.5727 [hep-th]].

\bibitem{Zupnik2010}
  B.~M.~Zupnik,
  ``Three-dimensional $\mathcal {N}=4 $ supersymmetry in harmonic $\mathcal {N}=3$ superspace,''
  Theor.\ Math.\ Phys.\  {\bf 165}, 1315 (2010)
  [Teor.\ Mat.\ Fiz.\  {\bf 165}, 97 (2010)]
  [arXiv:1005.4750 [hep-th]].

\bibitem{ZH}
B.~M.~Zupnik and D.~V.~Hetselius,
``Three-dimensional extended supersymmetry in harmonic superspace,''
Sov.\ J.\ Nucl.\ Phys.\  {\bf 47}, 730 (1988)
  [Yad.\ Fiz.\  {\bf 47}, 1147 (1988)].

\bibitem{KaoLee}
  H.~C.~Kao and K.~M.~Lee,
  ``Self-dual Chern-Simons systems with an N=3 extended supersymmetry,''
  Phys.\ Rev.\ D {\bf 46}, 4691 (1992)
  [hep-th/9205115].

\bibitem{Kao}
  H.~C.~Kao,
  ``Self-dual Yang-Mills Chern-Simons Higgs systems with an N=3 extended supersymmetry,''
  Phys.\ Rev.\ D {\bf 50}, 2881 (1994).

\bibitem{KN14}
S.~M.~Kuzenko and J.~Novak,
``Supergravity-matter actions in three dimensions and Chern-Simons terms,''
JHEP {\bf 1405}, 093 (2014) [arXiv:1401.2307 [hep-th]].


\bibitem{BrooksG}
  R.~Brooks and S.~J.~Gates Jr.,
  ``Extended supersymmetry and super-BF gauge theories,''
  Nucl.\ Phys.\  B {\bf 432}, 205 (1994)
  [arXiv:hep-th/9407147].


\bibitem{Zupnik98}
  B.~M.~Zupnik,
 ``Harmonic superspaces for three-dimensional theories,''
 in: {\it Supersymmetries and Quantum Symmetries}, J. Wess and E. Ivanov (Eds.),
 Springer, Berlin, 1999, pp. 116--123, arXiv:hep-th/9804167.

 \bibitem{Zupnik99}
 B.~Zupnik,
 ``Harmonic superpotentials and symmetries in gauge theories
with eight  supercharges,''
  Nucl.\ Phys.\ B {\bf 554},  365 (1999)
  [Erratum-ibid.\ B {\bf 644},  405  (2002)]
{[hep-th/9902038]}.

\bibitem{GW}
  D.~Gaiotto and E.~Witten,
  ``Janus configurations, Chern-Simons couplings,
  and the $\theta$-angle in N=4 super Yang-Mills theory,''
  JHEP {\bf 1006}, 097 (2010)
  [arXiv:0804.2907 [hep-th]].

\bibitem{HLLLP}
  K.~Hosomichi, K.~M.~Lee, S.~Lee, S.~Lee and J.~Park,
  ``N=4 superconformal Chern-Simons theories with hyper and twisted hyper multiplets,''
  JHEP {\bf 0807}, 091 (2008)
  [arXiv:0805.3662 [hep-th]].

\bibitem{ABJM}
  O.~Aharony, O.~Bergman, D.~L.~Jafferis and J.~Maldacena,
  ``N=6 superconformal Chern-Simons-matter theories, M2-branes and their gravity duals,''
  JHEP {\bf 0810}, 091 (2008)
  [arXiv:0806.1218 [hep-th]].

\bibitem{HLLLP2} 
  K.~Hosomichi, K.~M.~Lee, S.~Lee, S.~Lee and J.~Park,
  ``N=5,6 superconformal Chern-Simons theories and M2-branes on orbifolds,''
  JHEP {\bf 0809}, 002 (2008)
  [arXiv:0806.4977 [hep-th]].


\bibitem{ABJ}
  O.~Aharony, O.~Bergman and D.~L.~Jafferis,
  ``Fractional M2-branes,''
  JHEP {\bf 0811}, 043 (2008)
  [arXiv:0807.4924 [hep-th]].


\bibitem{BLG1}
  J.~Bagger and N.~Lambert,
  ``Modeling Multiple M2's,''
  Phys.\ Rev.\ D {\bf 75}, 045020 (2007)
  [hep-th/0611108].

\bibitem{BLG2}
  J.~Bagger and N.~Lambert,
  ``Gauge symmetry and supersymmetry of multiple M2-branes,''
  Phys.\ Rev.\ D {\bf 77}, 065008 (2008)
  [arXiv:0711.0955 [hep-th]].

\bibitem{Gustavsson}
A.~Gustavsson,
``Algebraic structures on parallel M2-branes,''
Nucl.\ Phys.\ B {\bf 811}, 66 (2009)
[arXiv:0709.1260 [hep-th]].

\bibitem{Park3}
J.-H.~Park, ``Superconformal symmetry in three dimensions,''
  J.\ Math.\ Phys.\  {\bf 41}, 7129 (2000) [arXiv:hep-th/9910199].

\bibitem{MackSalam}
  G.~Mack and A.~Salam,
  ``Finite component field representations of the conformal group,''
  Annals Phys.\  {\bf 53}, 174 (1969).

 \bibitem{OP}
H.~Osborn and A.~C.~Petkou,
``Implications of conformal invariance in field theories for general dimensions,''
  Annals Phys.\  {\bf 231}, 311 (1994)
  [hep-th/9307010].

\bibitem{Mack}
  G.~Mack, ``Convergence of operator product expansions on the vacuum
  in conformal invariant quantum field theory,''
  Commun.\ Math.\ Phys.\  {\bf 53}, 155 (1977).


\bibitem{FGG}
S.~Ferrara, A.~F.~Grillo and R.~Gatto,
  ``Manifestly conformal-covariant expansion on the light cone,''
  Phys.\ Rev.\ D {\bf 5}, 3102 (1972);
  ``Tensor representations of conformal algebra and conformally covariant operator product expansion,''
  Annals Phys.\  {\bf 76}, 161 (1973).

\bibitem{Koller}
  K.~Koller, ``The significance of conformal inversion in quantum field theory,''
 Commun. Math. Phys. {\bf 40}, 15 (1975).

\bibitem{TMP} I. T. Todorov, M. C. Mintchev and V. P. Petkova,
{\it Conformal Invariance in Quantum Field Theory},
Pisa, Scuola Normale Superiore, 1978.

\bibitem{FP}
  E.~S.~Fradkin and M.~Y.~Palchik,
  ``Recent developments in conformal invariant quantum field theory,''
  Phys.\ Rept.\  {\bf 44}, 249 (1978).

\bibitem{Stanev} 
  Y.~S.~Stanev,
``Stress-energy tensor and U(1) current operator product expansions in conformal QFT,''   Bulg.\ J.\ Phys.\  {\bf 15}, 93 (1988).

\bibitem{Osborn}
H.~Osborn,   ``N=1 superconformal symmetry in four-dimensional
 quantum field theory,'' Annals Phys.\  {\bf 272}, 243 (1999) [hep-th/9808041].

\bibitem{Park6}
J.-H.~Park,
``Superconformal symmetry in six-dimensions and its reduction to four,''
Nucl.\ Phys.\ B {\bf 539}, 599 (1999) [hep-th/9807186].

\bibitem{Park4}
 J.-H.~Park,  ``Superconformal symmetry and correlation functions,''
 Nucl.\ Phys.\ B {\bf 559}, 455 (1999) [hep-th/9903230].


\bibitem{KLT-M11}
S.~M.~Kuzenko, U.~Lindstr\"om and G.~Tartaglino-Mazzucchelli,
  ``Off-shell supergravity-matter couplings in three dimensions,''
  JHEP {\bf 1103}, 120 (2011)
  [arXiv:1101.4013 [hep-th]].


\bibitem{KT}
S.~M.~Kuzenko and S.~Theisen,
``Correlation functions of conserved currents in N = 2 superconformal
theory,''  Class.\ Quant.\ Grav.\  {\bf 17}, 665 (2000)  [hep-th/9907107].

\bibitem{BKT-M} 
  D.~Butter, S.~M.~Kuzenko and G.~Tartaglino-Mazzucchelli,
  ``Nonlinear sigma models with AdS supersymmetry in three dimensions,''
  JHEP {\bf 1302}, 121 (2013)
  [arXiv:1210.5906 [hep-th]].

\bibitem{BKNT-M}
  D.~Butter, S.~M.~Kuzenko, J.~Novak and G.~Tartaglino-Mazzucchelli,
  ``Conformal supergravity in three dimensions: New off-shell formulation,''
  JHEP {\bf 1309}, 072 (2013)
[arXiv:1305.3132 [hep-th]].

\bibitem{KNT-M}
  S.~M.~Kuzenko, J.~Novak and G.~Tartaglino-Mazzucchelli,
  ``N=6 superconformal gravity in three dimensions from superspace,''
  JHEP {\bf 1401}, 121 (2014)
  [arXiv:1308.5552 [hep-th]].


\bibitem{Sohnius}
  M.~F.~Sohnius, ``Supersymmetry and central charges,''
  Nucl.\ Phys.\  B {\bf 138}, 109 (1978).

\bibitem{Zupnik2009}
  B.~M.~Zupnik,
  ``Three-dimensional N=4 superconformal superfield theories,''
  Theor.\ Math.\ Phys.\  {\bf 162}, 74 (2010)
  [arXiv:0905.1179 [hep-th]].

\bibitem{GGRS}
S. J.~Gates Jr., M.~T.~Grisaru, M.~Ro\v{c}ek and W.~Siegel,
{\it Superspace, or One Thousand and One Lessons in Supersymmetry},
Benjamin/Cummings (Reading, MA),  1983, hep-th/0108200.  

\bibitem{HKLR}
N.~J.~Hitchin, A.~Karlhede, U.~Lindstr\"om and M.~Ro\v cek,
``Hyperk\"ahler metrics and supersymmetry,''
Commun.\ Math.\ Phys.\  {\bf 108}, 535 (1987).  

\bibitem{ZP}
B.~M.~Zupnik and D.~G.~Pak,
``Superfield formulation of the simplest three-dimensional gauge theories and
conformal supergravities,''  Theor.\ Math.\ Phys.\  {\bf 77}, 1070 (1988) 
[Teor.\ Mat.\ Fiz.\  {\bf 77}, 97 (1988)].


\bibitem{GIKOS}
  A.~Galperin, E.~Ivanov, S.~Kalitzin, V.~Ogievetsky and E.~Sokatchev,
  ``Unconstrained N=2 matter, Yang-Mills and supergravity theories in harmonic
  superspace,''
  Class.\ Quant.\ Grav.\  {\bf 1}, 469 (1984).


\bibitem{GIOS}
A.~S.~Galperin, E.~A.~Ivanov, V.~I.~Ogievetsky and E.~S.~Sokatchev,
{\it Harmonic Superspace}, Cambridge University Press,  2001.


\bibitem{KLR}
A. Karlhede, U. Lindstr\"om and M. Ro\v cek,
``Self-interacting tensor multiplets in N = 2 superspace,''
Phys.\ Lett.\ B {\bf 147}, 297 (1984).

\bibitem{LR}
U.~Lindstr\"om and M.~Ro\v{c}ek,
``New hyperk\"ahler  metrics  and new supermultiplets,''
  Commun.\ Math.\ Phys.\  {\bf 115}, 21 (1988);
 ``N = 2 super Yang-Mills theory in projective superspace,''
Commun.\ Math.\ Phys.\  {\bf 128}, 191 (1990).
 
\bibitem{K-Lectures}
  S.~M.~Kuzenko,
  ``Lectures on nonlinear sigma-models in projective superspace,''
  J.\ Phys.\ A {\bf 43}, 443001 (2010)
  [arXiv:1004.0880 [hep-th]].


\bibitem{BILPSZ}
  I.~L.~Buchbinder, E.~A.~Ivanov, O.~Lechtenfeld, N.~G.~Pletnev, I.~B.~Samsonov and B.~M.~Zupnik,
  ``Quantum N=3, d=3 Chern-Simons matter theories in harmonic superspace,''
  JHEP {\bf 0910},  075 (2009) 
  [arXiv:0909.2970 [hep-th]].


\bibitem{Buchbinder:2008vi}
  I.~L.~Buchbinder, E.~A.~Ivanov, O.~Lechtenfeld, N.~G.~Pletnev, I.~B.~Samsonov and B.~M.~Zupnik,
  ``ABJM models in N=3 harmonic superspace,''
  JHEP {\bf 0903}, 096 (2009)
  [arXiv:0811.4774 [hep-th]].

\bibitem{BKNT-M2}
D.~Butter, S.~M.~Kuzenko, J.~Novak and G.~Tartaglino-Mazzucchelli,
``Conformal supergravity in three dimensions: Off-shell actions,''
JHEP {\bf 1310}, 073 (2013)
  [arXiv:1306.1205 [hep-th]].



\bibitem{K-double}
S.~M.~Kuzenko,
``Projective superspace as a double-punctured harmonic superspace,''
Int.\ J.\ Mod.\ Phys.\  A {\bf 14}, 1737 (1999) [hep-th/9806147].


\bi{RS} A.~A.~Rosly and A.~S.~Schwarz,
  ``Supersymmetry in a space with auxiliary dimensions,''
 Commun.\ Math.\ Phys.\  {\bf 105}, 645 (1986).

\bibitem{Rosly}
 A.~A.~Rosly,
``Super Yang-Mills  constraints as integrability conditions,''
in {\it Proceedings of the International
Seminar on Group Theoretical Methods in Physics},'' (Zvenigorod, USSR, 1982),
M. A. Markov  (Ed.), Nauka, Moscow, 1983, Vol. 1, p. 263 (in Russian);
English translation: in {\it Group Theoretical 
Methods in Physics},'' M. A. Markov, V. I. Man'ko 
and A. E. Shabad  (Eds.), Harwood Academic Publishers, 
London, Vol. 3, 1987, pp. 587--593.



\bibitem{GIOS1}
  A.~Galperin, E.~A.~Ivanov, V.~Ogievetsky and E.~Sokatchev,
  ``Harmonic supergraphs. Green functions,''
  Class.\ Quant.\ Grav.\  {\bf 2}, 601 (1985).


\bibitem{Sokatchev}
  E.~Sokatchev,
  ``3-point functions in N=4 Yang-Mills in harmonic superspace,''
  in {\it Supersymmetries and quantum symmetries}, J. Wess and
  E. A. Ivanov (Eds.),  Lecture Notes in Physics 524,
Springer, Berlin,   1999,  pp. 106--115.

 


\end{thebibliography}
\end{document}